\documentclass{aa}

\usepackage{graphicx}
\usepackage{subcaption}
\usepackage{mwe}

\usepackage{txfonts}
\usepackage{hyperref}
\usepackage{url}
\usepackage{xcolor}
\hypersetup{
colorlinks,
linkcolor={red!80!black},
citecolor={blue!80!black},
urlcolor={blue!80!black}
}

\newenvironment{myfont}{\fontfamily{ppl}\selectfont}{\par}
\DeclareTextFontCommand{\textmyfont}{\myfont}

\newcommand{\code}[1]{\texttt{#1}}

\def\nifs{\iso{56}Ni}

\def\cm3{cm$^{-3}$}
\def\kms{\mbox{km~s$^{-1}$}}
\def\msunyr{$M_{\odot}$\,yr$^{-1}$}
\def\lsun{$L_{\odot}$}
\def\rsun{$R_{\odot}$}
\def\mdot{$\dot{M}$}

\def\msun{$M_{\odot}$}

\def\one{\ts {\,\sc i}}
\def\two{\ts {\,\sc ii}}
\def\three{\ts {\,\sc iii}}
\def\four{\ts {\,\sc iv}}
\def\five{\ts {\sc v}}
\def\six{\ts {\sc vi}}
\def\sev{\ts {\sc vii}}
\def\beq{\begin{equation}}
\def\eeq{\end{equation}}

\def\lesssim{\mathrel{\hbox{\rlap{\hbox{\lower4pt\hbox{$\sim$}}}\hbox{$<$}}}}
\def\gtrsim{\mathrel{\hbox{\rlap{\hbox{\lower4pt\hbox{$\sim$}}}\hbox{$>$}}}}

\def\one{{\,\sc i}}
\def\two{{\,\sc ii}}
\def\three{{\,\sc iii}}
\def\four{{\,\sc iv}}
\def\five{{\sc v}}
\def\six{{\sc vi}}
\def\sev{{\sc vii}}

\def\v1d{{\code{V1D}}}

\def\mesa{{\code{MESA}}}
\def\cmfgen{{\code{CMFGEN}}}
\def\heracles{{\code{HERACLES}}}
\def\stella{{\code{STELLA}}}

\def\ergs{erg\,s$^{-1}$}

\def\bw{$\beta_{\rm w}$}
\def\vinf{$V_\infty$}
\def\rcsm{$R_{\rm CSM}$}
\newcommand{\iso}[2]{\ensuremath{^{#1}\rm{#2}}}

\begin{document}

   \title{Probing red-supergiant atmospheres and winds with early-time high-cadence high-resolution type II supernova spectra}
   \titlerunning{High-resolution spectra of SNe II}

   \author{
   Luc Dessart\inst{1}
}

   \institute{
 Institut d'Astrophysique de Paris, CNRS-Sorbonne Universit\'e, 98 bis boulevard Arago, F-75014 Paris, France\\
 \email{dessart@iap.fr}
             }

   \date{}

  \abstract{High-cadence high-resolution spectroscopic observations of infant Type II supernovae (SNe) represent an exquisite probe of the atmospheres and winds of exploding red-supergiant (RSG) stars. Using radiation hydrodynamics and radiative transfer calculations, we study the gas and radiation properties during and after the phase of shock breakout, considering RSG star progenitors enshrouded within a circumstellar material (CSM) that varies in extent, density, and velocity profile. In all cases, the original, unadulterated CSM structure is probed only at the onset of shock breakout, visible in high-resolution spectra as narrow, often blueshifted emission components, possibly with an additional absorption trough. As the SN luminosity rises during breakout, radiative acceleration of the unshocked CSM starts, leading to a broadening of the ``narrow'' lines by several 100 and up to several 1000\,\kms, depending on the CSM properties. This acceleration is maximum close to the shock, where the radiative flux is greater, and thus typically masked by optical-depth effects. Generally, narrow-line broadening is greater for more compact, tenuous CSM because of the proximity to the shock where the flux is born, and smaller in denser and more extended CSM. Narrow-line emission should show a broadening that slowly increases first (the line forms further out in the original wind), then sharply rises (the line forms in a region that is radiatively accelerated), before decreasing until late times (the line forms further away in regions more weakly accelerated). Radiative acceleration should inhibit X-ray emission during the early, IIn phase. Although high spectral resolution is critical at the earliest times to probe the original slow wind, radiative acceleration and associated line broadening may be captured with medium resolution allowing a simultaneous view of narrow, Doppler-broadened as well as extended, electron-scattering broadened line emission.
  }

   \keywords{Radiative transfer -- Hydrodynamics -- supernovae : general
               }

   \maketitle


\section{Introduction}
\label{sect_intro}

High-cadence observations of young Type II supernovae (SNe) may hold critical information on the environment of their red-supergiant progenitors. With sufficient resolution, early-time spectra can reveal the slow velocity of this atmospheric material, as documented for SN\,1993J \citep{fransson_93j_96} or in exploding stars surrounded by dense and extended circumstellar material (CSM; \citealt{leonard_98S_00}; \citealt{fassia_98S_01}; \citealt{shivvers_98S_15}). This possibility has become more routinely exploited with the high cadence of modern sky surveys, allowing the discovery of SNe during the first day after shock emergence, starting with objects like SN\,2013fs \citep{yaron_13fs_17} but extended recently to larger samples \citep{bruch_csm_21,bruch_csm_22,wynn_pap2_24}. Events particularly noteworthy having such early-time observations with moderate or high spectral resolution include SN\,1998S \citep{shivvers_98S_15}, SN\,2020pni \citep{terreran_20pni_22}, SN\,2023ixf \citep{jacobson_galan_23ixf_23,bostroem_23ixf_24,smith_23ixf_23}, or SN\,2024ggi \citep{wynn_24ggi_24,pessi_24ggi_24,shrestha_24ggi_24,zhang_24ggi_24}. These observations suggest that narrow absorption and emission may be present in lines like H$\alpha$ for weeks after explosion (all examples above) but typically with values of order 100--1000\,\kms\ and thus much greater than the Doppler width of order 50\,\kms\ expected for a supergiant outflow.

It was already suggested a long time ago, in the case of SN\,1993J, that the unshocked CSM could be significantly accelerated by the SN radiation following shock breakout from the progenitor surface \citep{fransson_93j_96}. This phenomenon was confirmed in the context of SN\,1998S by \citet{chugai_98S_02}. A larger sample of Type II SNe with evidence for interaction were found to exhibit suspiciously large CSM velocities \citep{boian_groh_20}. Detailed radiation hydrodynamics calculations of SNe interacting with CSM routinely predict the presence of such radiative acceleration (see, e.g., \citealt{moriya_rsg_csm_11}; \citealt{D15_2n}). \citet{tsuna_grad_23} studied this process more specifically in the context of erupting massive stars immediately before core collapse, finding that this radiative acceleration can lead to CSM velocities of order 1000\,\kms\ for CSM masses of order 0.1\,\msun.

In this work, we complement these previous studies by carrying out calculations for Type II SNe exploding within CSM, employing the configurations already studied and that demonstrated a fair agreement with SNe II-P/CSM. These are the models published by \citet{d17_13fs}, \citet{dessart_wynn_23}, \citet{jacobson_galan_20tlf_22,jacobson_galan_23ixf_23,wynn_24ggi_24} for SNe 2013fs, 2020tlf, 2023ixf, or 2024ggi. Thus, we combine radiation hydrodynamics calculations and post-treatment with radiative transfer to document how the radiative acceleration of the unshocked CSM affects the morphology of narrow emission line components. This work thus complements the previous study of \citet{boian_groh_20}, who carried out radiative transfer calculations of SNe IIP/CSM, or the study of \citet{tsuna_grad_23} that focused on the radiation hydrodynamics of these interactions. Our simulations are still 1D and steady-state so they might exacerbate some features absent from turbulent, asymmetric CSM around red-supergiant stars. In some sense, this can help the interpretation since distinct regions leave clearer signatures in the emergent spectra. The model spectra also have infinite signal-to-noise ratio, allowing us to inspect flux variations down to the 1\% level. This is something typically not possible in observations, especially in echelle spectra and their wavelength-restricted orders.

In the next section, we describe the numerical setup for our simulations, including the progenitor models used and the CSM properties covered, the radiation-hydrodynamics calculations performed, and the post-treatment with radiative transfer. We then study in detail one model in Section~\ref{sect_ref}, describing first the properties of the unshocked CSM and how the SN radiation affects its dynamical properties (Section~\ref{sect_ref_grad}), and then describing the spectral properties of that model at multiple epochs from a fraction of a day to 15\,d (Section~\ref{sect_ref_lines}). We focus primarily on the properties at small velocity scales in order to connect to the information obtained in high-resolution spectral observations. We then discuss the dependency of these results for different CSM properties, covering a range of CSM extent (Section~\ref{sect_dep_rcsm}), density (Section~\ref{sect_dep_mdot}), and velocity profiles (Section~\ref{sect_dep_wind}).  In Section~\ref{sect_broad_vs_narrow}, we connect the evolution of spectral lines at small and large velocity scales, to confront the spectrum formation from regions below the photosphere (where Doppler and electron-scattering broadening operate) and above the photosphere (where only Doppler broadening acts, although at a very small level due to the relatively small velocities). Finally, in Section~\ref{sect_conc}, we discuss the implications of our results and present our conclusions.\footnote{All simulations in this work will be uploaded at \url{https://zenodo.org/communities/snrt}.}

\section{Numerical setup}
\label{sect_setup}

The present study uses a similar methodology as that used in \citet{D15_2n}, \citet{d17_13fs}, and more recently \citet{dessart_wynn_23}. We take a massive star model at the onset of core collapse (the preSN evolution is computed with \mesa; \citealt{mesa1,mesa2,mesa3,mesa4}), simulate the explosion with the 1D Lagrangian hydrodynamics code \v1d\ \citep{livne_93,dlw10b,dlw10a}, remap this exploding star into the radiation-hydrodynamics code \heracles\ \citep{gonzalez_heracles_07,vaytet_mg_11} just before the shock reaches $R_\star$, stitch some CSM atop $R_\star$, and evolve that structure for 15\,d of physical time. The goal of this work is to investigate how the presence of CSM may affect the outcome of the explosion. In particular, we are interested in the morphology and the evolution of the narrow emission lines that are revealed by high-resolution spectra (see, e.g., \citealt{shivvers_98S_15}; \citealt{smith_23ixf_23}; \citealt{pessi_24ggi_24}). Thus, at selected epochs, we post-process snapshots from the \heracles\ simulation with the 1D nonlocal thermodynamic equilibrium radiative transfer code \cmfgen\ \citep{HD12,D15_2n}, assuming steady state but treating accurately the complicated, nonmonotonic velocity structure. Additional details are provided in the following subsections such as the initial model, the CSM properties used, or the numerical setup with \cmfgen.

\subsection{Initial conditions}
\label{sect_init}

Our progenitor model was a 15\,\msun\ star on the zero-age main sequence, evolved with \mesa\ (version 10108) at solar metallicity and without rotation. This star reaches core collapse as a red supergiant star with an effective temperature $T_{\rm eff} = 3980$\,K, a surface bolometric luminosity $L_\star = 9.6 \times 10^{4}$\,\lsun, a surface radius $R_\star = 652$\,\rsun, and a surface composition for H, He, C, N, and O given by mass fractions $X_{\rm H}=$\,0.6987, $X_{\rm He}=$\,0.287, $X_{\rm C}=$\,1.52\,$\times$\,10$^{-3}$, $X_{\rm N}=$\,2.28\,$\times$\,10$^{-3}$, and $X_{\rm O}=$\,5.94\,$\times$\,10$^{-3}$. The explosion is performed with \v1d\ by depositing $1.47 \times 10^{51}$\,erg within 0.05\,\msun\ of a mass cut at  1.65\,\msun\ and at a constant rate during 0.1\,s. Corrected for the envelope binding energy to be shocked, the (asymptotic) total energy is $1.2 \times 10^{51}$\,erg.

This \v1d\ structure (radius, velocity, density, temperature, and composition) is remapped into \heracles\ when the shock is within a few 10$^{12}$\,cm below $R_\star$. This corresponds to a time of 1.04\,d since the explosion trigger. Since \heracles\ is a Eulerian code, we extend the original \v1d\ grid out to a maximum radius $R_{\rm max}=$\,$4 \times 10^{15}$\,cm so that we can track for 15\,d the evolution of the ejecta as they expand. The \heracles\ grid employs 14336 radial grid points, with a constant spacing out to  $5 \times 10^{14}$\,cm and a constant spacing in the log beyond. Five tracer particles are used to track the location of the H, He, O, Si, and \nifs-rich shells but because the preSN model has a massive H-rich envelope, the composition is essentially uniform (and given by that at the surface of the preSN star) in the spectrum formation region at all times studied here.

\begin{figure}
\centering
\includegraphics[width=\hsize]{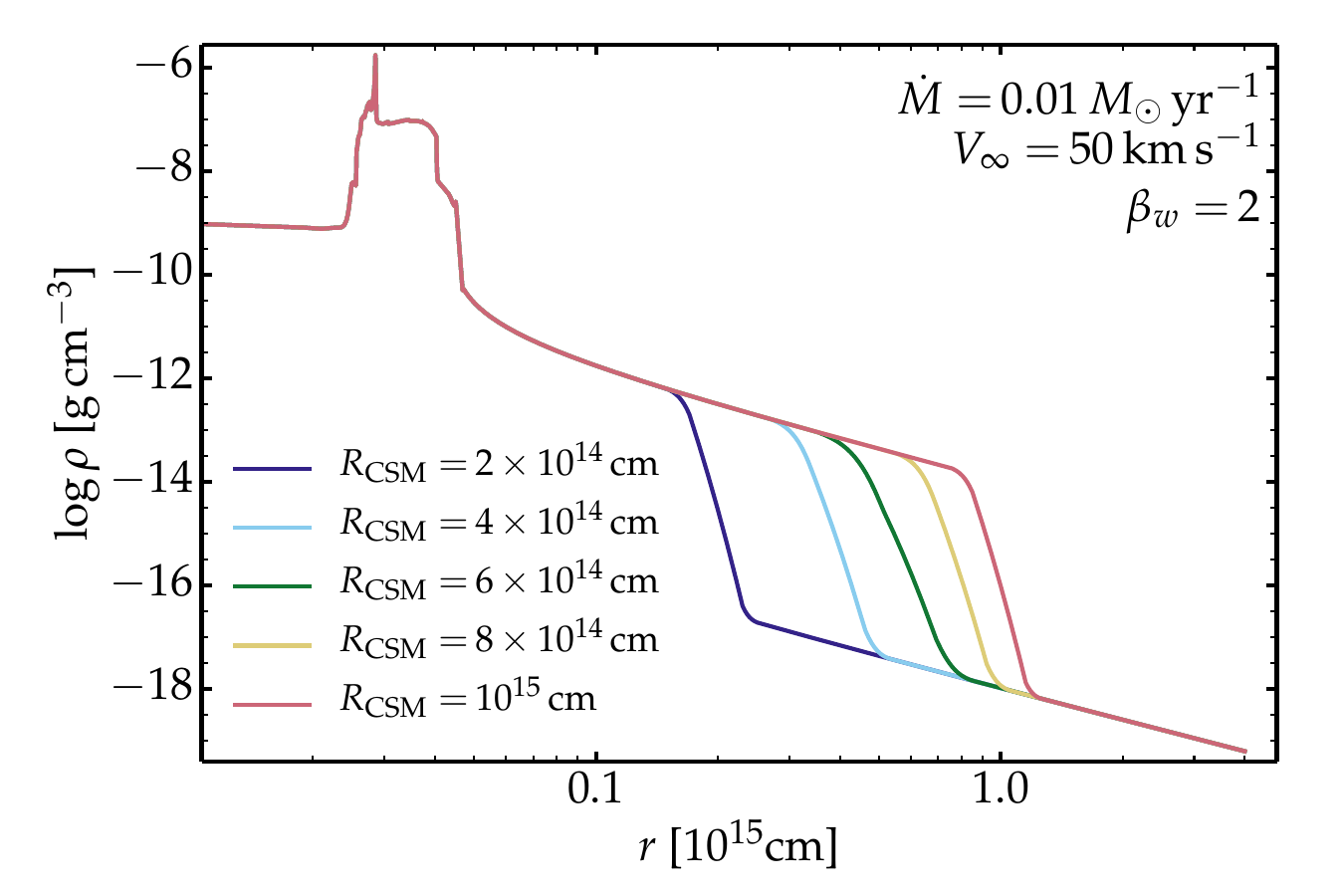}
\includegraphics[width=\hsize]{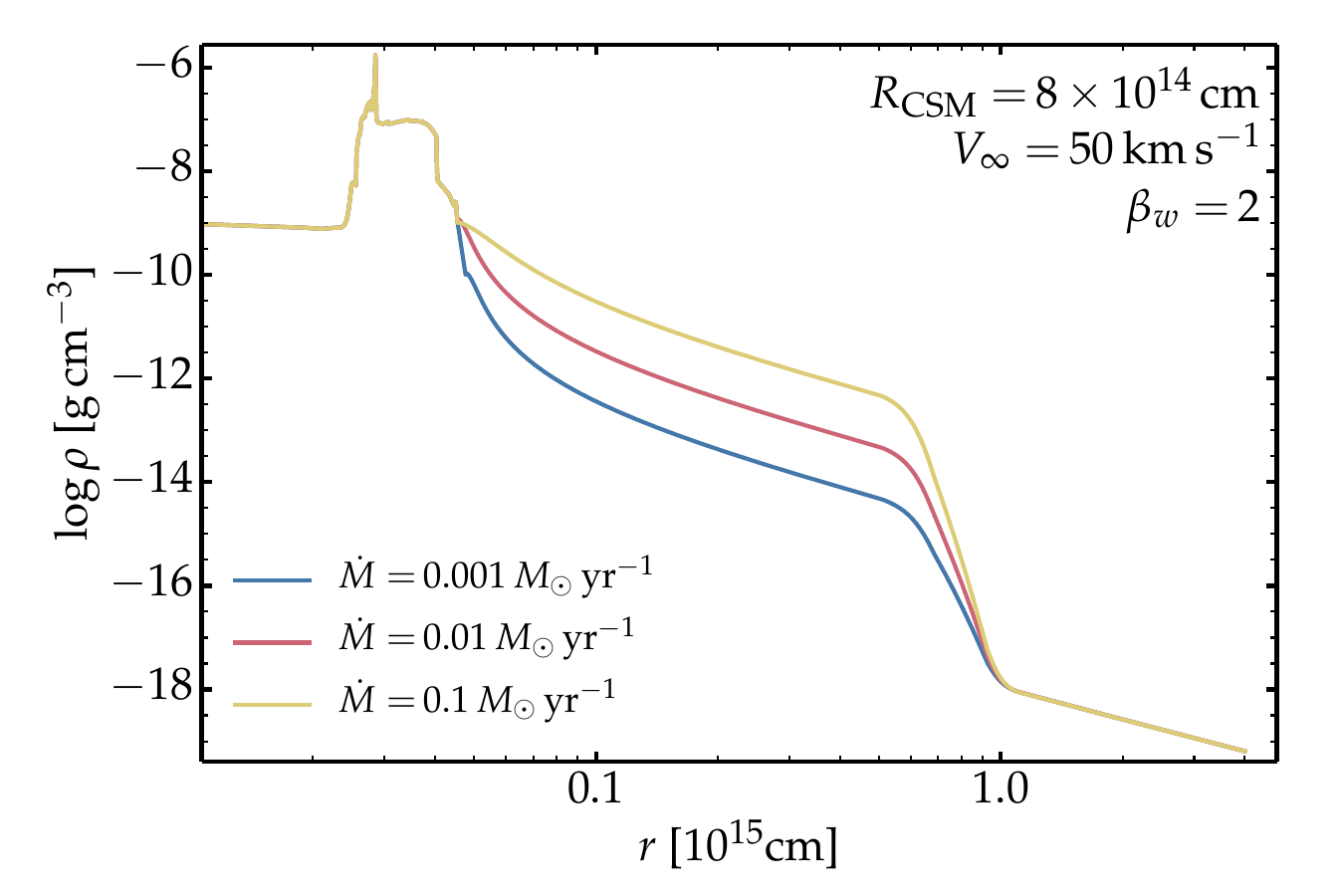}
\includegraphics[width=\hsize]{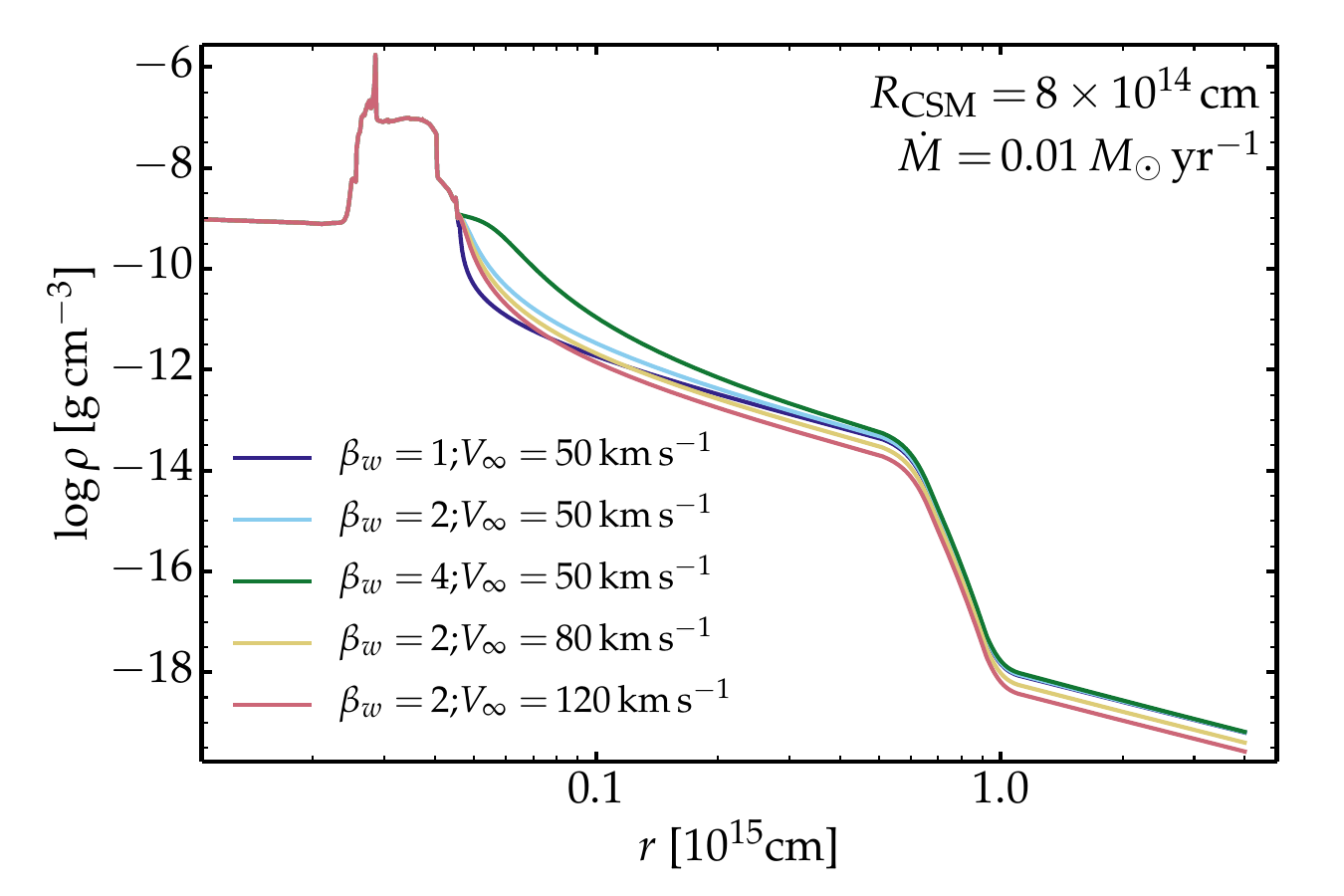}
\caption{Radial profiles of the mass density used as initial conditions for our radiation-hydrodynamics calculations. From top to bottom, we show groups of models varying in \rcsm, \mdot, and wind velocity profile (i.e., \bw\ and \vinf). Radial profiles of the velocity are shown for the last model group in the appendix, in Fig.~\ref{fig_init_wind_vel}. [See Section~\ref{sect_init} and Table~\ref{tab_init} for additional information.]
\label{fig_init}
}
\end{figure}

In this spherical volume between $R_\star$ and $R_{\rm max}$, we introduce some CSM of diverse properties (Fig.~\ref{fig_init}). For simplicity, the CSM density corresponds to that of two steady-state wind mass loss rates, chosen in the range 0.001 to 0.1\,\msunyr\ within a radius \rcsm\ and 10$^{-6}$\,\msunyr\ beyond -- the transition between the two regions occurs over a length scale of a few 10$^{14}$\,cm. This prescribed density is chosen to reflect a phase of high mass loss in the years or decades prior to core collapse and is crafted rather than physically consistent (we make no claim on the physical origin of this mass loss, which is a separate issue). For a given mass loss rate value \mdot, the wind density is given by $\rho = \dot{M} / 4 \pi r^2 V(r)$, and thus depends on the velocity profile. In all cases, the wind velocity profile is parameterized by a ``hydrostatic'', base velocity $V_0$ at $R_\star$, a terminal or asymptotic velocity \vinf, and an exponent \bw\ controling the acceleration length scale, such that  $V(r) = V_0 + (V_\infty-V_0)(1-R_\star/r)^{\beta_{\rm w}}$. The value of $V_0$ is chosen so that the density connects smoothly at $R_\star$. Obviously, by assuming a progressive acceleration of the CSM material with radius, the CSM density above $R_\star$ is much greater than for a prompt acceleration to \vinf, by as much as a factor of order a thousand. 

In Section~\ref{sect_ref}, we describe results from radiation-hydrodynamics and radiative-transfer calculations for one model, our reference, which corresponds to a dense CSM with \mdot$=$\,0.01\,\msunyr, \bw$=$\,2,  \vinf$=$\,50\,\kms, and \rcsm$= 8 \times 10^{14}$\,cm. We also explore other models that vary in the extent of the dense CSM (\rcsm\ between $2 \times 10^{14}$ and 10$^{15}$\,cm), in mass loss rate (\mdot\ of 0.001 and 0.1\,\msunyr), or in wind velocity profile (\bw\ of 1, 2, and 4; \vinf\ of 20, 50, 80, and 120\,\kms), as summarized in Table~\ref{tab_init}. This table also gives the CSM mass, optical depth (assuming an opacity of 0.34\,cm$^2$\,g$^{-1}$ representative of an ionized solar-metallicity gas), and the photospheric radius that would result for that same opacity. Prior to shock breakout, this CSM is cold and optically thin to electron scattering (we ignore any presence of dust or molecules in that nearby CSM -- it should be promptly destroyed by the SN radiation). 

In Fig.~\ref{fig_init}, the labels give the explicit value for these various model parameters but in the text we will also use a more compact nomenclature. For example, model mdot0p01/rcsm8e14/bw2/vinf5e6 stands for a model whose initial CSM properties correspond to a high mass-loss rate value of 0.01\,\msunyr, \rcsm\ of $8 \times 10^{14}$\,cm, $\beta_{\rm w}=$\,2, and $V_\infty=$\,50\,\kms. When comparing models that differ in only one parameter, we will refer to these models through that varying parameter only, omitting parameters that are kept fixed among models in the sample. For example, for those differing in \mdot\ only, we will refer to models mdot0p001, mdot0p01, and mdot0p1 (see Table~\ref{tab_init}).

\begin{table*}
  \caption{Summary of initial conditions for our radiation-hydrodynamics calculations. Three groups of models are explored corresponding to variations in \rcsm, \mdot, and wind velocity profile.
\label{tab_init}
}
\begin{center}
\begin{tabular}{
l@{\hspace{5mm}}c@{\hspace{5mm}}c@{\hspace{5mm}}c@{\hspace{5mm}}c@{\hspace{5mm}}c@{\hspace{5mm}}c@{\hspace{5mm}}c@{\hspace{5mm}}c@{\hspace{5mm}}
}
\hline
Model name                         &      \mdot  & \rcsm\          &   \bw\   & \vinf\ & $M_{\rm CSM}$  & $\tau_{\rm CSM}$    & $R_{\rm phot}$ &   \cmfgen\ \\
                                   &   [\msunyr] & [10$^{14}$\,cm]  &          & [\kms] & [\msun]      &         & [10$^{14}$\,cm]    &  \\
\hline
mdot0p01/rcsm2e14                  & 0.01        &  2               &    2     & 50  &  0.0231  &   322.7  &    1.68   & Yes \\
mdot0p01/rcsm4e14                  & 0.01        &  4               &    2     & 50  &  0.0347  &   334.5  &    3.06   & Yes\\
mdot0p01/rcsm6e14                  & 0.01        &  6               &    2     & 50  &  0.0435  &   337.9  &    4.14   & Yes\\
mdot0p01/rcsm8e14                  & 0.01        &  8               &    2     & 50  &  0.0576  &   340.6  &    5.78   & Yes\\
mdot0p01/rcsm1e15                  & 0.01        & 10               &    2     & 50  &  0.0710  &   342.0  &    7.32   & Yes\\
\hline                                                                                                               
mdot0p001/rcsm8e14/bw2/vinf5e6     & 0.001       &  8               &    2     & 50  &  0.0249  &   447.0  &    3.19   & Yes \\
mdot0p01/rcsm8e14/bw2/vinf5e6      & 0.01        &  8               &    2     & 50  &  0.1155  &  1313.8  &    5.53   & No \\
mdot0p1/rcsm8e14/bw2/vinf5e6       & 0.1         &  8               &    2     & 50  &  0.7187  &  4424.5  &    6.39   & Yes\\
\hline                                                                                                               
mdot0p01/rcsm8e14/bw1/vinf5e6      & 0.01        &  8               &    1     & 50  &  0.0634  &   515.7  &    5.50   & No \\
mdot0p01/rcsm8e14/bw2/vinf5e6      & 0.01        &  8               &    2     & 50  &  0.1155  &  1313.8  &    5.53   & No \\
mdot0p01/rcsm8e14/bw4/vinf5e6      & 0.01        &  8               &    4     & 50  &  0.3063  &  3606.9  &    5.63   & No \\
mdot0p01/rcsm8e14/bw2/vinf2e6      & 0.01        &  8               &    2     & 20  &  0.2332  &  2049.2  &    6.02   & Yes\\
mdot0p01/rcsm8e14/bw2/vinf8e6      & 0.01        &  8               &    2     & 80  &  0.0813  &  1032.6  &    5.22   & No \\
mdot0p01/rcsm8e14/bw2/vinf1p2e7    & 0.01        &  8               &    2     & 120 &  0.0612  &   855.5  &    4.90   & Yes \\
\hline
\end{tabular}
\end{center}
    {\bf Notes:} All models correspond to the explosion of a 15\,\msun\ star initially, evolved to core collapse at solar metallicity, and exploded to yield an ejecta with a kinetic energy of $1.2 \times 10^{51}$\,erg. The mass loss rate indicated in this table refers to the high-density CSM at the surface of the star, which drops rapidly to 10$^{-6}$\,\msunyr\ beyond \rcsm\ (see Fig.~\ref{fig_init}). The wind velocity profile is characterized by parameters \bw\ and \vinf, and an adjustable base velocity $V_0$ to smoothly connect to the density at $R_\star$. The next three columns indicate the CSM mass, optical depth $\tau_{\rm CSM}$, and the corresponding photospheric radius. For $\tau_{\rm CSM}$, we integrate from $R_\star$ to infinity and adopt an opacity of 0.34\,cm$^2$\,g$^{-1}$. With this approach, the bulk of the optical depth arises from the denser, deeper regions near $R_\star$. The last column indicates whether the \heracles\ simulation of the model was post-processed with \cmfgen. Model mdot0p01/rcsm8e14 is a close analog of model mdot0p01/rcsm8e14/bw2/vinf5e6, for which we therefore skipped the calculation with \cmfgen.
\end{table*}

\subsection{Radiation hydrodynamics with \heracles}

We perform multi-group radiation-hydrodynamics calculations with \heracles\ exactly as discussed in detail and for similar configurations in \citet{d17_13fs}. We use eight energy groups spanning from the Lyman to the Balmer continuum. We account for bound-free opacity of all species as well as electron scattering (the opacity sources implemented in \heracles\ are discussed in \citealt{D15_2n}). For simplicity, we ignore the contribution from bound-bound transitions in the \heracles\ opacity tables so that the radiative acceleration we obtain in our simulations is an underestimate of the true acceleration, although probably not by much \citep{chugai_98S_02}. Because the influence of lines is a very complex phenomenon, it is not clear if their simplistic treatment would be a great addition. 

\subsection{Radiative transfer with \cmfgen}

A critical aspect of this work is the post-treatment of the \heracles\ simulations with the radiative transfer code \cmfgen. Although this method has been very fruitful for predicting the early-time properties of SN ejecta interacting with CSM, it is not fully physically consistent. While the simulation in \heracles\ is time dependent, the post-treatment with \cmfgen\ assumes steady state. Practically, we ignore time delays affecting photons emitted from different regions (equivalent to assuming an infinite speed of light), and in particular different depths along the line of sight (even though the \heracles\ simulation captures the progressive, time dependent, diffusion of radiation, or the heating and the acceleration of the unshocked CSM). The different travel time for these photons arises from the large value of \rcsm/$c$, which may be augmented by the necessary diffusion of these photons across the optically-thick CSM. Unfortunately, this is not solved by merely giving a delay to the escaping photons, as done in \citet{grafener_vink_13cu_16}. Photons emitted deeper at earlier times cross an ever changing medium (i.e., in temperature and ionization but also in velocity etc) as they progress along our line of sight. This physical inconsistency is most problematic at the earliest times when the CSM conditions change fast. This includes the rising part of the bolometric light curve, or when one witnesses a rising ionization, as observed during the first day in SN\,2024ggi (see, e.g.,  \citealt{wynn_24ggi_24}, \citealt{shrestha_24ggi_24}; \citealt{zhang_24ggi_24}). The rising ionization implies that light-travel time delays may allow low-ionization lines to persist during a CSM diffusion time (emission lines would first appear blueshifted and progressively become redshifted), although their strength would be considerably reduced relative to lines from higher-ionization lines (e.g., He\one\ relative to He\two\ lines). These light-travel time effects should become secondary when the luminosity and ionization reach a steady state or exhibit a slow evolution. 

A second inconsistency arises from using the temperature computed in \heracles\ (which assumes LTE for the gas) for the \cmfgen\ calculation (which solves the NLTE rate equations). For H-rich compositions where electron scattering and bound-free opacities dominate, this has proven satisfactory. The main benefit from this approach is that this temperature imported from \heracles\ reflects the complex radiation hydrodynamics of these interactions, the influence of the shock, the shock breakout signal, the heat- or ionization-wave propagating through the CSM at breakout. Assuming radiative equilibrium and steady state, as \citet{groh_13cu} or \citet{grafener_vink_13cu_16}, ignores this time dependence and yields a temperature structure that conflicts with the radiation hydrodynamics of shock breakout.

An essential asset of the steady-state radiative transfer calculations with \cmfgen\ is the direct treatment of the nonmonotonic velocity structure of these interaction configurations. This approach accounts for all emission, absorption, scattering processes occurring in the unshocked ejecta, shocked ejecta, shocked CSM, and unshocked CSM, together with the interplay between the gas and the wavelength- and depth-dependent radiation field. In practice, for each \heracles\ snapshot, we select the electron-scattering optical-depth region between 10$^{-4}$ and $\sim$\,50. We sample this region with about 100 points, ensuring a good resolution around the dense shell but also of the unshocked CSM, including the region beyond the photosphere where narrow emission lines form.

Most simulations performed with \heracles\ are post-processed with \cmfgen\ -- some are excluded to avoid redundancy (see Table~\ref{tab_init}). We try to select snapshots that cover from the initial rise of the luminosity out to 15\,d after the onset of shock breakout (i.e., when the shock crosses $R_\star$). We use a uniform composition for the CSM corresponding to the surface composition of the preSN model, and treat all important species including H, He, C, N, O, Mg, Al, Si, S, Ca, Ti, Cr, and Fe. The model atom covers multiple ionization stages to track the ionization stratification of the gas (typically cooler and less ionized further away from the shock). Specifically, we included H\one, He\one-\two, C\one-\five, N\one-\five, O\one-\six, Mg\two-\five, Al\three-\five, Si\two-\five, S\two-\five, Ca\two-\six, Ti\two-\four, Cr\two-\six, and Fe\two-\sev. Because the simulations exhibit different overall ionization depending on the CSM properties or time, some unimportant, low or high ionization stages may be omitted for specific snapshots. In this work, we focus the discussion to optical lines and in particular H$\alpha$, but we also touch upon lines of He\one, He\two, C\three, C\four, N\three, and N\four\ as they provide interesting probes of different CSM regions. We discuss how these different lines compare with each other as well as how they evolve in time. Our model set contains information in the ultraviolet and in the infrared for numerous other lines not discussed in this work. This may prove useful for future, high-resolution and high-cadence observations of objects with low or high mass loss rates, and with data in or out of the optical ranges.  

The emergent \cmfgen\ spectra are computed at high resolution, adopting a turbulent velocity of 5--10\,\kms\ and a similar frequency-grid spacing. To capture the influence of electron scattering and the associated frequency redistribution, we iterate ten times in order to obtain a converged mean intensity during the spectral calculation (for details, see \citealt{d09_94w}). With fewer iterations, the wings of most emission lines would be weaker and narrower. Because of the radiative acceleration of the unshocked CSM, the high resolution is essential only at the earliest times -- the narrowest emission lines may exhibit a width of 100\,\kms\ or more a few days after the onset of shock breakout.  


\section{Results for the reference model mdot0p01/rcsm8e14 (aka r1w6b)}
\label{sect_ref}

\subsection{Radiative acceleration of the unshocked CSM}
\label{sect_ref_grad}

\begin{figure*}
\centering
\includegraphics[width=0.49\hsize]{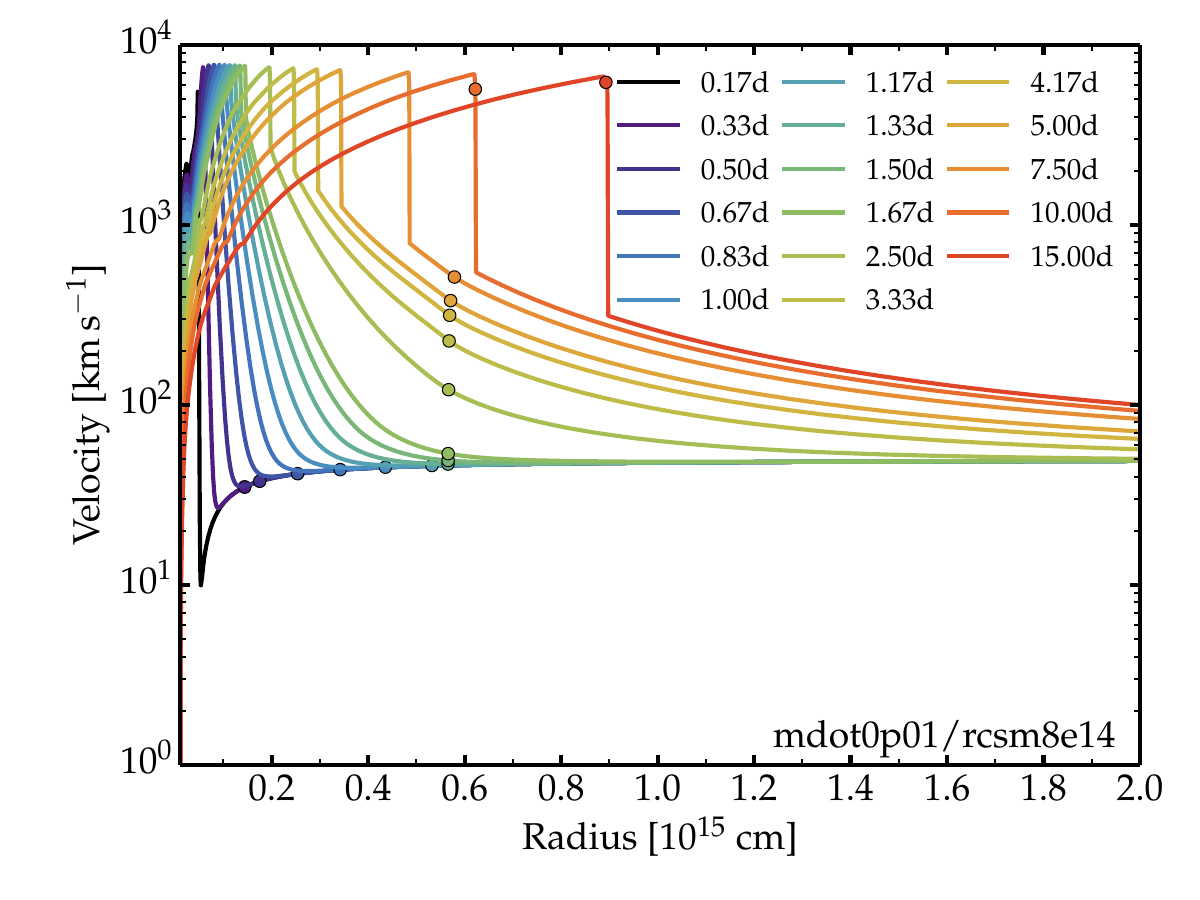}
\includegraphics[width=0.49\hsize]{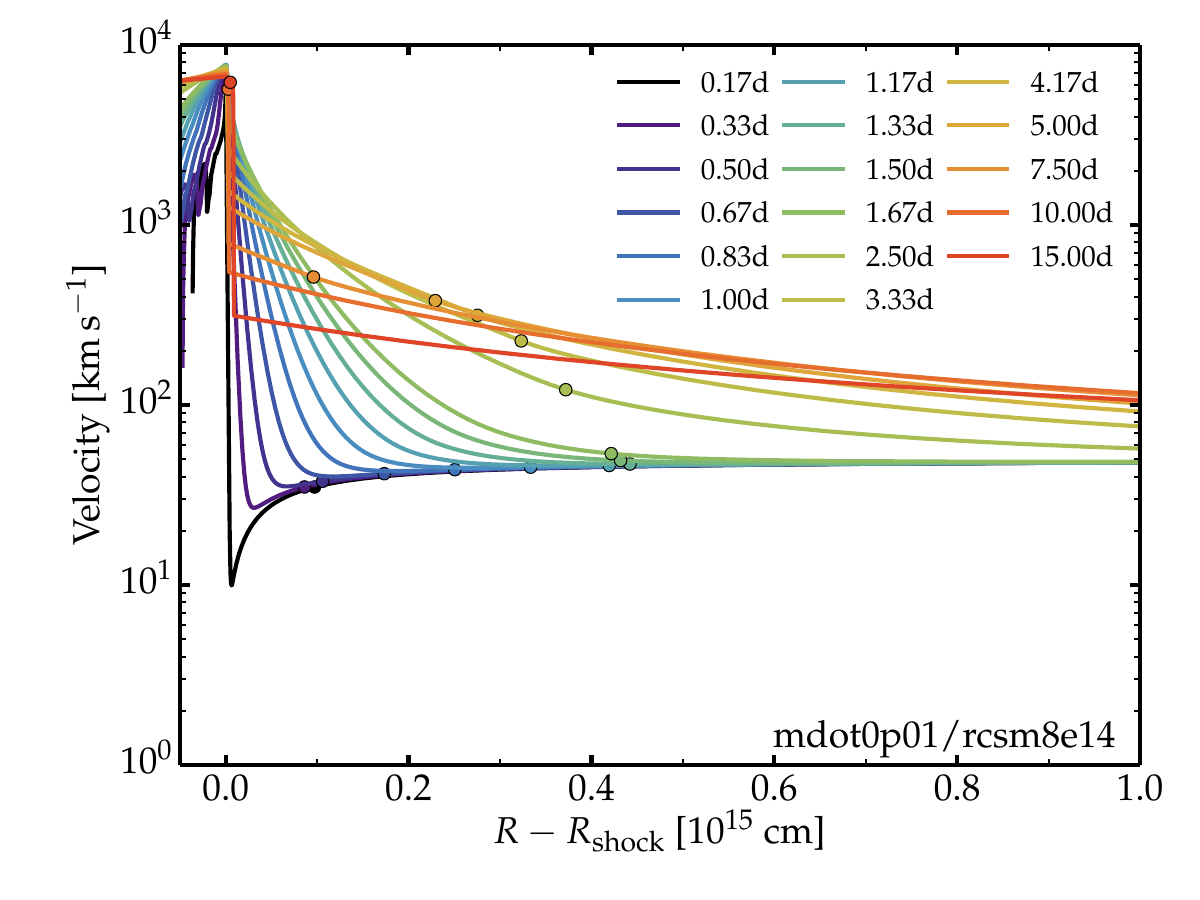}
\includegraphics[width=0.49\hsize]{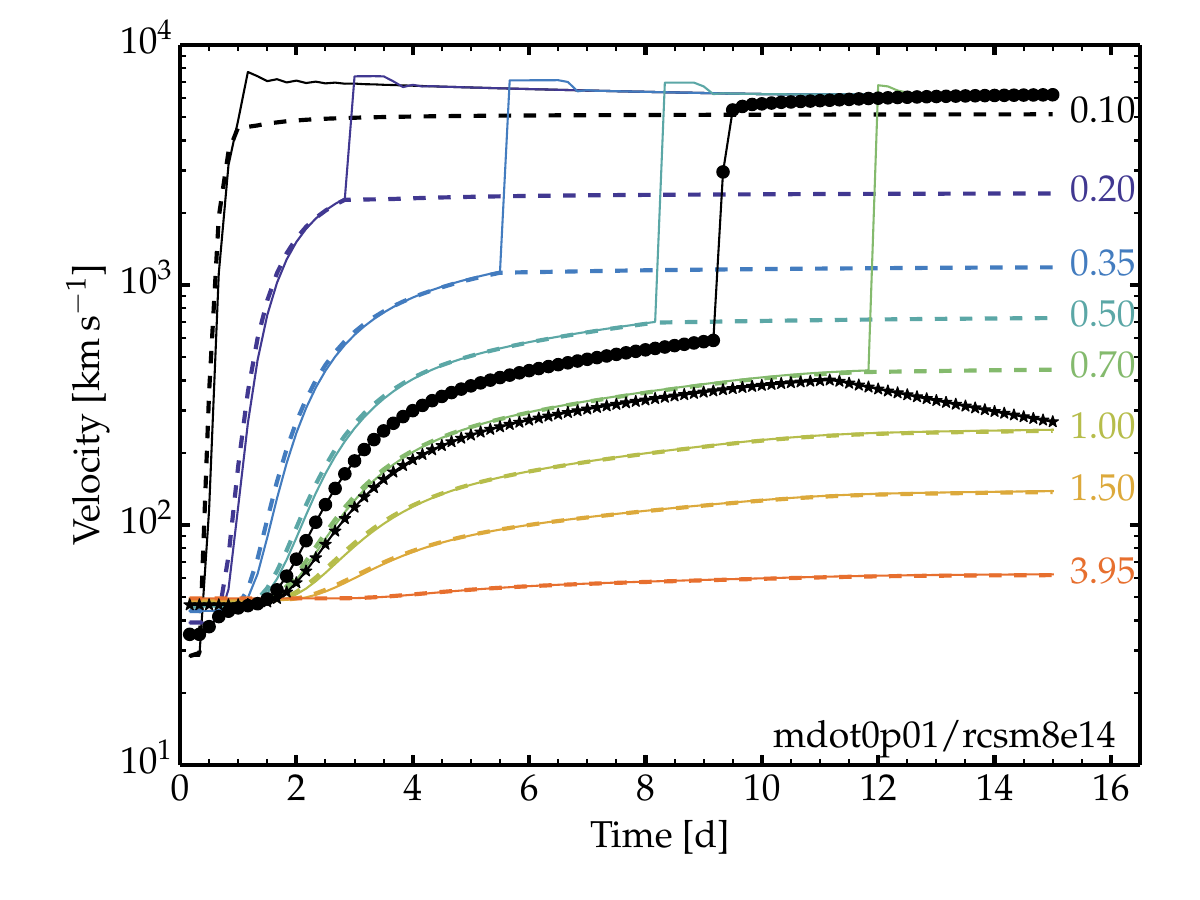}
\includegraphics[width=0.49\hsize]{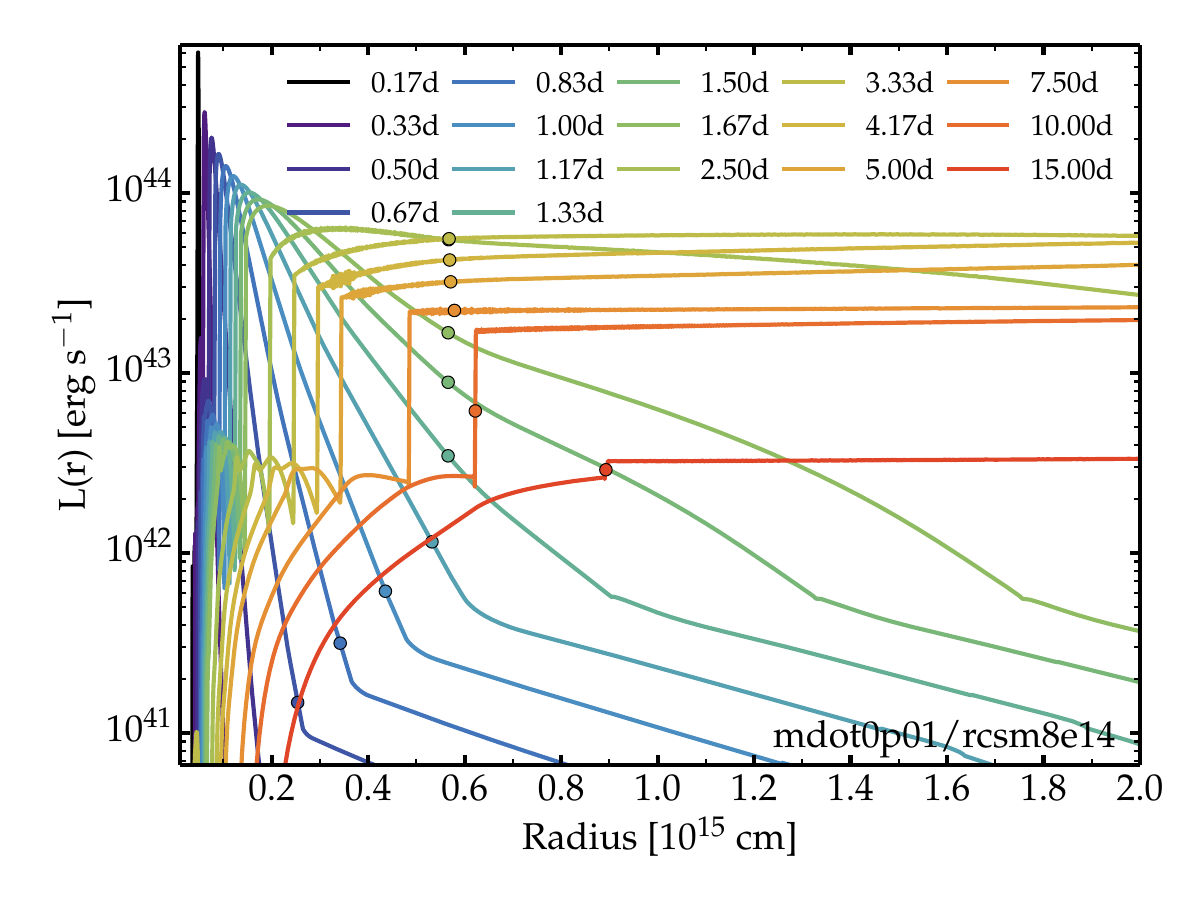}
\caption{Illustration of radiation and hydrodynamical properties computed by \heracles\ for model mdot0p01/rcsm8e14. Top row: Variation of the velocity at multiple epochs versus radius (top left) and versus the distance above the shock (top right). Bottom left: Evolution of the velocity of ``mass parcels" initially at radii 0.1, 0.2, 0.35, 0.5, 0.7, 1.0, 1.5, and 3.95$\times$\,10$^{15}$\,cm (the last radius is near the outer grid boundary) and over the full duration of 15\,d of the \heracles\ simulation. The solid line shows the value of this velocity as read in directly from the simulation. The dashed line shows the same value but estimated through a time integral of the radiative acceleration (see discussion in Section~\ref{sect_ref_grad}). In all cases, the two curves agree closely until the mass parcel is shocked, which corresponds to the near-vertical jump of the solid lines.  The black circles and the stars indicate the velocity at the location where the electron-scattering optical depth is 2/3 (i.e., the photosphere) and at 0.01.
Bottom right: Counterpart of the top-left panel but for the luminosity (i.e., the quantity $L_r = 4 \pi r^2 F_r$). Only the inner half of the grid is shown. 
\label{fig_evol_v_at_r}
}
\end{figure*}

Figure~\ref{fig_evol_v_at_r} shows some results from the radiation-hydrodynamics calculations for our reference model mdot0p01/rcsm8e14, which is also known as model r1w6b \citep{jacobson_galan_23ixf_23}. The top-left panel gives the multiepoch velocity profile versus radius. This evolving structure is generic for such interactions and has been shown multiple times in the past, both for similar \heracles\ simulations (see, e.g., \citealt{D15_2n,d17_13fs}) or for \stella\ simulations (see, .e.g., \citealt{moriya_rsg_csm_11}). The sharp jump in velocity (i.e., the shock) separates the shocked ejecta and CSM from the unshocked CSM. At the start of the simulation, the region below the shock has a complex structure (not shown in detail here) because of the equivalently complicated density structure of the progenitor star (e.g., leading to the formation a reverse and forward shock etc). At shock breakout, about 50\,\%\ of the total energy is stored in radiation and so only 50\% of the explosion energy has been turned into kinetic energy. The region behind the shock is therefore radiation dominated and the escape of that radiation forms the core component of the SN light curve, even in the absence of interaction with CSM. As time goes on, this inner region accelerates and eventually reaches homologous expansion. With the presence of CSM, a reverse shock propagates inwards and decelerates the outer regions of the (already shocked) stellar envelope while a forward shock propagates outwards and crosses the inner regions of the CSM, leading to the formation of a dense shell. 

In this work, we are mostly interested in what takes place beyond the forward shock, thus in the unshocked CSM. Close to the shock, the unshocked material is accelerated to velocities of several 1000\,\kms\ and the magnitude of this acceleration drops steeply with radius. At the photosphere (indicated by the filled dots in Fig.~\ref{fig_evol_v_at_r}), the velocity initially rises as it progresses outwards in the CSM because of the photoionization of the CSM caused by the SN radiation (i.e., this reflects the rising velocity with radius of the original CSM), so from 30 to 40\,\kms. This is followed by a strong increase from 40 to 500\,\kms\ over the time span of $\sim$\,1 to $\sim$\,7\,d. After shock passage, the photosphere is located in the fast-moving dense shell, with a typical velocity in this model of about 6300\,\kms. In contrast, the outer CSM is modestly accelerated, reaching at 15\,d a velocity of 100\,\kms\ at $2\times 10^{15}$\,cm and 62\,\kms\ at $4\times 10^{15}$\,cm, compared to its original velocity of $\sim$\,50\,\kms\ prior to the SN explosion. 

Overall, the acceleration of the unshocked CSM is greatest in optically thick regions below the photosphere, thus largely obscured to a distant observer. An alternate way of showing the acceleration of the unshocked CSM is to plot the multiepoch velocity profile relative to the location of the shock (top-right panel of Fig.~\ref{fig_evol_v_at_r} -- this is equivalent to showing the velocity in a frame moving with the shock). We can see more clearly that just ahead of the shock, the acceleration of the unshocked CSM is enormous with velocities rising from a few 10 to a few 1000\,\kms\ at 1-2\,d, but progressively decreasing at later times to about 300\,\kms\ at 15\,d.  

The bottom-left panel illustrates the physical origin of the acceleration of the unshocked CSM. Here, we show the evolution of the velocity of mass parcels located initially at radii spanning the full extent of the CSM beyond $R_\star$. Specifically, we place these Lagrangian ``tracers'' at radii 0.1, 0.2, 0.35, 0.5, 0.7, 1.0, 1.5, and $3.95 \times 10^{15}$\,cm (the last value is just below the outermost grid radius). The solid lines give the results from \heracles, that is they show the velocity of each mass parcel as it advects. Starting with a mass parcel at $(r,t)$ with a velocity $V(r,t)$, we look for the velocity at $t + \delta t$ at the new location $r + V(r,t) \times \delta t$. The results from \heracles\ indicate that the acceleration is indeed huge for the mass parcels located close to the shock and much more modest at large distances, although some acceleration is visible even near the outer grid location. They also show that the acceleration is naturally truncated when the shock reaches the location of a given mass parcel, as evidenced by the sudden jump in velocity (over the 15\,d of the simulation, the shock reaches out to a maximum radius of $8.9 \times 10^{14}$\,cm). After being shocked, a mass parcel moves here at about 6300\,\kms. 

The dashed-lines shown in the bottom-left panel of Fig.~\ref{fig_evol_v_at_r} correspond to the velocity of the mass parcels estimated from the radiative acceleration computed in \heracles. Now, starting with a mass parcel at $(r,t)$ with a velocity $V(r,t)$, we compute the new velocity at $t + \delta t$ and at the new location $r + V(r,t) \times \delta t$ as $V(r,t)  + g_{\rm rad} \delta t$, with the radiative acceleration $g_{\rm rad} = \kappa_{\rm es} F_r / c$ ($\kappa_{\rm es}$ is the electron scattering opacity, $F_r$ is the radiative flux in the radial direction, and $c$ is the speed of light). Here, both $F_r$ and $\kappa_{\rm es}$ are taken at the relevant location for each (advected) mass parcel and thus reflect the changes in ionization and luminosity with time and depth (see bottom-right panel of Fig.~\ref{fig_evol_v_at_r}). Both solid and dashed curves overlap at most epochs (the curves sharply diverge if and when a given mass parcel is shocked), confirming that the mass parcels are radiatively accelerated, as has been suggested in the past by \citet{fransson_93j_96} in the context of SN\,1993J and by \citet{chugai_98S_02} in the context of SN\,1998S.

The strong radial dependence of the acceleration results from the time and radial evolution of the radiative flux. CSM closer to the shock is influenced by a greater flux (in proportion to $r^2$) but it is also subject to the flux from earlier times because it takes a diffusion time for the radiation to reach the outer, optically-thin regions of the CSM. For an optically-thin CSM, this time delay would be the free-flight time through the CSM, which is approximately 0.39\,$R_{\rm 15,CSM}$\,d, where $R_{\rm 15,CSM} = R_{\rm CSM}/10^{15}$\,cm. For an optically-thick CSM, this delay is augmented by a factor equal to the electron-scattering optical depth $\tau_{\rm CSM}$ of the CSM. Hence, a more compact and more optically-thin CSM should be more strongly and promptly accelerated, although this radiative acceleration will be eventually limited by the reduced luminosity (i.e., caused by the smaller extraction of kinetic energy from the ejecta). With our model grid of CSM properties, we indeed recover a wide range of CSM acceleration (see Section~\ref{sect_dep}).

The question is now to evaluate how this radiative acceleration may be inferred from SN spectra. The formation of the spectrum in these interactions is complex. At large optical depth, both continuum and lines form. If the CSM is sufficiently optically thick, most or all photons arising from the fast-moving ejecta or shock are reprocessed (i.e., absorbed and reemitted) by the CSM. Thus, the photons emitted in this unshocked CSM will scatter multiple times with free electrons before escaping. Upon each scattering, the photons are redistributed in frequency or velocity space. For continuum photons, this produces no obvious signature since their original distribution with wavelength is smooth. For line photons, which are originally emitted over a few Doppler widths (i.e. few \kms) around the line rest wavelength, this redistribution leads to the transfer of core photons into the wings and the formation of the notorious IIn line morphology with a narrow core and broad symmetric wings. Once a photon is outside of the line core, it is subject to continuum (essentially electron scattering) opacity and thus behaves just like a continuum photon.  

Emergent photons located in the narrow cores of emission lines must therefore arise from regions of low electron scattering optical depth, so typically beyond the photosphere (it is by virtue of this property that these narrow line cores should be strongly depolarized; see discussion in \citealt{dessart_98S_24}). Because line emission is greater at high density (particularly true for recombination lines like H$\alpha$), these narrow lines will tend to form close to the photosphere. In our calculations, we find that a good proxy for that formation region is where $\tau_{\rm es}$ is about 0.01-0.1. With time passing, this region may eventually be crossed by the shock so when that occurs, any residual narrow line emission would tend to form just exterior to the shock. In the bottom-left panel of Fig.~\ref{fig_evol_v_at_r}, we indicate with star symbols the evolution of the velocity at that location in our reference model. It shows a very slow rise by 10-20\,\kms\ initially as the radiation from shock breakout ionizes and turns the CSM optically thick. There is then a substantial rise up to 400\,\kms\ caused by radiative acceleration. Eventually, when the shock reaches the outer CSM, the narrow-line region is limited to just outside the shock and thus probes ever more distant locations where the radiative acceleration has been weaker and thus that velocity should decrease. Eventually, far away from the progenitor surface, the velocity should be the original \vinf\ of the CSM. Thus, the evolution of the narrow line emission should exhibit a similar pattern, which we discuss in the next section.


\subsection{Evolution of individual spectral lines}
\label{sect_ref_lines}

\begin{figure}
\centering
\includegraphics[width=\hsize]{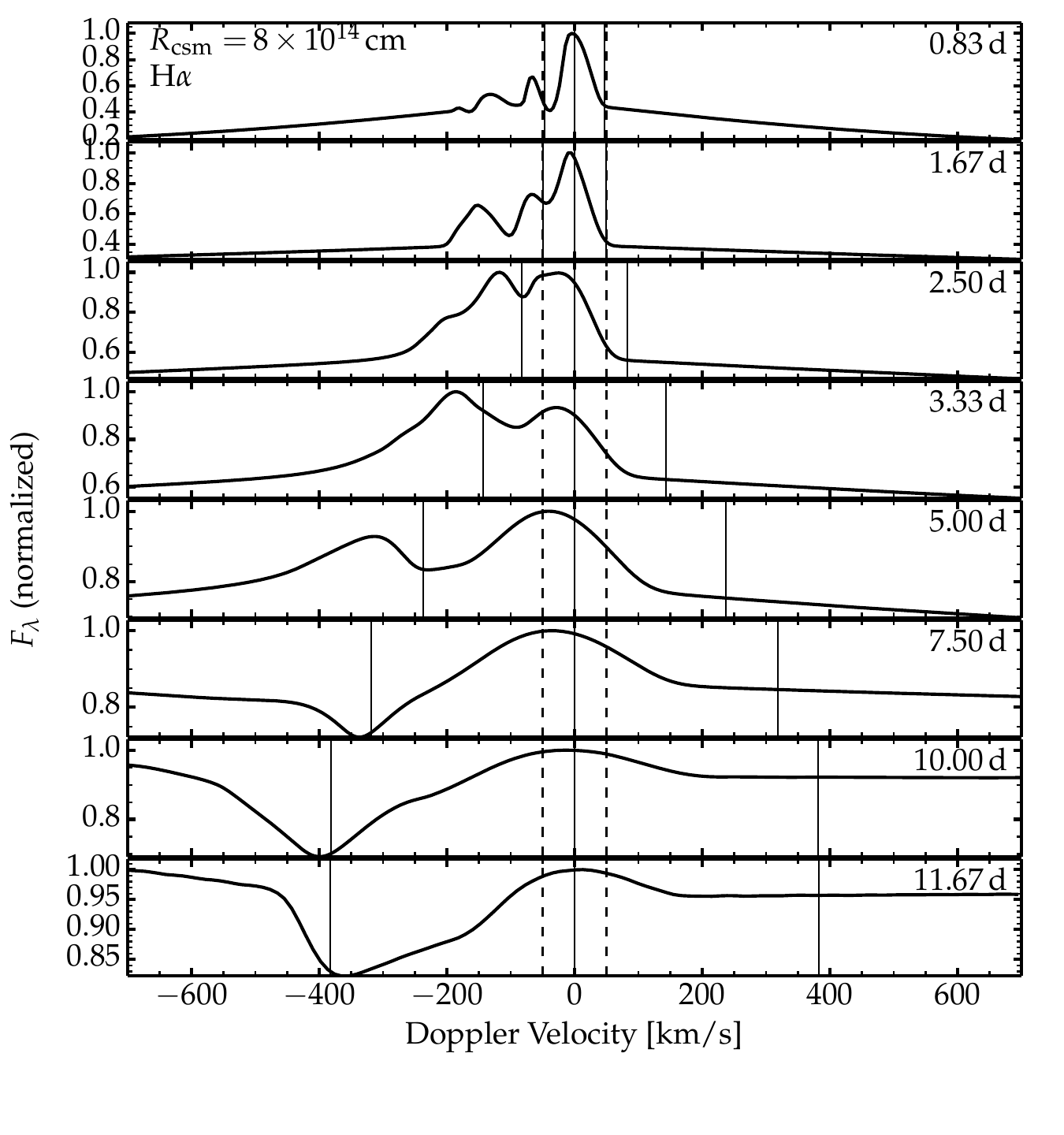}
\caption{Evolution of the spectral region centered on the rest wavelength of H$\alpha$ for model mdot0p01/rcsm8e14. The symmetric solid vertical lines on each side of the origin correspond to the velocity at the location in the unshocked CSM where the electron-scattering optical depth is 0.01, thus exterior to the location of the photosphere. The dashed vertical lines correspond to $\pm$\,\vinf\ of the original, unadulterated CSM. The spectral evolution in other H\one\ Balmer lines is analogous to that shown here for H$\alpha$ (see the case of H$\beta$ shown in Fig.~\ref{fig_4861} in the appendix). Note that He\two\,6560.09\,\AA\ contributes some flux on the blue edge of H$\alpha$, well separated from it initially but overlapping with the H$\alpha$ emission at 2--5\,d. The ordinate range changes in each panel for better visibility.
\label{fig_6562}
}
\end{figure}

\subsubsection{Evolution of H$\alpha$}

Figure~\ref{fig_6562} shows the spectral evolution from 0.83\,d until 11.67\,d and within 700\,\kms\ of the rest wavelength of H$\alpha$ for our reference model mdot0p01/rcsm8e14 discussed in the preceding section (the full radiative transfer model computed with \cmfgen\ is not shown since such spectra have been presented and confronted to observations on numerous occasions in the past). The impact of radiative acceleration is immediately visible: the narrow emission line remains bounded within 50\,\kms\ (i.e., the original value of \vinf) for nearly two days, followed by a considerable broadening until about 10\,d, and a hint for a reduction in the extent of the line absorption at the last epoch of 11.67\,d, as expected from the discussion in the preceding section. For clarity, we add a vertical bar indicating the velocity at $\tau_{\rm es}$ of 0.01 and indicated by star symbols in the bottom-left panel of Fig.~\ref{fig_evol_v_at_r}.

What is shown in Fig.~\ref{fig_6562} is, however, a little more complex than what one would have anticipated. First, there is a He\two\ transition on the blue side of H$\alpha$ at 6560.09\,\AA. This emission starts weak, strengthens as the temperature of the CSM rises in the initial few days and then disappears around 5\,d as the conditions become too cool for He\two\ emission. It is also impacted by the radiative acceleration of the CSM so He\two\,6560.09\,\AA\ progressively broadens until it disappears. Because of this broadening, the contribution from He\two\,6560.09\,\AA\ and H$\alpha$ overlap at 2--5\,d and the ``contaminant'' is not easily identified. 

A second aspect is that H$\alpha$ starts roughly symmetric but becomes progressively blueshifted in both absorption and emission. At 0.83\,d, the H$\alpha$ line is well centered on the rest wavelength, with roughly symmetric emission blueward and redward, but it exhibits excess emission on the blueside beyond 50\,\kms. All the CSM beyond $2 \times 10^{14}$\,cm moves at 40--50\,\kms\ at that time (see bottom left panel of Fig.~\ref{fig_evol_v_at_r}), so this likely arises from the escape into the line wings of photons trapped in the highly optically thick line core. This feature affects all optically-thick lines and is visible too here in He\two\,6560.09\,\AA. After 5\,d, the ``narrow'' H$\alpha$ line has a blueward extent out to about 400\,\kms\ with a clear P-Cygni profile morphology. Unlike stellar winds that increase in velocity from the hydrostatic base to their asymptotic velocity, the P-Cygni trough here probes the region at $\sim$\,500\,\kms\ just outside the photosphere (blue edge of the trough) out to the slow moving outer CSM with velocity of $\sim$\,100\,\kms. In this model, this P-Cygni trough disappears at later times because the CSM density is too low (see alternate results for model mdot0p01/rcsm1e15 for the case where the CSM is even more extended).

The fact that most of the changes in emission and absorption are seen blueward of the rest wavelength of H$\alpha$ is a consequence of optical depth effects. Regions near the mid-plane or from the receding hemisphere suffer greater absorption. For optically-thin conditions, the line would exhibit a symmetric emission relative to the rest wavelength and no absorption. Optical-depth effects here may be exacerbated by the assumption of spherical symmetry and steady-state constant velocity at large distances (there could for example be more turbulence, or some nonmonotonicity of the flow at large distances etc). 

\subsubsection{Evolution of other lines}

\begin{figure}
\centering
\includegraphics[width=\hsize]{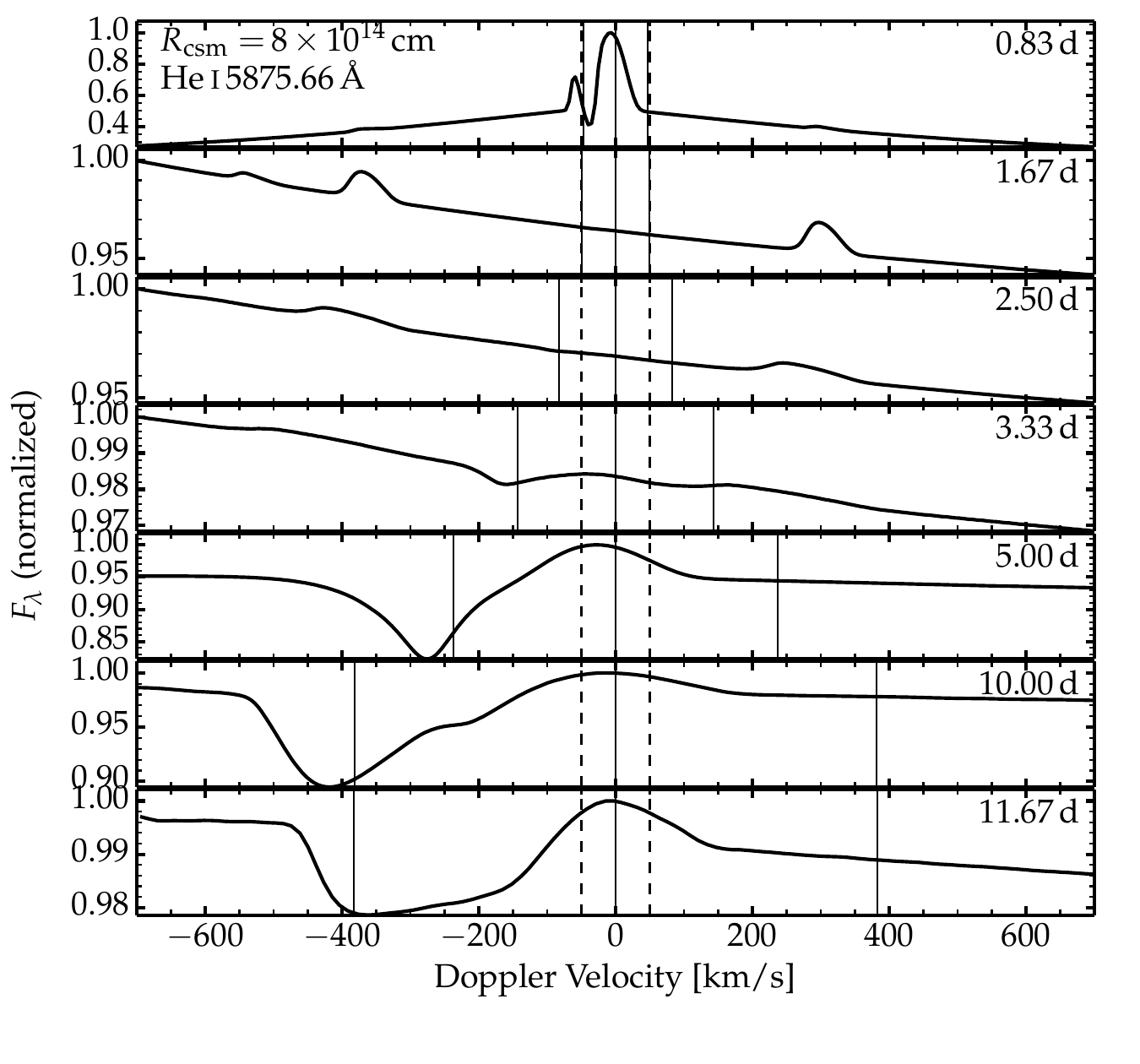}
\caption{Same as Fig.~\ref{fig_6562} but now for He\one\,5875.66\,\AA. This line is present at early times until about 1\,d and then again after $\sim$\,3\,d but absent in between because the ionization of the unshocked CSM is too high at those intermediate epochs.
\label{fig_5875}
}
\end{figure}

\begin{figure}
\centering
\includegraphics[width=\hsize]{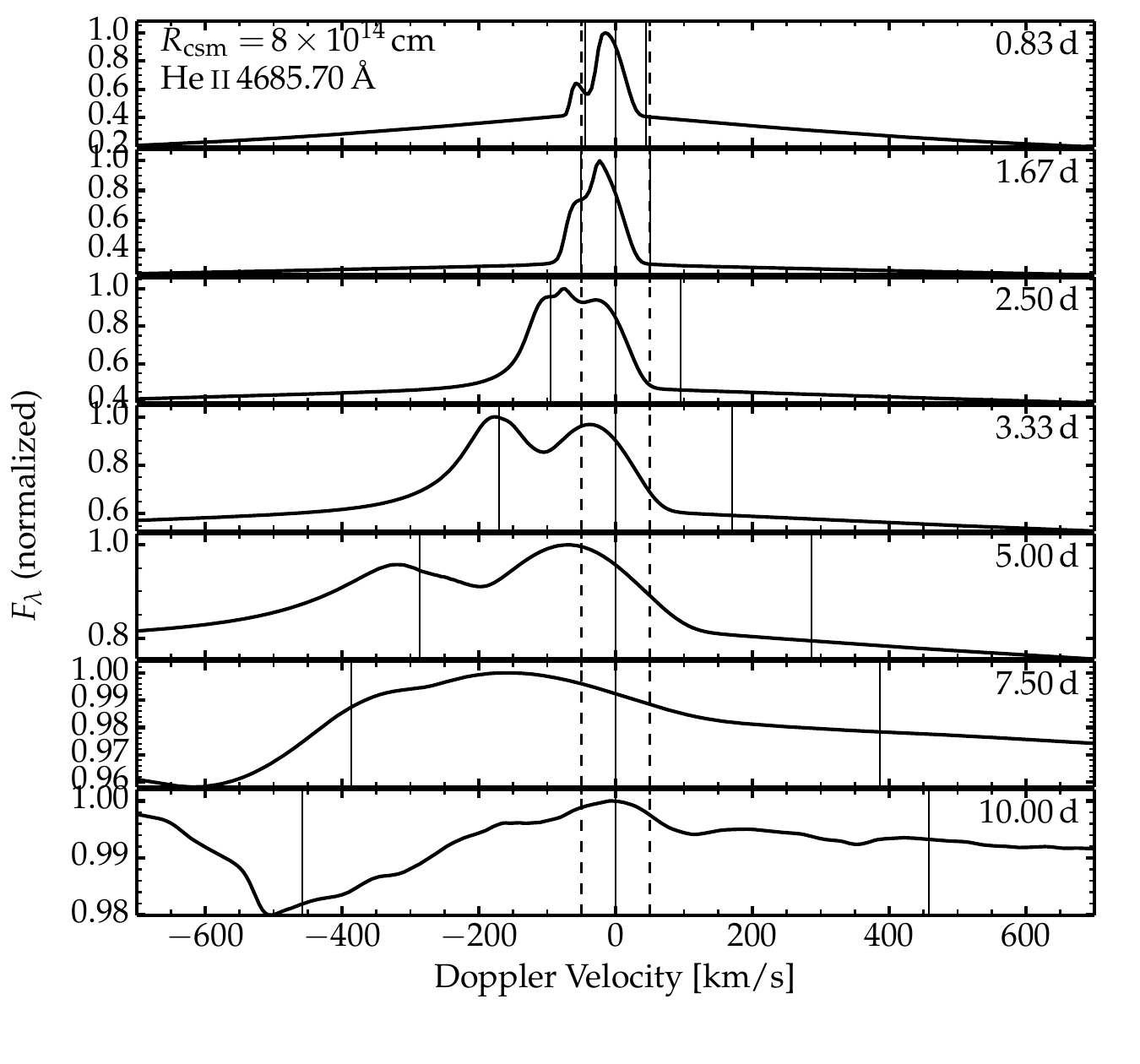}
\caption{Same as Fig.~\ref{fig_6562} but now for He\two\,4685.70\,\AA. This line is essentially unblended and present until $\sim$\,5\,d, after which the He ionization is too low. The symmetric solid vertical lines on each side of the origin correspond here to the velocity at the location in the unshocked CSM where the electron-scattering optical depth is 0.1.
\label{fig_4685}
}
\end{figure}

\begin{figure}
\centering
\includegraphics[width=\hsize]{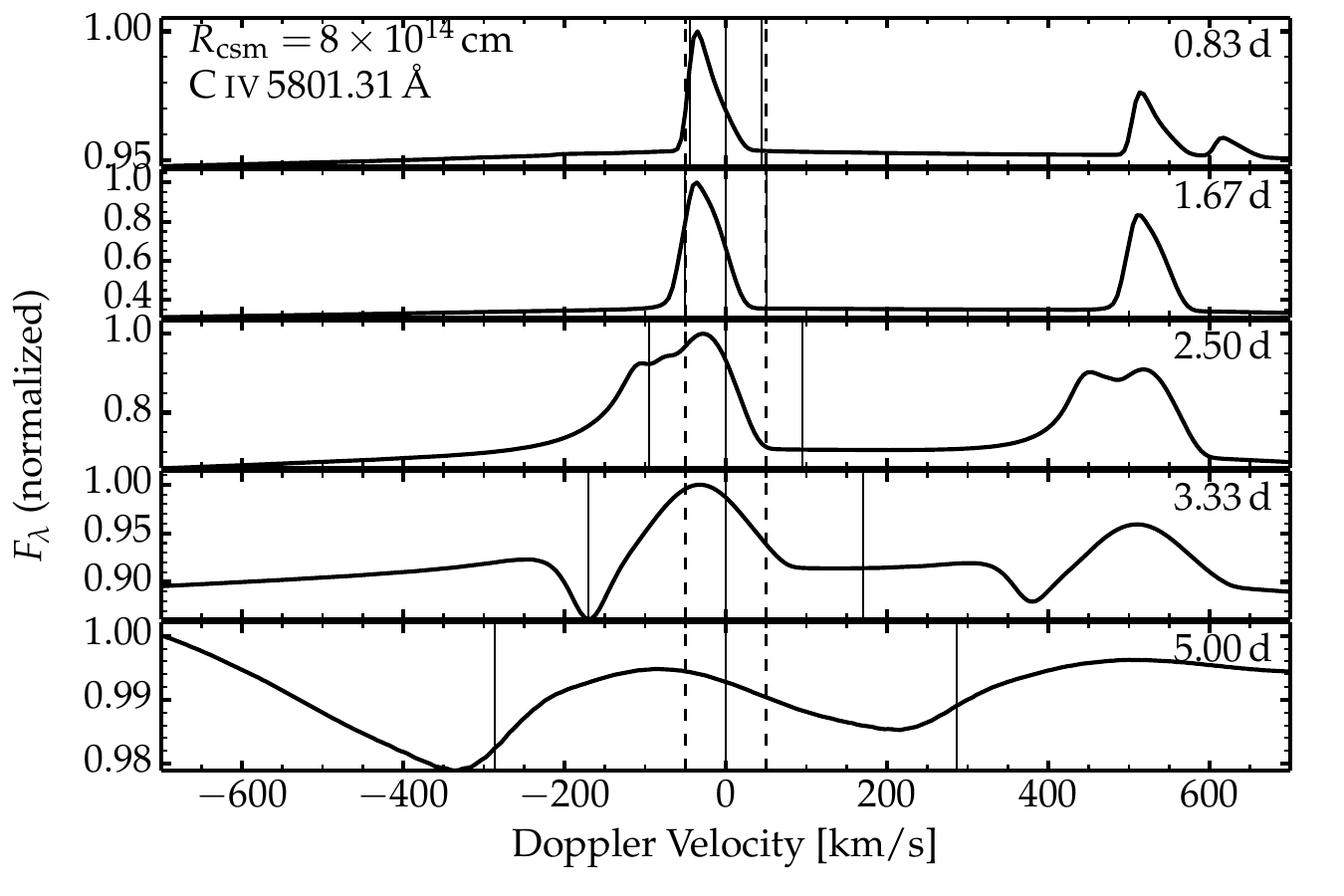}
\caption{Same as Fig.~\ref{fig_6562} but now for C\four\,5801.31\,\AA. Here, the solid vertical bars on each side of the origin correspond to the velocity at the location in the unshocked CSM where the electron-scattering optical depth is 0.1.
\label{fig_5801}
}
\end{figure}

\begin{figure}
\centering
\includegraphics[width=\hsize]{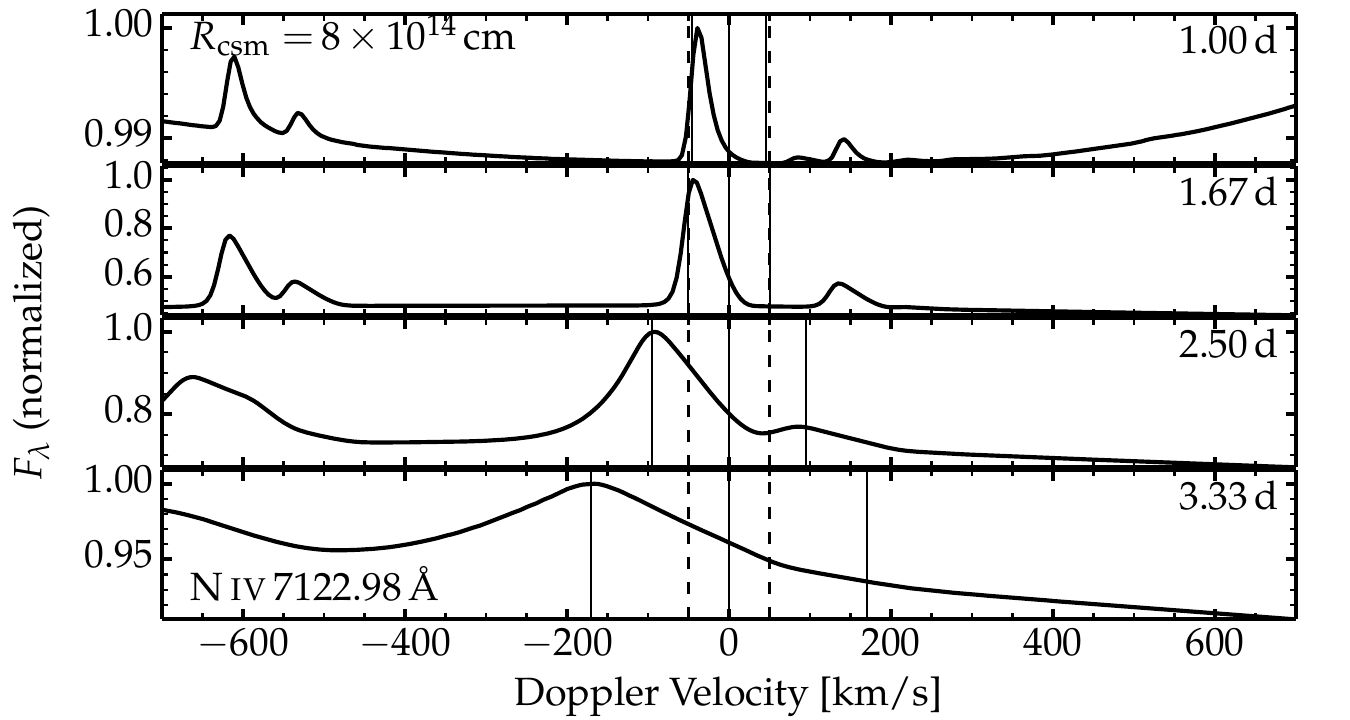}
\caption{Same as Fig.~\ref{fig_6562} but now for N\four\,7122.98\,\AA. This line is absent prior to $\sim$\,1\,d and after $\sim$\,5\,d, because of the unsuitable ionization state of the unshocked CSM at those epochs. This spectral region contains other N\four\ lines at 7109.35, 7111.28, and 7127.25\,\AA. Here, the solid vertical bars on each side of the origin correspond to the velocity at the location in the unshocked CSM where the electron-scattering optical depth is 0.1.
\label{fig_7122}
}
\end{figure}

\begin{figure}
\centering
\includegraphics[width=\hsize]{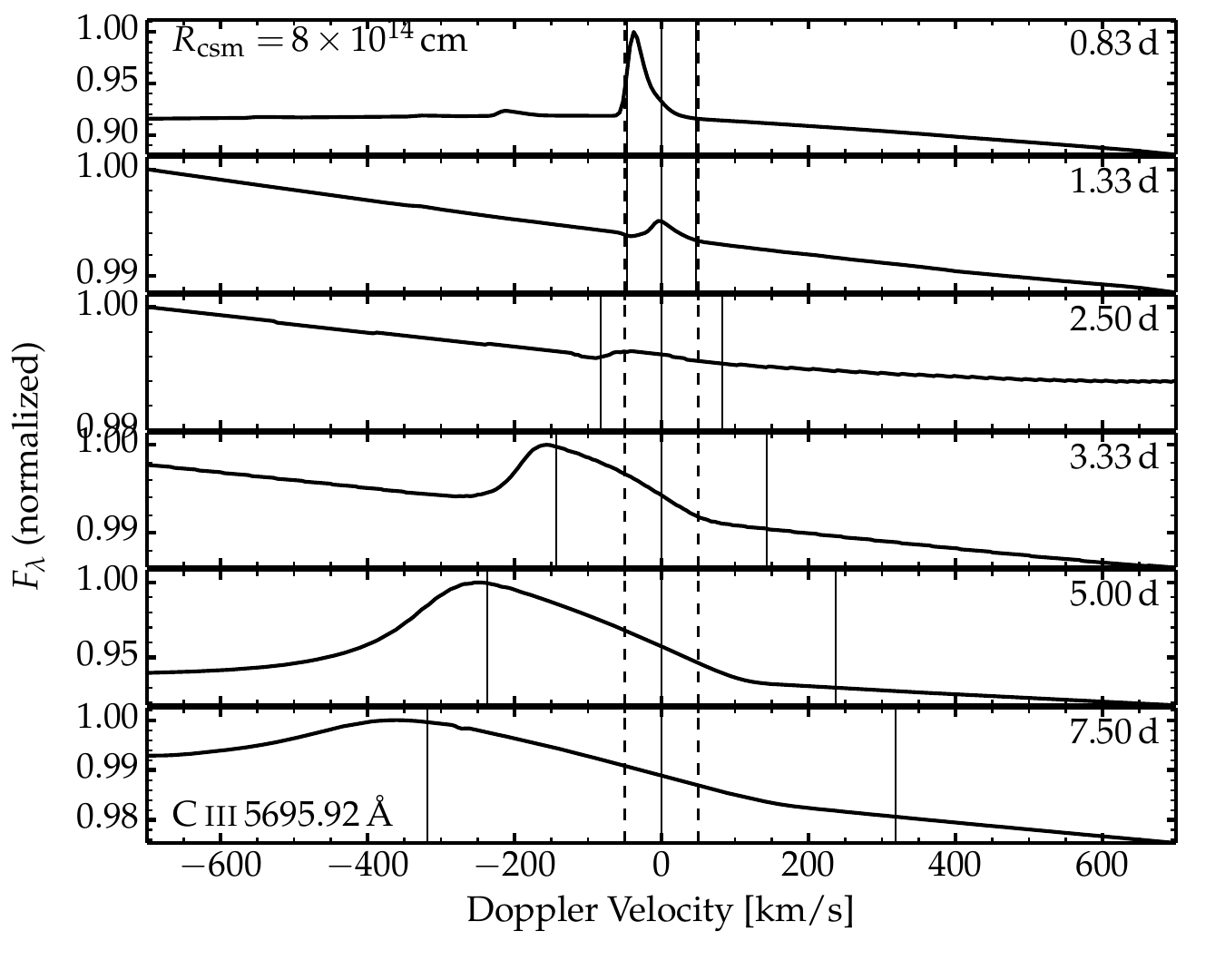}
\caption{Same as Fig.~\ref{fig_6562} but now for C\three\,5695.92\,\AA. Besides the evolution of that line in width caused by the radiative acceleration of the unshocked CSM, its strength also varies in time with a minimum around 1.5\,d caused by the high ionization of the unshocked CSM (at those epochs, C\four\ is the dominant ion).
\label{fig_5695}
}
\end{figure}

Figures~\ref{fig_5875}--\ref{fig_5695} show the evolution for other lines in the reference model mdot0p01/rcsm8e14, namely He\one\,5875.66\,\AA\ (Fig.~\ref{fig_5875}), He\two\,4685.70\,\AA\ (Fig.~\ref{fig_4685}), C\four\,5801.31\,\AA\ (Fig.~\ref{fig_5801}), N\four\,7122.98\,\AA\ (Fig.~\ref{fig_7122}), and C\three\,5695.92\,\AA\ (Fig.~\ref{fig_5695}). There is some redundancy in line evolution. For example, Balmer lines tend to show a similar evolution so we show the evolution for H$\beta$ in the appendix (Fig.~\ref{fig_4861}) but skip the discussion of additional H\one\ transitions. The same applies to He\two\ lines so we show only the evolution of He\two\,5411.52\,\AA\ in the appendix (Fig.~\ref{fig_5411}).

The narrow component of the He\one\,5875.66\,\AA\ emission line (Fig.~\ref{fig_5875}) shows at 0.83\,d a similar profile to H$\alpha$, roughly symmetric in emission around the rest wavelength (with a width reflecting a formation within the accelerating part of the original, unadulterated wind CSM) and with an excess emission in the blue related to photon escape in the line wings. In the first spectrum, we clearly see a P-Cygni trough (there is a marked absorption at about $-$40\,\kms). As the CSM is progressively photoionized during shock breakout, He becomes predominantly He$^+$ and He\one\ lines disappear, before reappearing again after 3\,d when the CSM cools and recombines. During this second phase, He\one\,5875.66\,\AA\ exhibits a P-Cygni profile morphology, with a strong absorption extending to about $-$450\,\kms\ at 10\,d and little emission on the blue side. As for H$\alpha$, the relatively narrow emission falling within $\pm$\,100\,\kms\ at 11.67\,d forms in the distant CSM whereas the extended trough forms much closer in, just outside of the shock. In other words, the P-Cygni profile forms within a radially declining velocity profile.

He\two\ lines show a reverse trend relative to He\one\ lines since they take some time to strengthen as the CSM heats up but survive only as long the CSM stays sufficiently hot. Indeed, the narrow component of the He\two\,4685.70\,\AA\ line emission progressively strengthens up until 3-5\,d before decreasing and vanishing at about 7\,d (Fig.~\ref{fig_4685}). The emission has a similar shape as H$\alpha$ at 0.83\,d but with a marked blueshift. Radiative acceleration becomes visible after about 2\,d when the line stretches out to 150\,\kms\ blueward of the rest wavelength (the red side of the profile remains narrow and hardly changes). At 7--10\,d, the model predicts a residual blueshifted absorption out to $-$500\,\kms\ but it is extremely weak (slight inflection at the percent level).  

Figure~\ref{fig_5801} shows the spectral region within $\pm$\,700\,\kms\ of the rest wavelength of C\four\,5801.31\,\AA. As for He\two\,4685.70\,\AA, this line disappears quickly after 5\,d because of the reduction in ionization of the unshocked CSM. There is no C\four\,5801.31\,\AA\ emission or absorption from the more distant, cooler, CSM so a narrow component is only present at the earliest times in the model. This C\four\ transition is a doublet, whose other component at 5811.97\,\AA\ is visible at right as a close replication of C\four\,5801.31\,\AA. Unlike the lines previously discussed, this C\four\ line is strongly blueshifted at all times but this is striking at the earliest epochs (prior to $\sim$\,2\,d) when the line is narrow. This occurs because the line tends to form closer to the photosphere where occultation (or optical-depth) effects are greater. After 2\,d, the line blue edge broadens and eventually forms a P-Cygni absorption trough extending to several 100\,\kms. Unlike H$\alpha$ or He\one\,5875.66\,\AA, this doublet C\four\ transition does not exhibit an absorption at early times, which may arise from the fact that the line forms over a restricted range of depths (i.e., where the optical depth is low but the density still high enough and importantly where the temperature is high enough).


Figure~\ref{fig_7122}--\ref{fig_5695} show the evolution for N\four\,7122.98\,\AA\ and C\three\,5695.92\,\AA. Both exhibit a similar behavior with strongly blueshifted emission at all times (due to optical-depth effects), a lack of P-Cygni absorption (due the relatively confined formation region of these lines). The relatively high ionization of these lines limits their presence to early times with a very modest strength at the earliest times when the CSM is still heating up and rising in ionization.

To summarize, the narrow component of the emission lines discussed above show both the influence of radiative acceleration, with line starting narrow as expected for the unadulterated CSM velocity but growing with time to reach about 500\,\kms. For low ionization lines still present at 10\,d, their narrow component starts showing signs of a reduction in velocity as the more distant CSM is being probed. Optical-depth effects tend to quench the emission from the receding part of the CSM, causing a blueshift of the line emission. This skewness is greater for lines of high ionization because they tend to form over a smaller volume and close to the photosphere. Although the lines are skewed, the line emission remains immediately adjacent to the rest wavelength. None of our model profiles shows the detached narrow emission reported by \citet{pessi_24ggi_24} for most lines observed in SN\,2024ggi -- this offset may be due to adopting an inadequate recession velocity to the SN. Finally, to facilitate the comparison of all lines at once, we present a mosaic of numerous lines but all at 1\,d after the onset of shock breakout in the reference model mdot0p01/rcsm8e14 in the appendix, in Fig.~\ref{fig_r1w6b_1d}.

\section{Dependencies on CSM velocity, density, and extent}
\label{sect_dep}

Having presented in detail the reference model mdot0p01/rcsm8e14 in terms of radiation-hydrodynamics properties (shock propagation, radiative acceleration of the unshocked CSM etc) and the associated characteristics of narrow emission lines versus time, we now explore the sensitivity of these results to variations in CSM properties. We start in the next section by addressing the impact of the extent of the CSM, followed by the impact of the CSM density (Section~\ref{sect_dep_mdot}) and finally the impact of the CSM velocity profile (Section~\ref{sect_dep_wind}).


\subsection{Impact of CSM extent}
\label{sect_dep_rcsm}

In this section, we discuss the change in the results discussed earlier when the CSM extent is varied while other CSM characteristics are kept the same. Namely, we study models in which the CSM mass loss rate is 0.01\,\msunyr, the wind velocity parameters are \bw$=$\,2 and \vinf$=$\,50\,\kms, but the value of \rcsm\ is increased from $2 \times 10^{14}$\,cm to $10^{15}$\,cm. Here, these models are named rcsm2e14, rcsm4e14, rcsm6e14, rcsm8e14, and rcsm1e15. 

\begin{figure}
\centering
\includegraphics[width=\hsize]{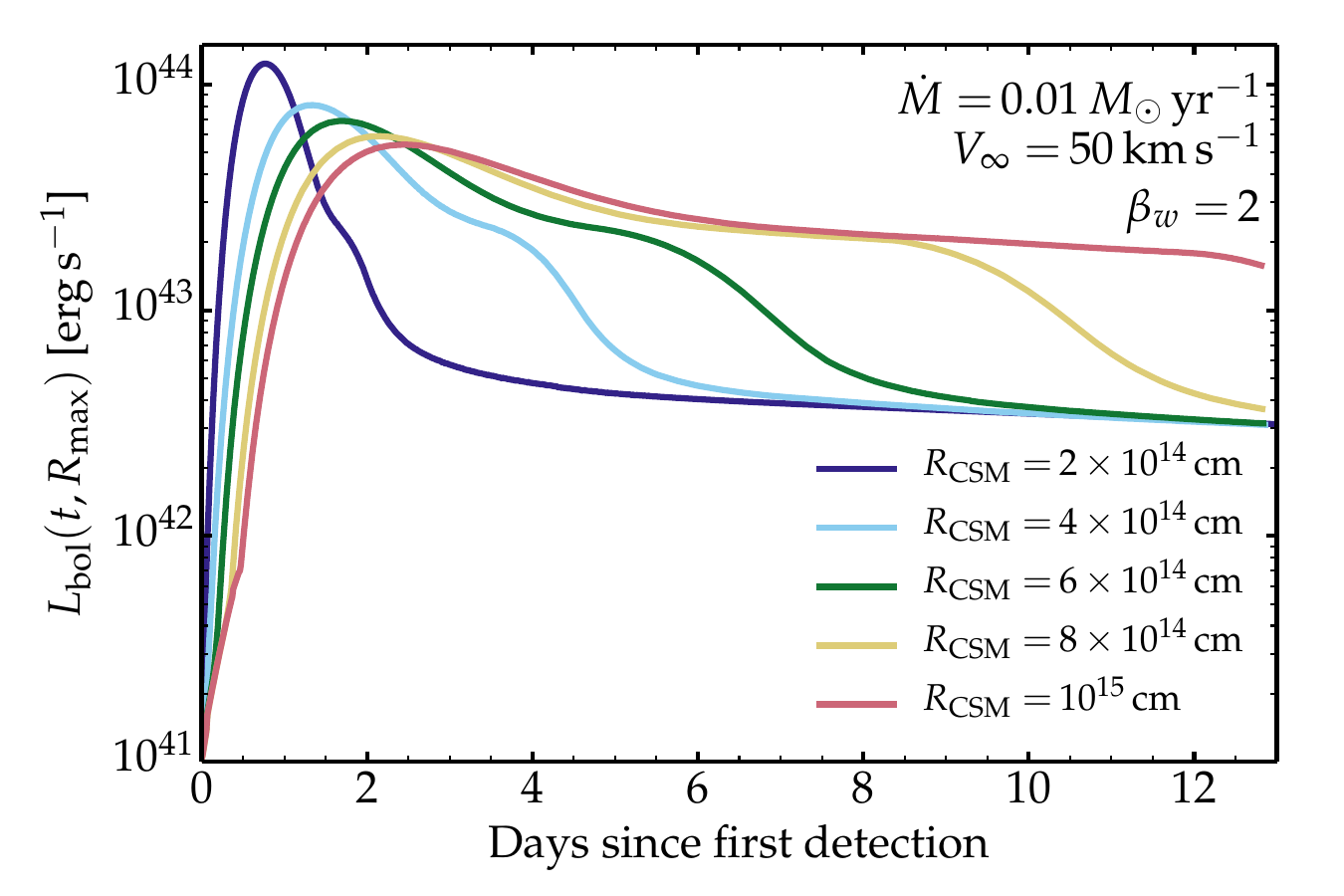}
\includegraphics[width=\hsize]{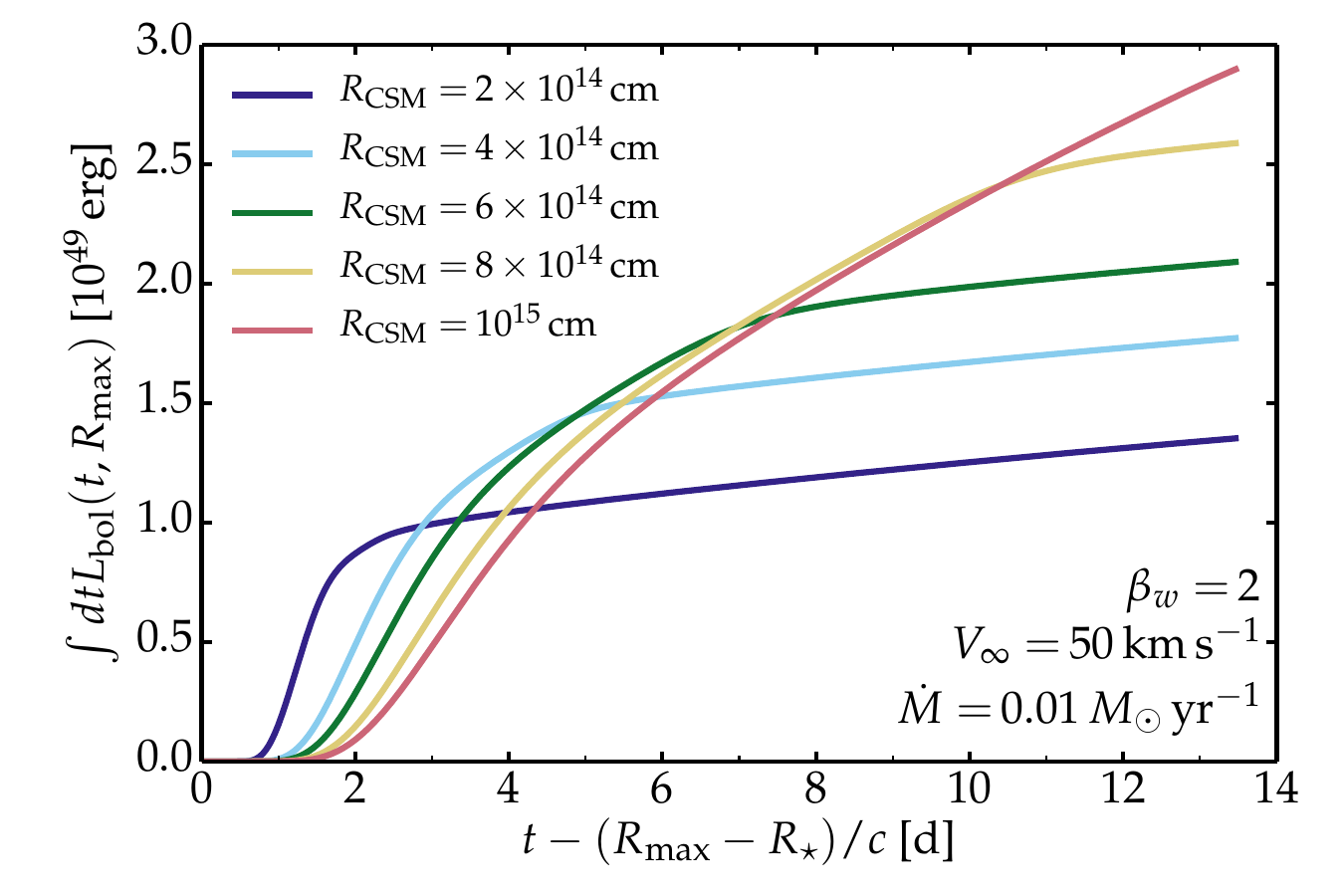}
\includegraphics[width=\hsize]{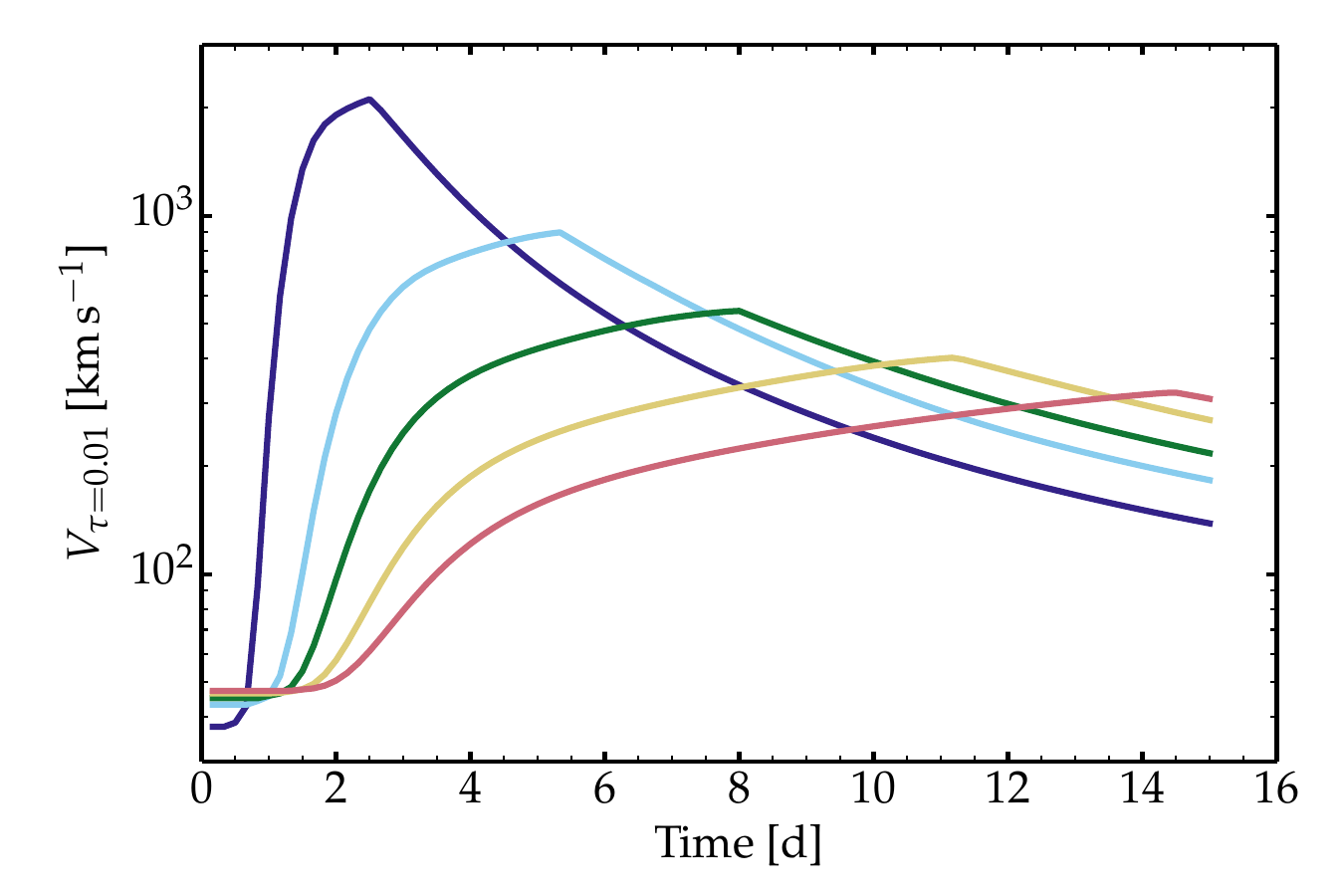}
\caption{Results from the radiation-hydrodynamics calculations for models differing in \rcsm. Top: Bolometric luminosity recorded at the outer boundary and starting when the power exceeds 10$^{41}$\,\ergs. Middle: Same as top for the cumulative bolometric luminosity. Here, we use the \heracles\ time modified for the light-travel time to the outer boundary. Bottom: Evolution of the velocity at the location in the unshocked CSM where the electron-scattering optical is 0.01. When that location is shocked, we report the velocity at the radius immediately exterior to the shock.
\label{fig_var_rcsm}
}
\end{figure}

Figure~\ref{fig_var_rcsm} shows some results from the radiation-hydrodynamics calculations for this model grid. With increasing \rcsm, the CSM optical depth increases (although the bulk of the optical depth arises from the inner CSM, which is the same in all five models) and the photosphere is located further out at 1.7, 3.1, 4.1, 5.7, and $7.1 \times 10^{14}$\,cm in the models rcsm2e14 to rcsm1e15. The diffusion time through the CSM is longer so the light curve rise time to maximum is greater for more extended CSM. Relative to a time of first detection at the outer grid boundary set to 10$^{41}$\,\ergs, the rise time increases from $\sim$\,1.5\,d (and a peak $L_{\rm bol}$ in excess of 10$^{44}$\,\ergs) in model rcsm2e14 up to $\sim$\,2.5\,d (and a peak $L_{\rm bol}$ of $5 \times 10^{43}$\,\ergs). The time-integrated bolometric luminosity over the first 15\,d varies from around 1.4 to $3 \times 10^{49}$\,erg in this model set. In model rcsm1e15, the shock is still embedded in the dense CSM at the end of the simulation and the luminosity is still very large.

The evolution of the velocity at the location where the unshocked CSM optical depth is 0.01 is shown in the bottom panel of Fig.~\ref{fig_var_rcsm} -- this is a good representation of the Doppler-velocity width of the narrow emission line component (see previous section). The greatest velocity is obtained in model rcsm2e14 with a peak at 2100\,\kms\ at only 2.5\,d after the start of the simulation. This occurs despite the relatively low luminosity (and its time integral) of that model. In contrast, the smallest velocity is obtained for the rcsm1e15 model with a velocity of 320\,\kms\ attained just before the end of the simulation at 14.5\,d, and thus in spite of the much larger luminosity (and time integral). Models with intermediate values of \rcsm\ yield values that lie in between these two models. 

We thus obtain the paradoxical result that the greater the CSM mass and extent, the smaller the recorded acceleration. This arises simply from the fact that for a more extended CSM, the region where the narrow emission line component forms is located further away from the deeply embedded shock or ejecta regions where the radiation decouples from the gas. Beyond that location, the luminosity is essentially constant, so the flux drops as $1/r^2$, and the associated radiative acceleration is considerably weaker at greater distances. Paradoxically, the radiative acceleration to be recorded from narrow emission lines would be maximum for more compact CSM although that acceleration would be short-lived because of the imminent shock passage.  

\begin{figure}
\centering
\includegraphics[width=\hsize]{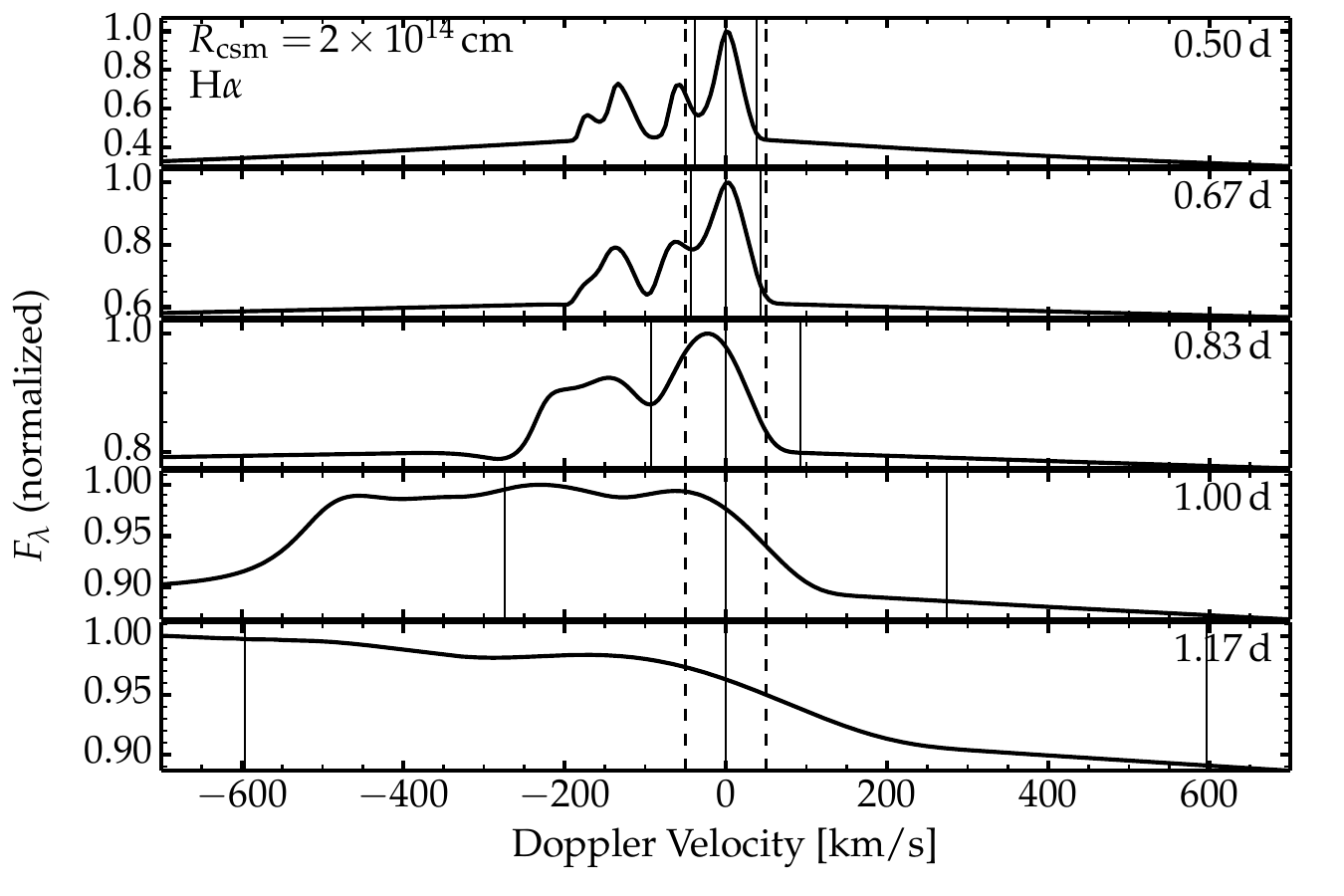}
\caption{Evolution of the spectral region centered on the rest wavelength of H$\alpha$ for model mdot0p01/rcsm2e14. The symmetric solid vertical lines on each side of the origin correspond to the velocity at the location in the unshocked CSM where the electron-scattering optical depth is 0.01. The dashed vertical lines correspond to $\pm$\,\vinf, where \vinf\ is the terminal wind velocity of the original, unadulterated CSM. 
\label{fig_rcsm2e14_halpha}
}
\end{figure}

\begin{figure}
\centering
\includegraphics[width=\hsize]{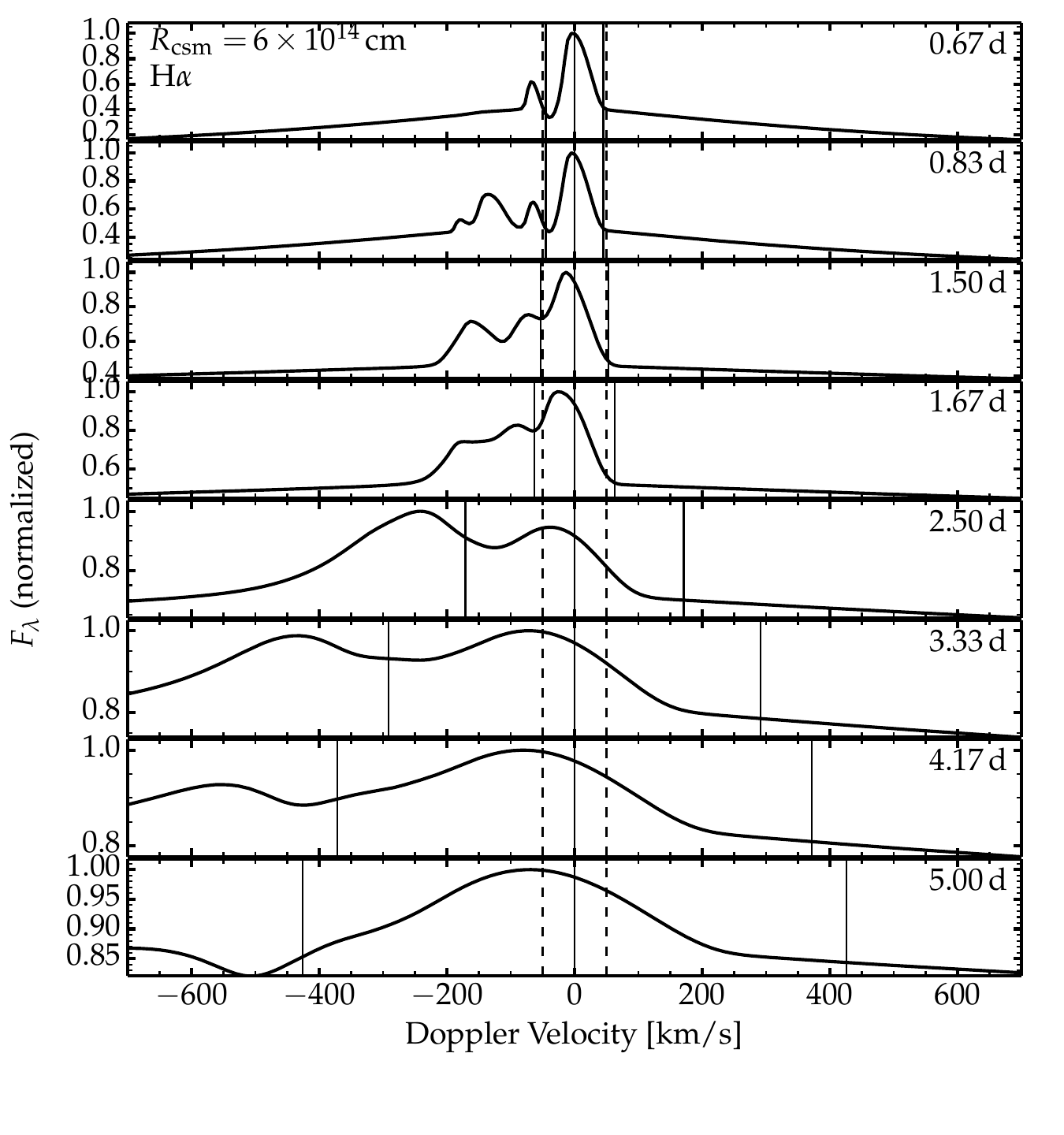}
\caption{Same as Fig.~\ref{fig_rcsm2e14_halpha}, but now for the model counterpart with a greater \rcsm\ value of $6 \times 10^{14}$\,cm.
\label{fig_rcsm6e14_halpha}
}
\end{figure}

\begin{figure}
\centering
\includegraphics[width=\hsize]{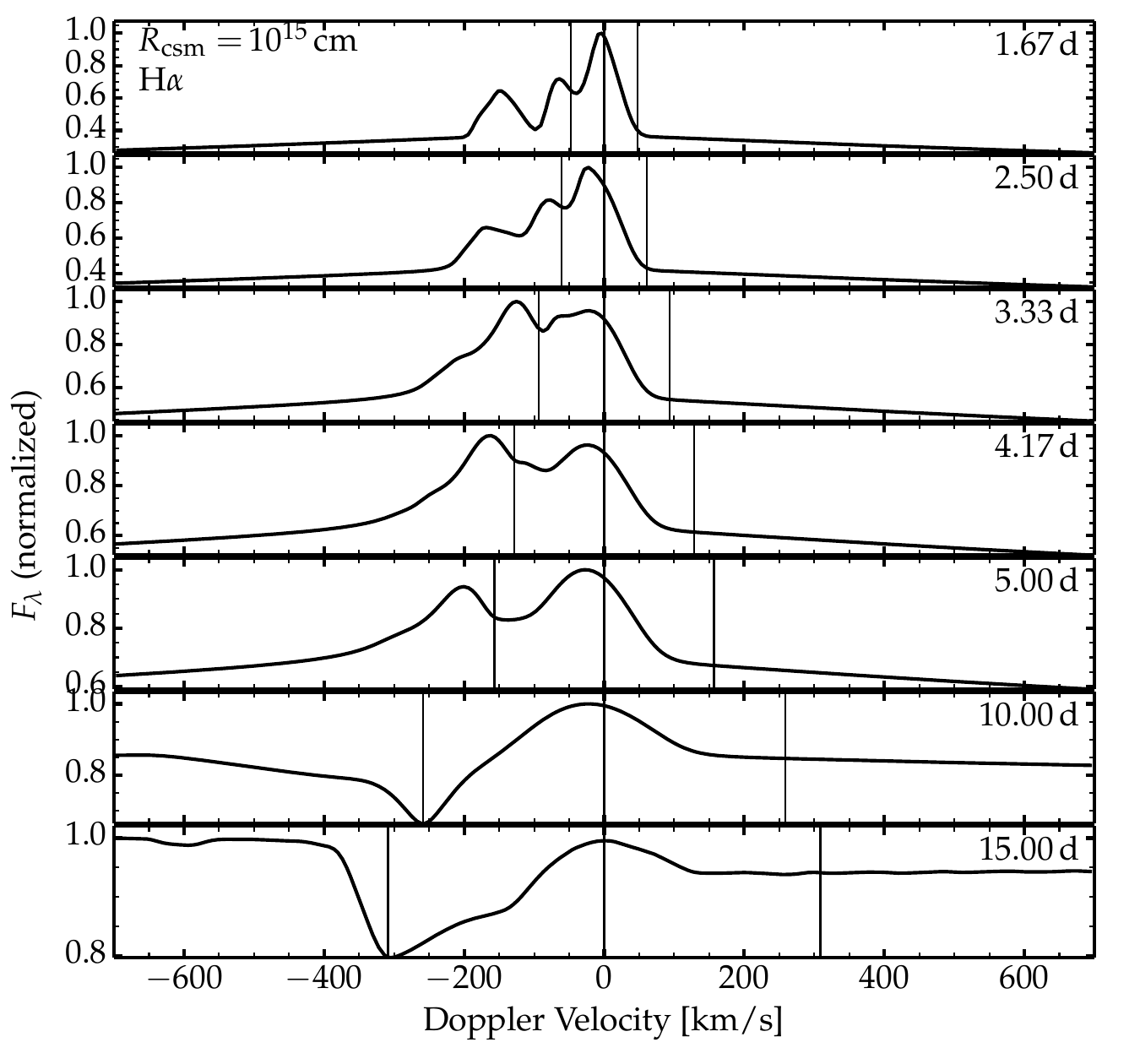}
\caption{Same as Fig.~\ref{fig_rcsm2e14_halpha}, but now for the model counterpart with the largest \rcsm\ value of $10^{15}$\,cm.
\label{fig_rcsm1e15_halpha}
}
\end{figure}

We post-processed these simulations with \cmfgen, covering all epochs from a fraction of a day until the end of the simulations at 15\,d. Figures~\ref{fig_rcsm2e14_halpha}--\ref{fig_rcsm6e14_halpha}--\ref{fig_rcsm1e15_halpha} show the evolution at a few, distinctive epochs in the evolution of H$\alpha$ within $\pm$\,700\,\kms\ from the rest wavelength for models rcsm2e14, rcsm6e14, and rcsm1e15 (in the appendix, we also show results for the intermediate model rcsm4e14 in Fig.~\ref{fig_rcsm4e14_halpha} -- corresponding results for model rcsm8e14 were already presented in Section~\ref{sect_ref_lines}). 

The drastic differences are immediately apparent. For model rcsm2e14, the H$\alpha$ profile is still narrow at 0.5\,d but broadens considerably within 0.5\,d and the profile extends beyond the plot limits at 1.17\,d. The H$\alpha$ profile emission is strongly blueshifted with essentially no flux redward of the rest wavelength although we stress that some line emission is present at the rest wavelength. In the model rcsm6e14, the broadening and blueshift of the line emission start at $\sim$\,2\,d, reach some plateau at about 4\,d, after which a weakening P-Cygni profile forms with a trough peaking at about $-$500\,\kms\ -- this is the value of $V_{\tau=0.01}$ at about 5\,d. In model rcsm1e15, the evolution is much reduced. The H$\alpha$ profile stays narrow for up to about 2.5\,d, after which a slow broadening appears leading to a maximum blueshift of the emission and absorption out to about 300\,\kms, again in agreement with $V_{\tau=0.01}$.

In all profile simulations, the narrow-line emission extends to large velocities on the blue side, up to values well in excess of the velocity at optical depth 0.01. This arises because there is not sharp boundary separating the narrow line emission from the region below the photosphere where line emission is both broader (the material is more strongly radiatively accelerated, and eventually shocked) and influenced by electron scattering (thus broadened because of the scattering with relatively high velocity, thermal electrons).

\subsection{Impact of CSM density}
\label{sect_dep_mdot}

In this section, we study the impact of the CSM density on the radiative acceleration of the unshocked CSM as well as the spectral signatures for the narrow emission line component. These results are analogous to those obtained in the previous section for variations in \rcsm. Specifically, we consider model counterparts to the reference model in which the mass loss rate of 0.01\,\msunyr\ is raised to 0.1 or reduced to 0.001\,\msunyr. All other model parameters are kept identical, i.e. \bw$=$\,2, \vinf$=$\,50\,\kms, and \rcsm$=$\,8$\times$\,10$^{14}$\,cm. We refer to these models as mdot0p001, mdot0p01, and mdot0p1.

\begin{figure}
\centering
\includegraphics[width=\hsize]{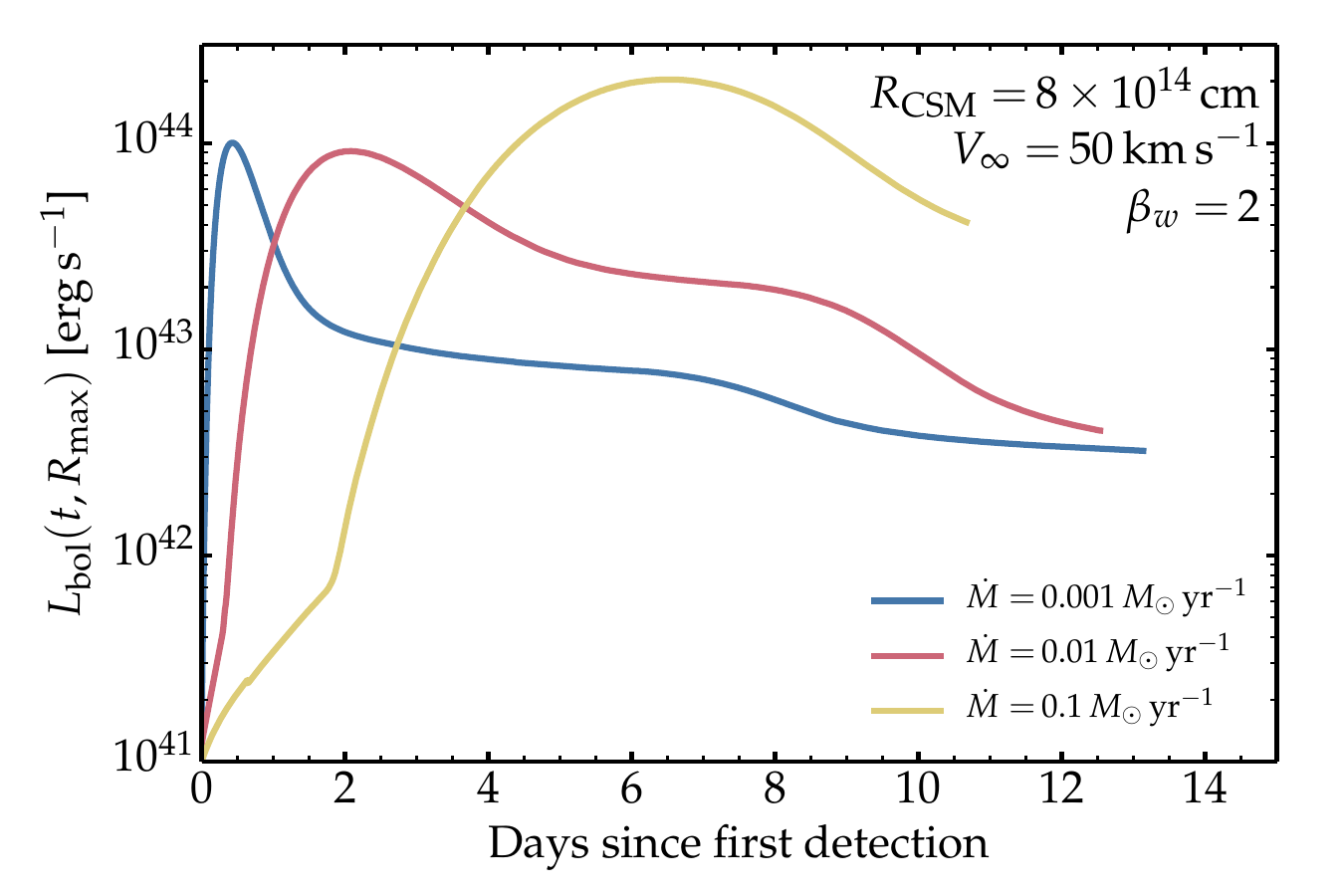}
\includegraphics[width=\hsize]{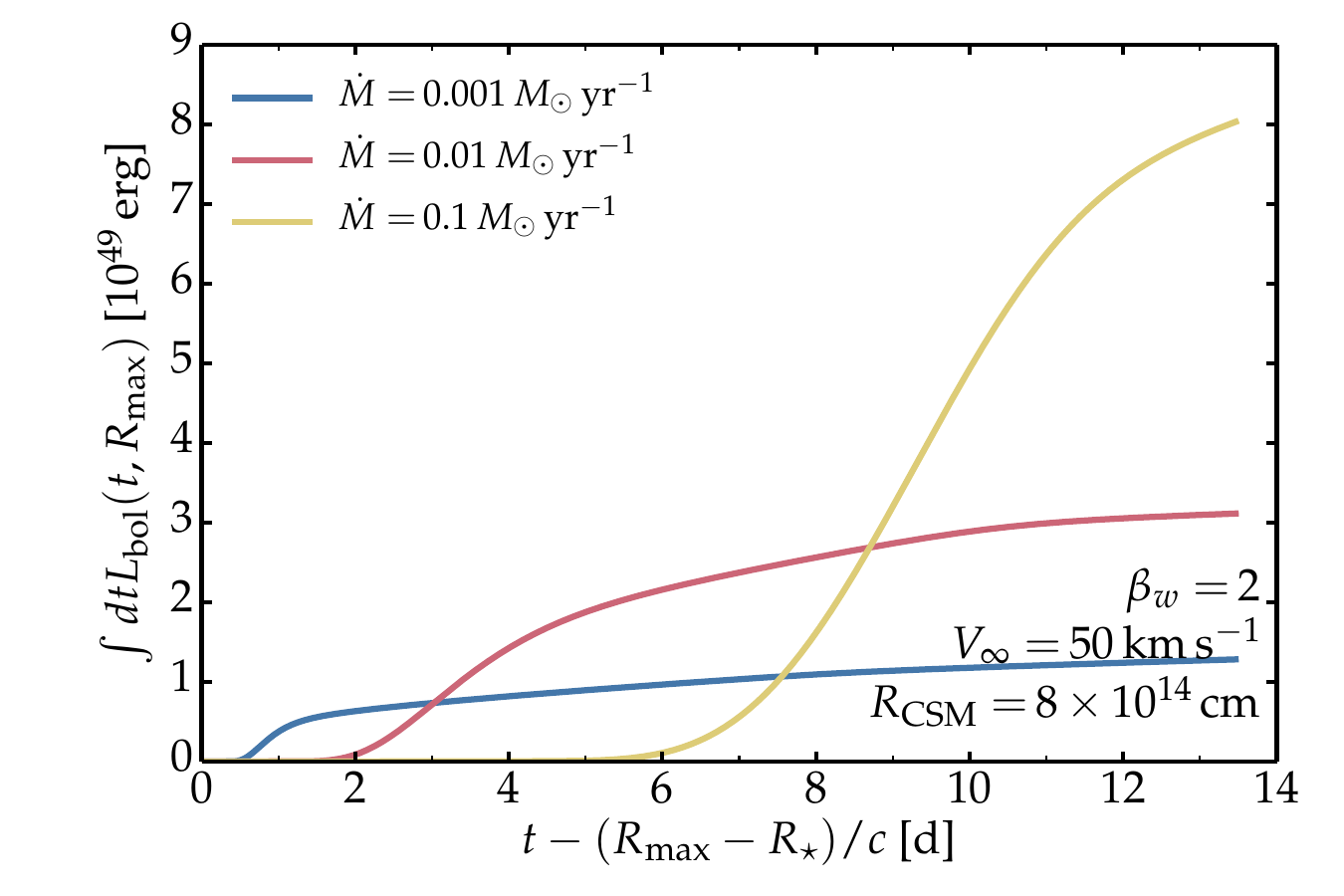}
\includegraphics[width=\hsize]{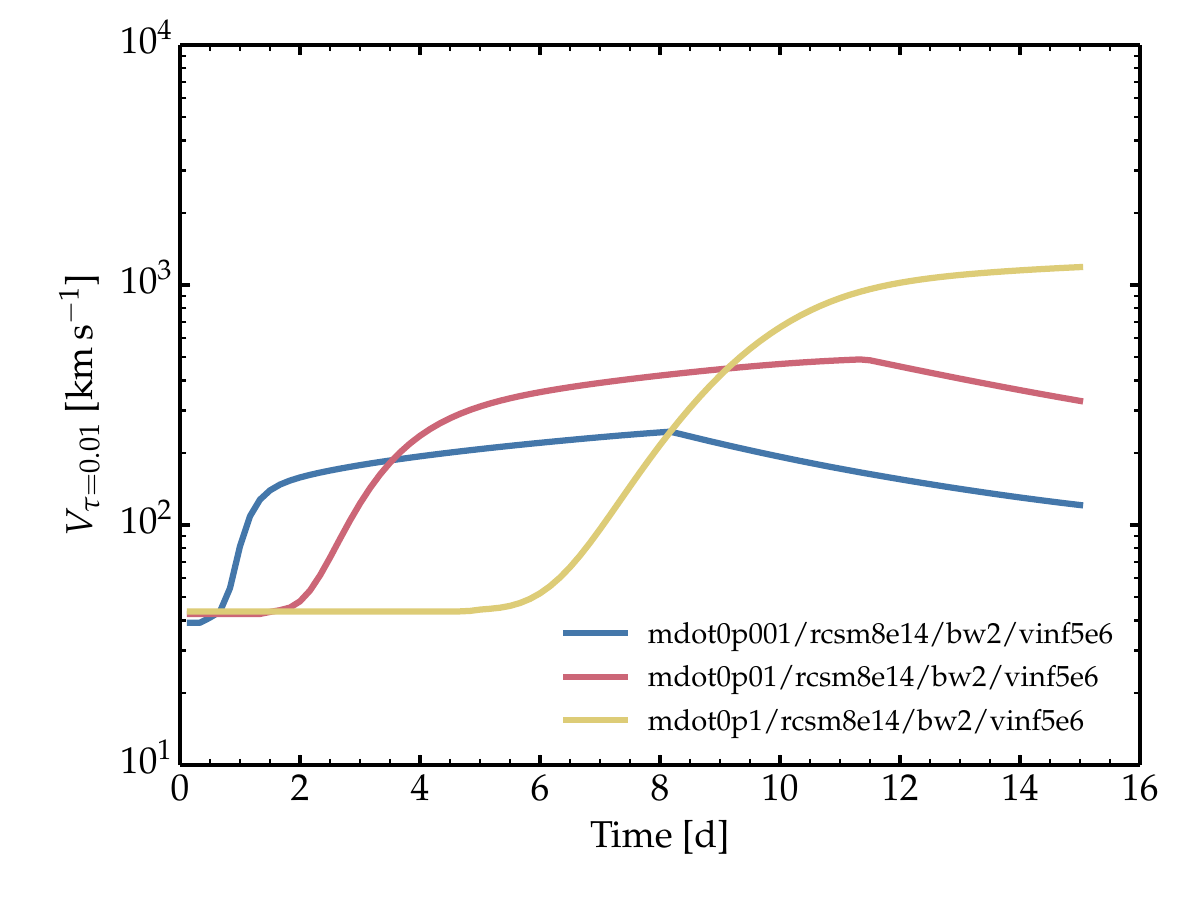}
\caption{Same as Fig.~\ref{fig_var_rcsm} but now for the model set differing in CSM density (i.e., \mdot).
\label{fig_var_mdot}
}
\end{figure}

Figure~\ref{fig_var_mdot} shows the evolution of $L_{\rm bol}$, its time integral, and $V_{\tau=0.01}$ for this model set. The factor of a hundred in mass loss rate implies a similar factor in CSM optical depth, which affects the location of the photosphere (at 3, 5.5, and $6.5 \times 10^{14}$\,cm) although not by as much as in the preceding section because the extent of the dense CSM is the same in all three models (with a dramatic drop in CSM density at $8 \times 10^{14}$\,cm). The extraction of kinetic energy from the CSM is modest in model mdot0p001 ($\sim$\,1.2\,$\times$\,10$^{49}$\,erg) but large in model mdot0p1  ($\sim$\,8\,$\times$\,10$^{49}$\,erg). 

The combination of a relatively low extraction of kinetic energy and the relative large radius of formation of the narrow-line emission region leads to a modest $V_{\tau=0.01}$ of at most $\sim$\,240\,\kms\ in model mdop0p001. For enhanced mass loss rate, $V_{\tau=0.01}$ reaches about 500\,\kms\ in model mdot0p01, and 1200\,\kms\ in model mdot0p1 (continuing that model mdot0p1 to later times would have yielded even higher velocities).

\begin{figure}
\centering
\includegraphics[width=\hsize]{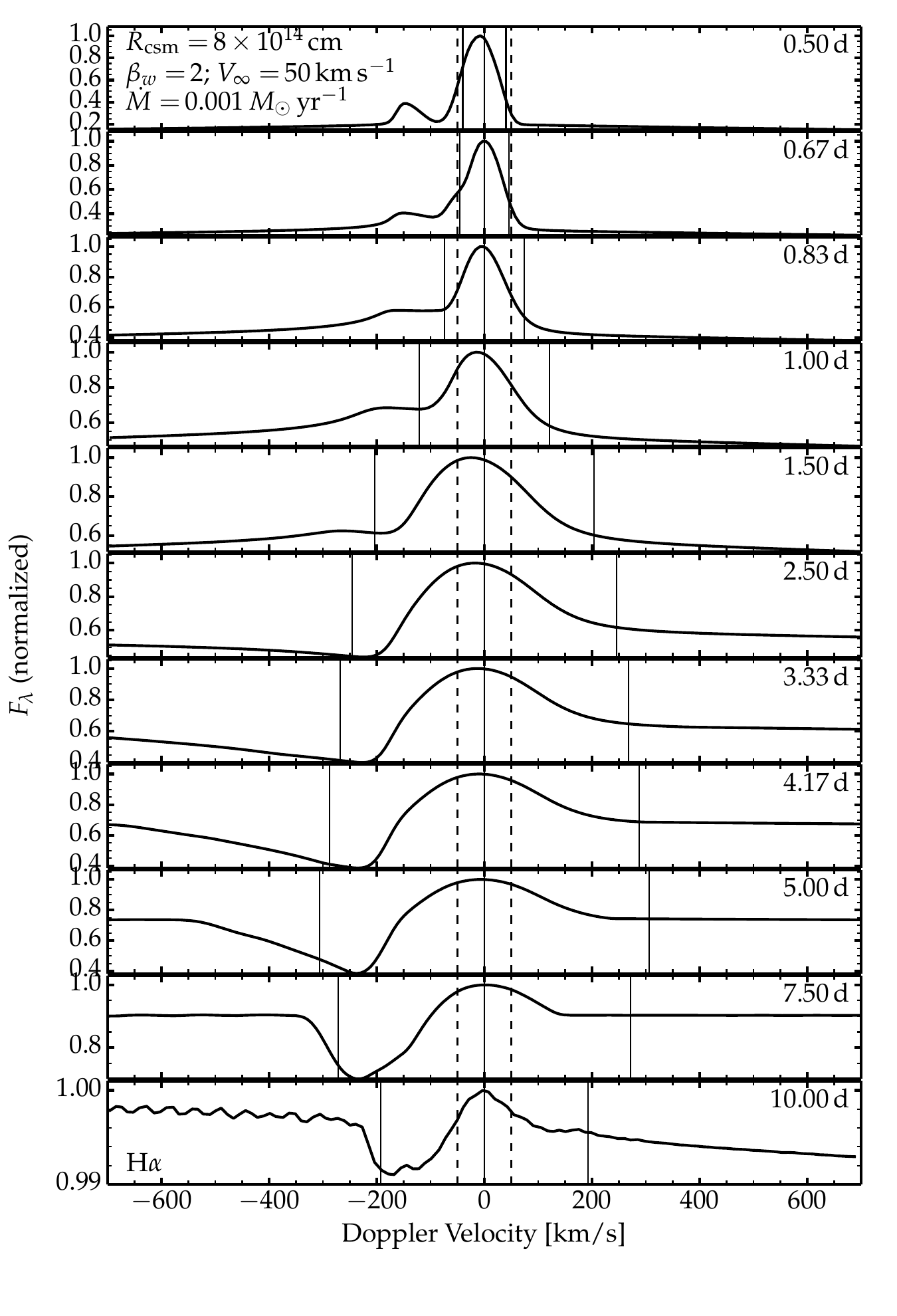}
\includegraphics[width=\hsize]{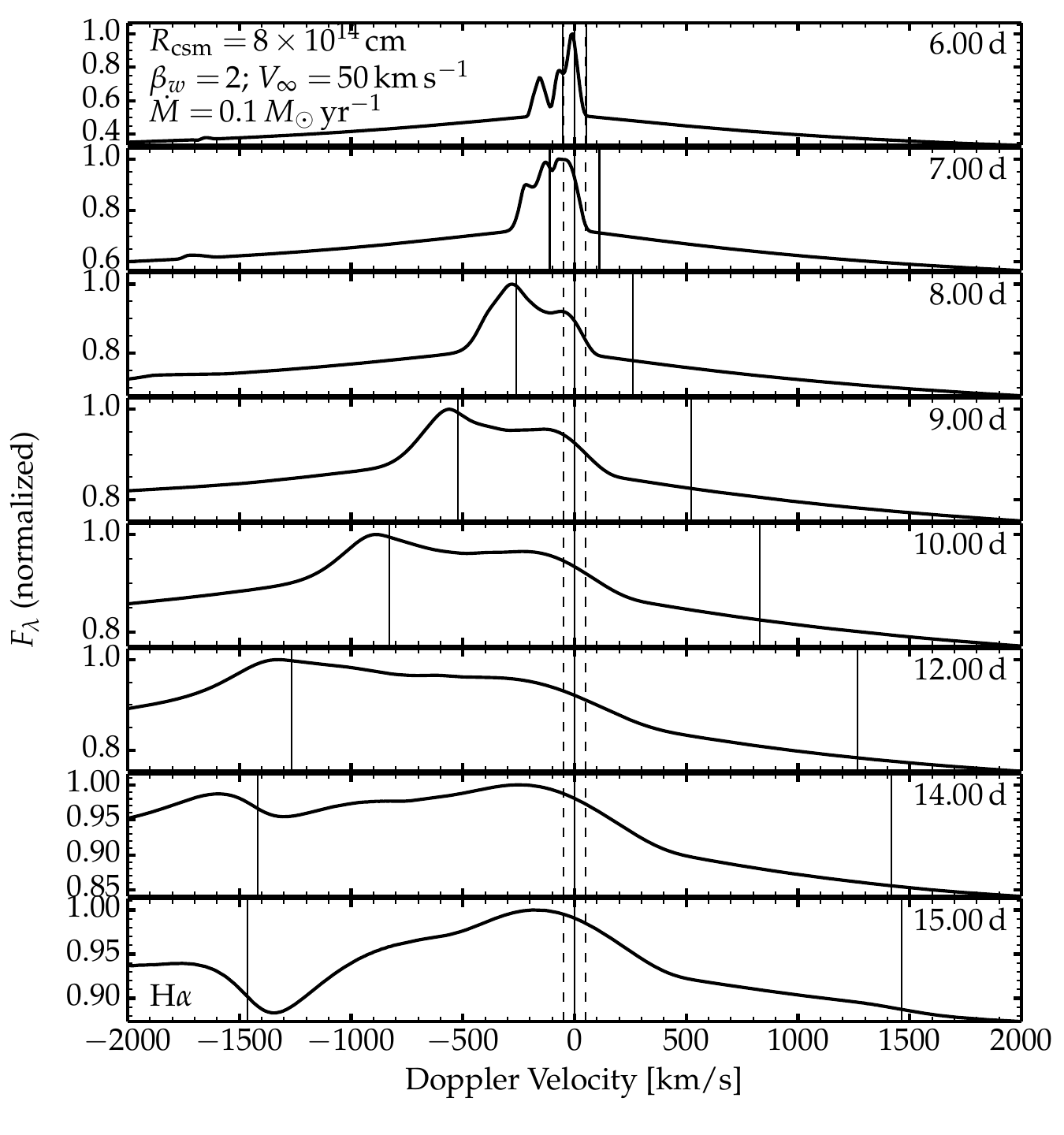}
\caption{Same as Fig.~\ref{fig_rcsm2e14_halpha} but now for models differing in \mdot, covering from low (top panel) to high mass loss rates (bottom panel).
\label{fig_var_mdot_halpha}
}
\end{figure}

Figure~\ref{fig_var_mdot_halpha} shows the evolution of the H$\alpha$ region for model mdot0p001 (bounds are $\pm$\,700\,\kms) and for model mdot0p1 (bounds are $\pm$\,2000\,\kms). In model mdot0p001, the influence of the radiative acceleration on H$\alpha$ is well identified, with a small and progressive broadening of the line after a fraction of a day, and the  development of a nearly symmetric and well defined P-Cygni profile. This lack of blueshift here is caused by the relatively low optical depth of the CSM, which affects little the distant, and in some sense detached regions where the narrow emission line component forms. The line also clearly exhibits the progressive broadening and subsequent narrowing at late times, as obtained in all model curves of $V_{\tau=0.01}$. Not surprisingly, the radiative acceleration in the model with high mass loss rate is considerably reduced. Model mdot0p1 shows the first signatures of H$\alpha$ broadening after 7\,d and this continues until the latest epoch at 15\,d when the line exhibits a P-Cygni profile with a maximum absorption in the trough at about 1300\,\kms. The narrow line emission exhibits a strong blueshift at all times because the CSM is now highly optically thick, causing a strong occultation effect of the mid-plane regions and those receding from the observer.

\subsection{Impact of wind velocity law}
\label{sect_dep_wind}

\begin{figure}
\centering
\includegraphics[width=\hsize]{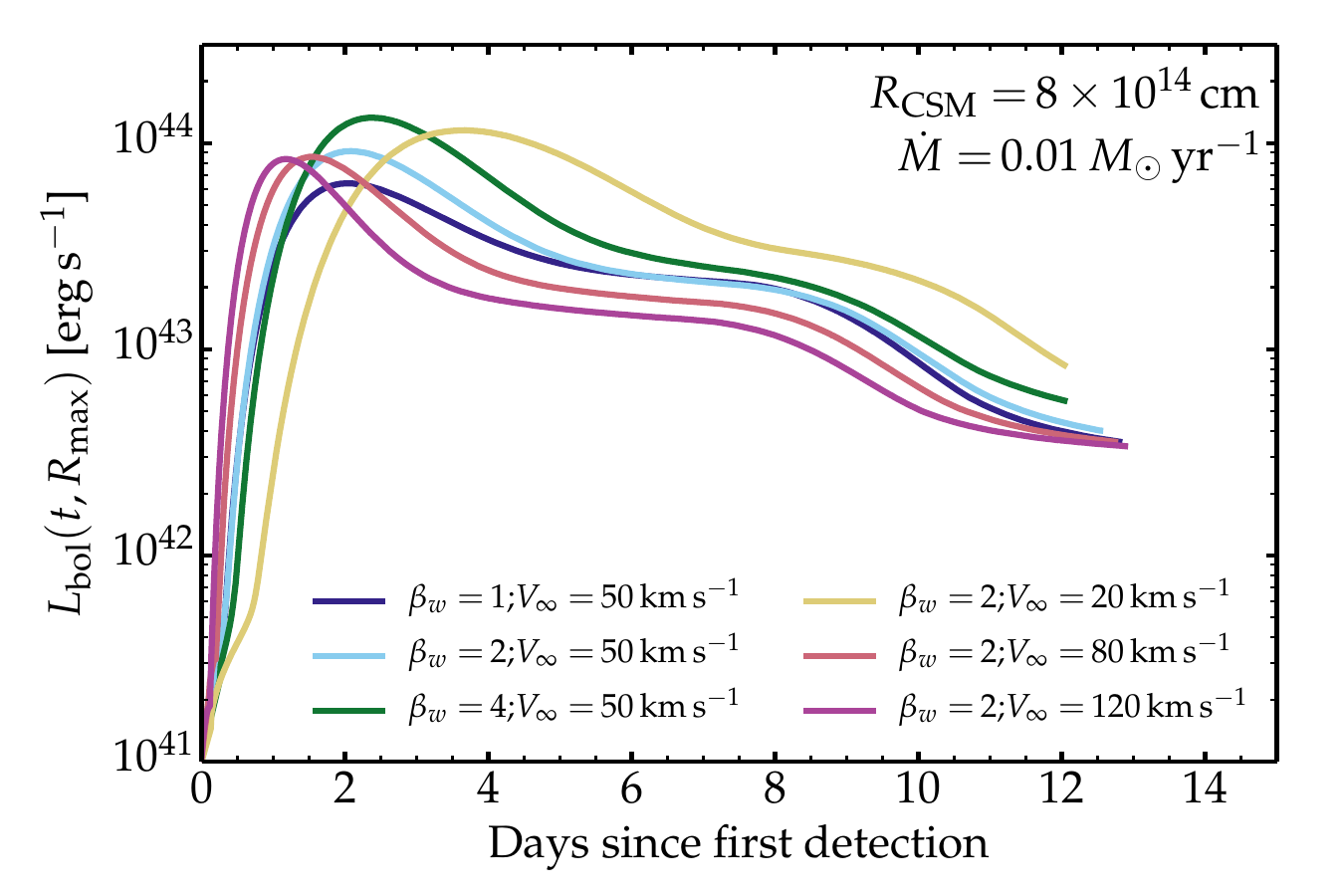}
\includegraphics[width=\hsize]{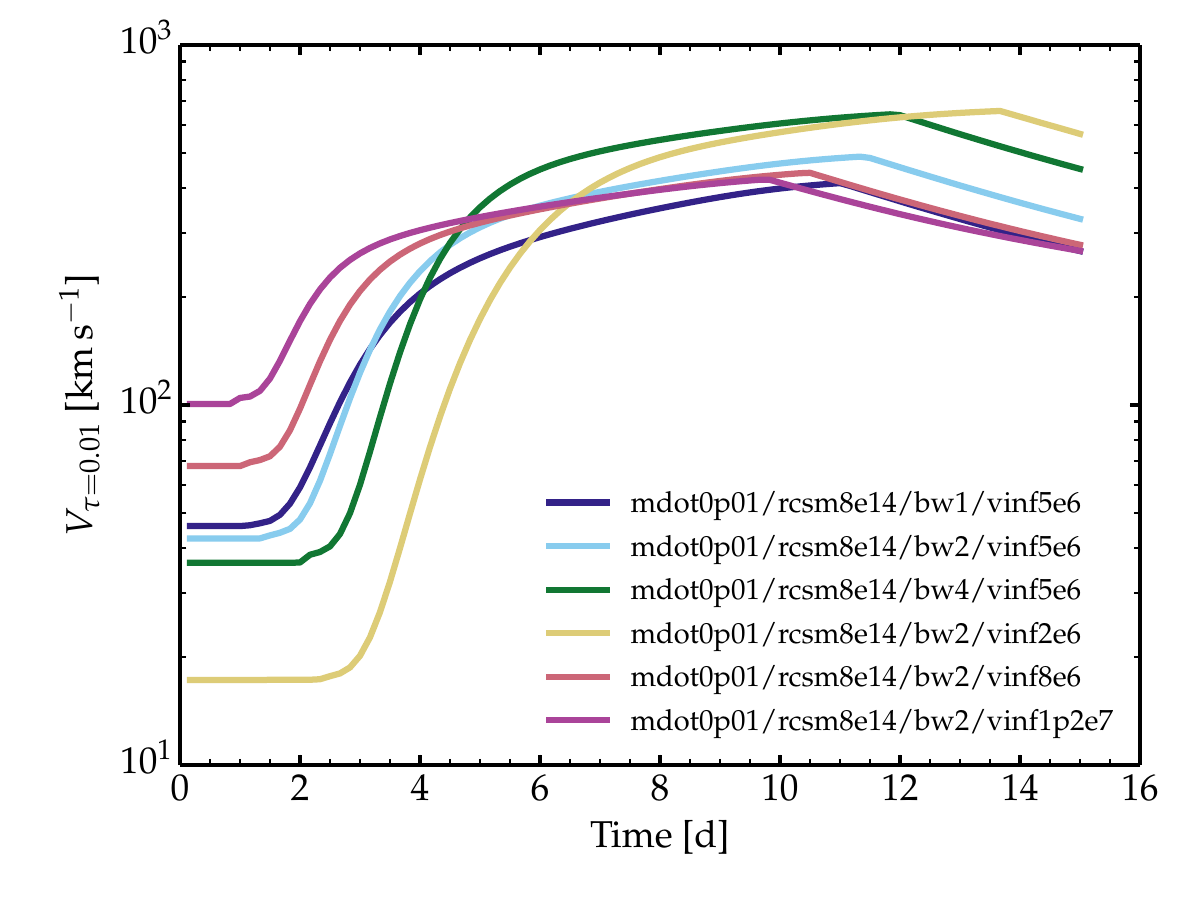}
\caption{Close counterpart to Fig.~\ref{fig_var_rcsm} but for models differing in \bw\ and \vinf. 
\label{fig_var_wind}
}
\end{figure}

In this final exploration, we consider CSM configuration that differ in wind velocity profile. All models use a mass loss rate of 0.01\,\msunyr\ and a value of $8 \times 10^{14}$\,cm for \rcsm\ for the high density CSM, but the wind velocity covers values of 1, 2, and 4 for \bw\ and 20, 50, 80, and 120\,\kms\ for the original CSM velocity. Because the CSM density varies as $\dot{M}/r^2V$, the different radial variation of the velocity imply different density profiles (even for the same \mdot). Consequently, the bolometric light curves shown in Fig.~\ref{fig_var_wind} (top panel) differ between models. In particular, winds with a long acceleration length scale (high values of \bw) or low terminal velocities initially (low \vinf) correspond to a CSM with a larger density, greater diffusion time, greater extraction of kinetic energy, and thus produce light curves with a longer rise time, possibly with a greater peak luminosity (but not necessarily), and a greater time-integrated bolometric luminosity.
 
The results for $V_{\tau=0.01}$ are different for every model. In all cases, $V_{\tau=0.01}$ initially indicates roughly the original value of \vinf, typically reduced by a few 10\,\% because this region is located in the accelerating part of the original, unadulterated wind. The value of  $V_{\tau=0.01}$ rises earlier in models with more tenuous CSM (high \vinf\ or low \bw). It reaches a large maximum value in models with a denser CSM, although that maximum is reached later -- the range of peak values covers roughly 400 to 700\,\kms\ at 10 to 14\,d after the onset of the simulation.
 
\begin{figure}
\centering
\includegraphics[width=\hsize]{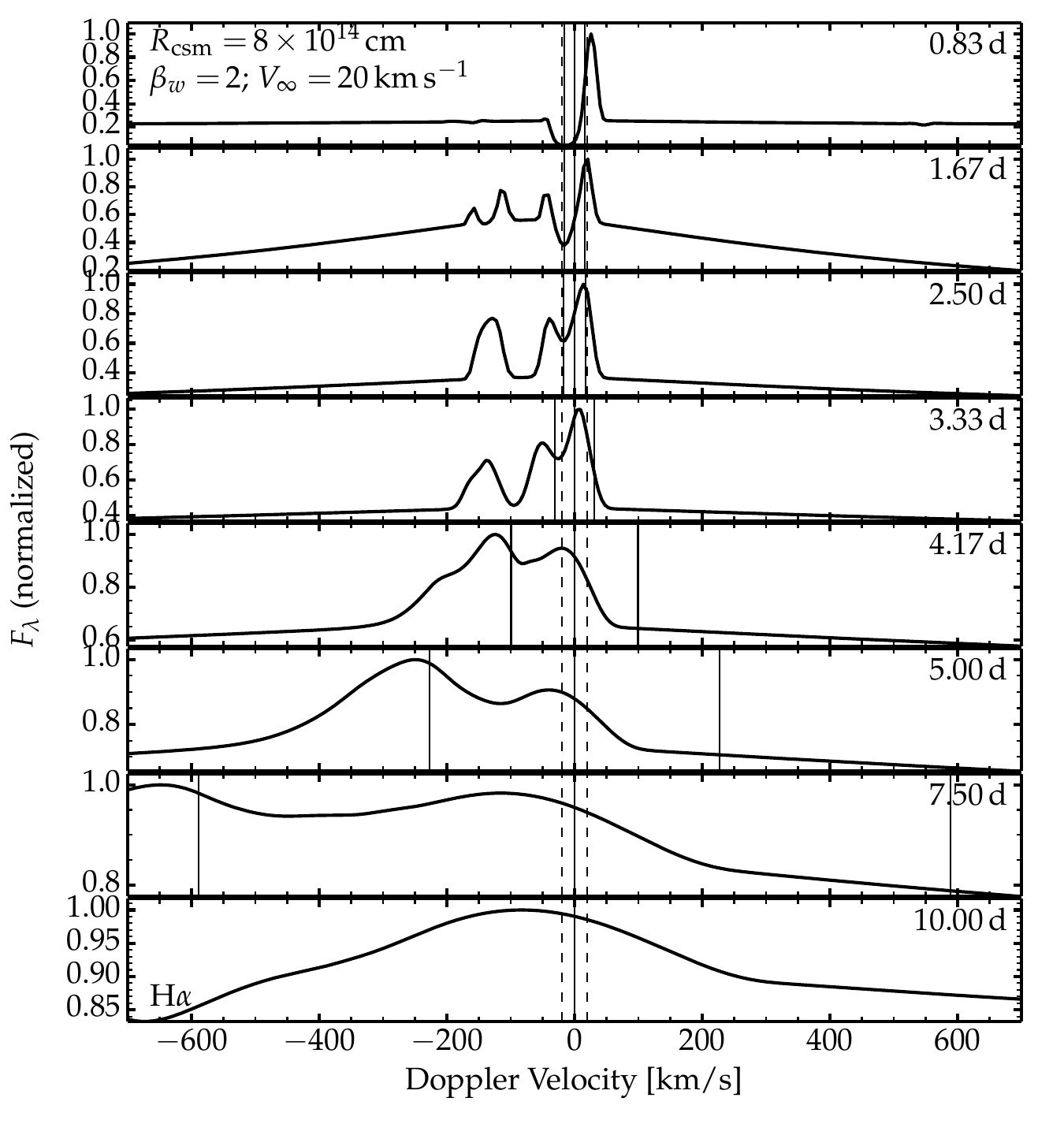}
\includegraphics[width=\hsize]{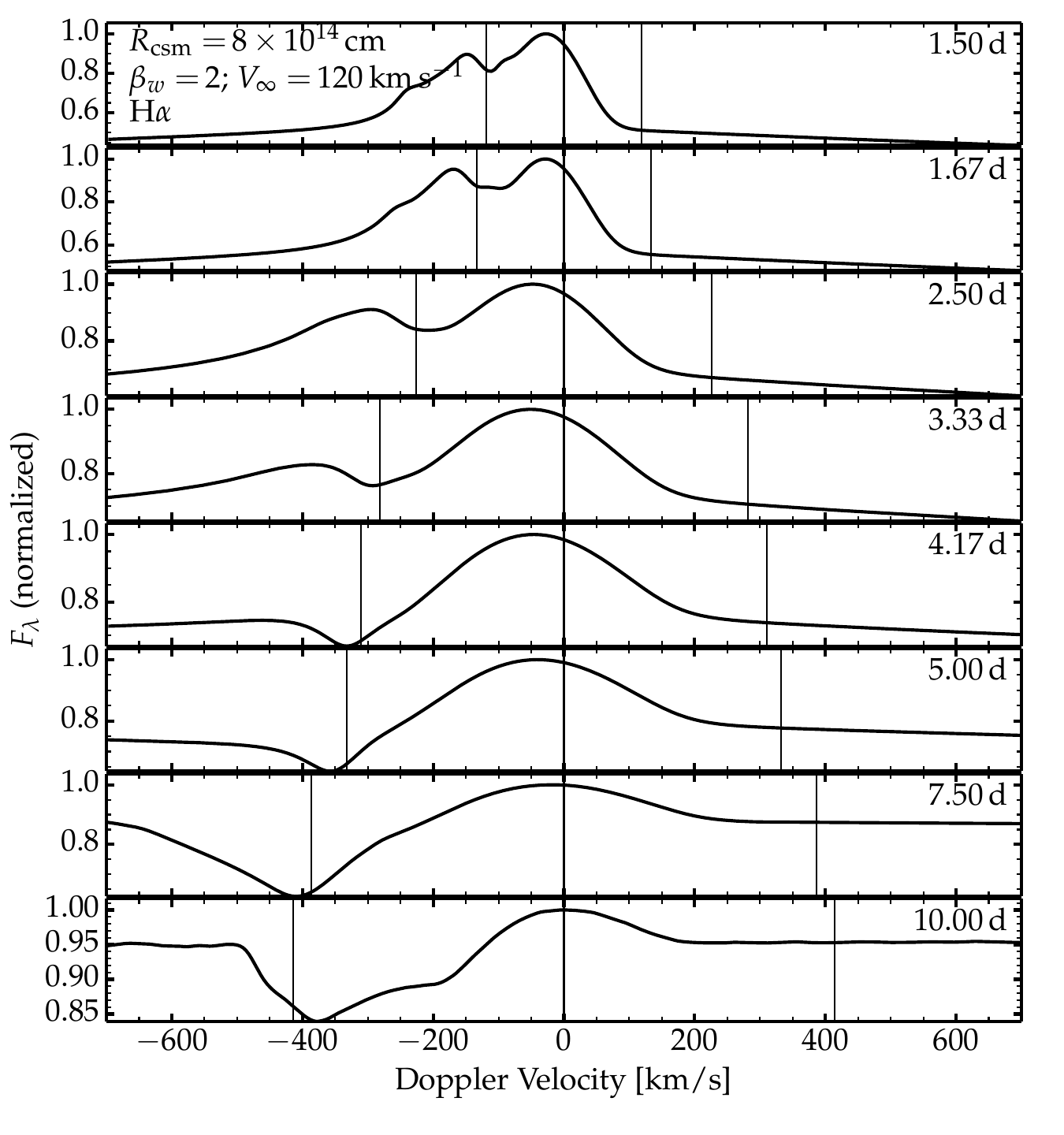}
\caption{Same as Fig.~\ref{fig_rcsm2e14_halpha} but now for two models differing in the original CSM velocity \vinf\ with a value of 20 (top panel) and 120\,\kms\ (bottom panel).
\label{fig_var_wind_halpha}
}
\end{figure}

Figure~\ref{fig_var_wind_halpha} shows the results for two specific cases differing in the value of \vinf, taking the extremes at 20 and 120\,\kms\ (\bw\ is two in both cases). For a \vinf\ of 20\,\kms, we were able to converge one model at very early times when the SN radiation has hardly crossed the CSM at all (the SN luminosity is still very small). At 0.83\,d, the model predicts a strongly blueshifted narrow H$\alpha$ component with a P-Cygni trough that extends to about $-$50\,\kms. We suspect this is due to photon leakage in the wings of H$\alpha$ where the line optical depth is lower. The line then changes in morphology as the CSM heats up but it remains relatively narrow until about 3\,d, after which the broadening starts in earnest until this broadening peaks at about 10\,d with a value of about 600\,\kms.

In the model with a larger initial value of \vinf\ of 120\,\kms, the narrow H$\alpha$ emission component starts broader 
and the broadening due to radiative acceleration kicks in already at 1.5\,d. The broadening is strongest in the blue part of the line and extends out to about 400\,\kms\ with a clear P-Cygni profile morphology at late times. As a reminder, this P-Cygni profile arises despite the decreasing velocity with radius. The central, narrow emission comes from the outer CSM regions whereas the extended blue absorption arises from deeper regions located nearer the photosphere.

It is interesting that the original velocity of the CSM is hard to infer from the emergent spectrum. The reason is that the radiative acceleration of the CSM starts immediately with shock breakout. The radiation from the embedded shock accelerates the CSM as is streams away from the shock and diffuses through the CSM. Thus, this radiative acceleration is intimately tied to the rising luminosity from the SN. Unless the SN is discovered at the earliest times when its luminosity is still very low, the CSM probed by high-resolution spectra will already be affected by radiative acceleration. 
However, because of radiative acceleration, the need for the highest spectral resolution is necessary only at the earliest times. Once the narrow emission line component has broadened, the main limiting factor may instead be the signal-to-noise ratio. The exception is for configurations in which distant CSM is still optically-thick in some lines like H$\alpha$ and produce a persisting, narrow, through likely weak P-Cygni profile (see for example the case of SN\,2020pni; \citealt{terreran_20pni_22}). 

\section{Broad- versus narrow-line emission evolution}
\label{sect_broad_vs_narrow}

The information contained in high-resolution spectra is complementary to that present in low-resolution spectra. Any information in high-resolution spectra and over a broad wavelength range should match that present in low-resolution spectra. This is a critical check that no erroneous normalization of high-resolution spectra has been done. With high resolution, we capture the narrowest line features, and because they are obtained with echelle spectra, broad lines that lie over multiple orders are compromised. Thus, high-resolution spectra are mostly useful to constrain the distant CSM where line broadening effects due to radiative acceleration and electron scattering are reduced. In contrast, low- or medium-resolution spectra provide critical information on the regions closer to the photosphere, where line broadening by electron scattering, by radiative acceleration, and eventually by the fast expansion of the shocked material are large. Hence, a combination of low-/medium- and high-resolution, high cadence observations is needed to build a complete picture of SN explosions. Such data are available at best for a few objects only.

\begin{figure*}
\centering
\includegraphics[width=0.49\hsize]{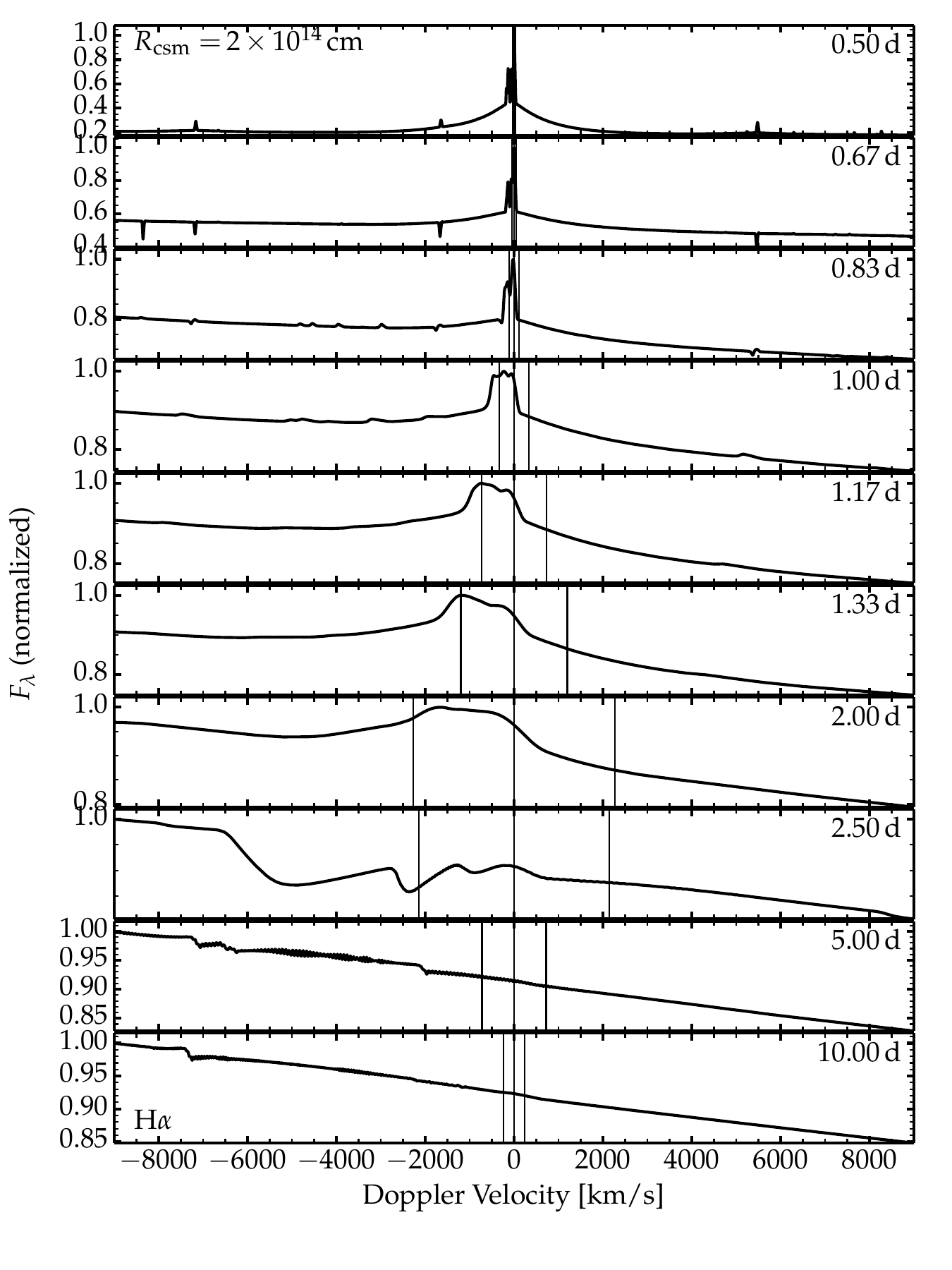}
\includegraphics[width=0.49\hsize]{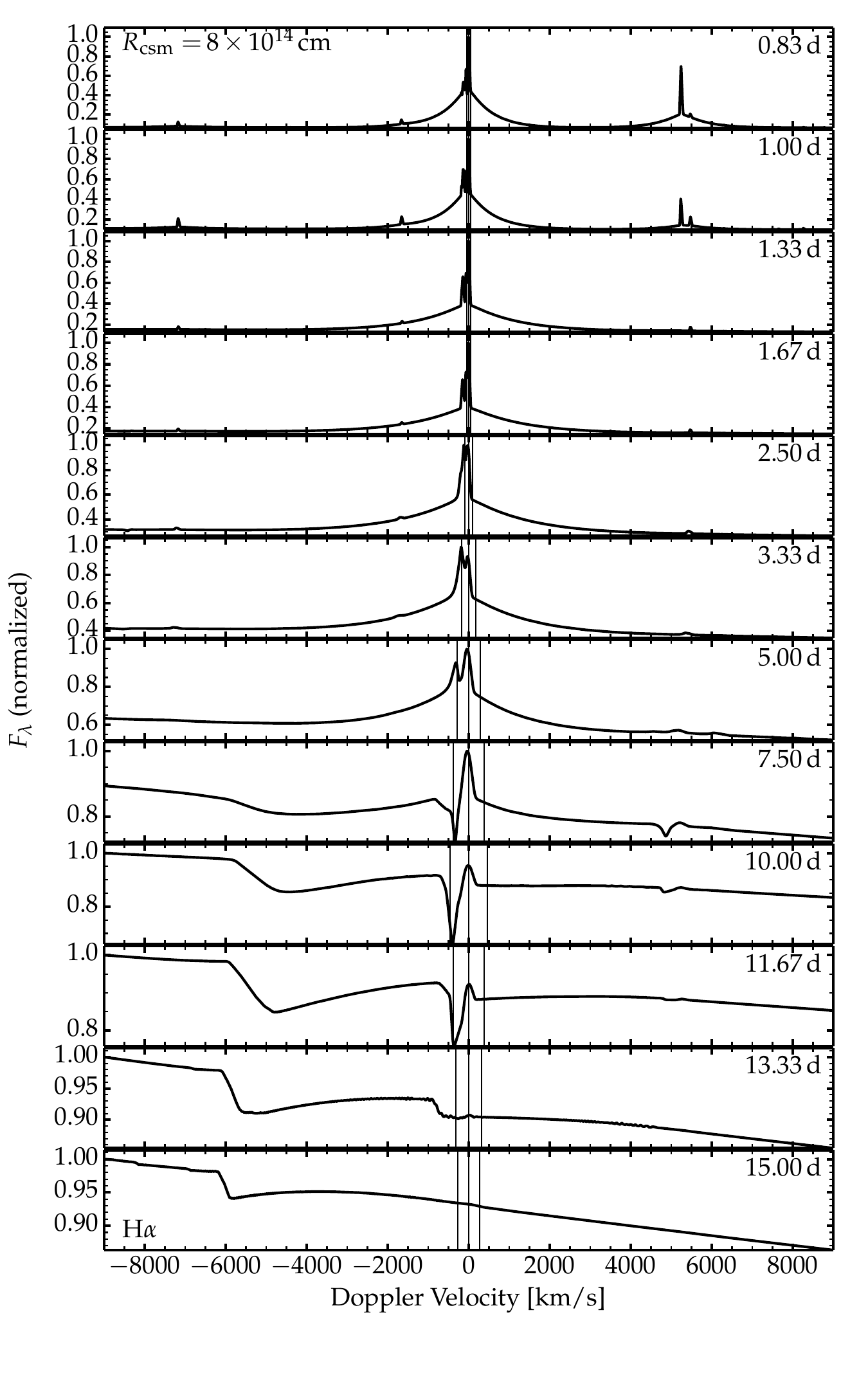}
\caption{Evolution of the broad and narrow emission components in the H$\alpha$ spectral region from $\sim$\,0.5\,d until $\sim$\,15\,d after shock breakout for model rcsm2e14 (left) and rcsm8e14 (right). The symmetric solid vertical lines on each side of the origin correspond to the velocity at the location in the unshocked CSM where the electron-scattering optical depth is 0.01. For the CDS velocity of $\sim$\,6000\,\kms\ in the rcsm8e14 model, the He\one\,6678\,\AA\ trough extends out to the H$\alpha$ rest wavelength and thus may overlap with a narrow H$\alpha$ component. 
\label{fig_broad_vs_narrow}
}
\end{figure*}

Figure~\ref{fig_broad_vs_narrow} illustrates the combined evolution of the narrow and broad emission (and absorption) components of H$\alpha$ for models rcsm2e14 and rcsm8e14 (the other model parameters for both are $\dot{M}=$\,0.01\,\msunyr, \bw$=$\,2, and \vinf$=$\,50\,\kms). In both models, the early phase with the narrowest line emission component is also the phase during which the broad component exhibits strong electron-scattering broadened wings extending out to 2000-3000\,\kms\ from line center. In model rcsm2e14, the wings weaken quickly and the broadened narrow component persists until about 2\,d. By 2.5\,d, the entire H$\alpha$ profile is in absorption, with multiple components reflecting absorption from the unshocked CSM around the photosphere and from the fast-moving dense shell. The spectrum then becomes quasi featureless when the entire CSM has been swept up inside the dense shell and a weak H$\alpha$ absorption appears at about $-$7500\,\kms. 

In model rcsm8e14, the evolution is comparable but the electron-scattering broadened wings survive until about 5\,d. During the next 5\,d, the photosphere is in the fast-moving dense shell and one sees a strong broad blueshifted absorption with a maximum absorption at about $-$6000\,\kms\ (this is the velocity of the dense shell). However, a narrow P-Cygni profile persists in this case until about 13\,d and suggestive of a residual unshocked CSM at large distances but relatively broadened due to radiative acceleration. By 15\,d, the narrow component has vanished.
 
\section{Discussion and conclusions}
\label{sect_conc}

In this paper, we have presented radiation-hydrodynamics simulations of red-supergiant stars exploding within CSM characterized by different extent (i.e., \rcsm), density (i.e., \mdot) and velocity profile (\bw\ and \vinf\ parameters). For many of these interaction configurations (five with different values of \rcsm; three with different \mdot; three with different \vinf), we computed NLTE radiative-transfer simulations at multiple epochs covering from the onset of shock breakout until the ``ejecta" phase when the CSM has been fully swept-up by the shock and when the spectrum forms primarily within the ejecta. By combining radiation hydrodynamics and detailed radiative transfer, we can document how the dynamical properties of the ejecta and CSM translate into spectral properties both for the broad and the narrow line components. Indeed, at all times, our synthetic spectra reveal emission and absorption from distinct regions, potentially including the unshocked CSM, the shocked CSM, the shocked ejecta, and the unshocked ejecta. Each of these regions has distinct velocity, temperature, and density and thus leaves a specific imprint on spectra. The subject of this study was to document such properties with a particular emphasis on the information contained on the smallest velocity scale and revealed in high-resolution spectra. The main limitations of this work are the assumption of steady-state, spherical symmetry, and the neglect of line opacity, although we surmise that these are  secondary relative to the importance of radiative acceleration and optical-depth effects.

In all simulations, we find that the unshocked CSM undergoes significant radiative acceleration prior to being shocked, as has been proposed in the past \citep{fransson_93j_96,chugai_98S_02,tsuna_grad_23}. This acceleration starts at the onset of shock breakout and essentially persists for as long as there is SN radiation. However, this acceleration is strongest when the SN luminosity is large and thus primarily during the early photospheric phase. The acceleration is caused by momentum transfer from radiation to the gas and scales with the flux (i.e., not the luminosity). The radiative acceleration is therefore maximum near the shock and declines as $1/r^2$ with increasing distance above the shock. Because the agent leading to radiative acceleration is the SN radiation, the original CSM velocity is altered immediately after shock breakout -- our ability to infer the original CSM velocity is thus strictly impossible since it would require obtaining high-resolution spectra of the exploding star before shock breakout, thus before the SN has even been discovered. The best that can be done is to obtain high-resolution spectra at the earliest epochs after shock breakout, when the SN luminosity is still modestly greater than the progenitor luminosity. And the earlier the better.

We also find that the radiative acceleration can give rise to very large velocities of the unshocked CSM, with maximum values of several 1000\,\kms\ and thus barely below the velocity of the shocked material (which is no less than about 5000\,\kms, whatever the CSM mass and extent). Such a strong radiative acceleration just ahead of the shock weakens the shock and is a strong limitation for X-ray emission at early times after shock breakout. In fact, the actual shock power is much weaker than the SN luminosity leaking from behind the shock at those early times and could explain the low X-ray luminosities observed in SNe II-P interacting with CSM at early times. Overall, the shock appears strong once it has cleared out of the dense CSM indicated by \rcsm\ in our simulations.

In interacting SNe, the narrow emission line component arises from regions external to the photosphere. Indeed, at and below the photosphere (where the radiative acceleration is greater), frequency redistribution caused by electron scattering is an important line broadening mechanism, depleting line core photons and boosting emission line wings. In contrast, beyond the photosphere, the electron-scattering optical depth is small and the main line broadening mechanism is Doppler broadening, which thus reflects the relatively low expansion velocities of the distant, unshocked CSM. A good proxy for this region is where $\tau_{\rm es}=$\,0.01 and we thus refer to $V_{\tau=0.01}$. 

With our large grid of simulations, we find that the radiative acceleration of such distant CSM (where narrow line emission forms) leads to radial velocities that exhibit a bell shape morphology. Initially, the velocity rises as the CSM becomes ionized by the SN radiation (i.e., $V_{\tau=0.01}$ rises to esentially \vinf). As more flux reaches those outer regions, $V_{\tau=0.01}$ slowly rises to a few 100\,\kms\ in extended or dense CSM or rapidly rises to several 1000\,\kms\ in cases of compact or tenuous CSM. The main reason for this difference is the scaling of the radiative acceleration with the flux, which is maximum for configurations in which the region with $\tau=$\,0.01 is close to the shock. This is paradoxical since it indicates that the radiative acceleration to be observed from high-resolution spectra will be maximum in SNe II with less CSM. In contrast, strong interactions will show much reduced acceleration because the regions with $\tau=$\,0.01 is at large distances where the radiative flux is strongly diluted.

By post-processing our radiation-hydrodynamics calculations, we can go beyond speculations and confirm the previous findings with detailed NLTE radiative-transfer calculations. Using a reference model with 0.01\,\msunyr, \rcsm$=8 \times 10^{14}$\,cm, $\beta_{\rm w}=$\,2, and $V_\infty=$\,50\,\kms, we documented the evolution of the narrow line component of H\one, He\one, He\two, C\three, C\four, and N\four\ lines in the optical. We find that their behavior mirrors what is obtained for the broad component in terms of overall strength because of the similar evolution in ionization of the optically thick and optically thin regions lying on each side of the photosphere. For example, He\one\ lines are strong initially, weaken as the CSM ionization increases after shock breakout, are eventually reborn after shock passage and further cooling of the spectrum formation region, and ultimately disappear as recombination kicks in. Lines of He\two\ or C\four\ have similarly very limited survival times since they require a higher ionization or temperature that is fundamentally short lived. 

In all cases, we find that the narrow emission line component is strongly impacted by radiative acceleration. These narrow spectral features are  therefore very sensitive probe of the unshocked CSM. We find that they reflect closely the evolution of $V_{\tau=0.01}$ reported above. The original, narrowest emission lines are present only at the earliest times, with broadening of the line emission as the SN luminosity ramps up. The lines can become significantly broad and no longer require the highest resolution in order to be revealed. They probe CSM velocities of several 100\,\kms\ and may thus be resolved with even standard spectral resolution. The signal-to-noise ratio may instead be a stronger requirement since the broadened lines may be weak relative to the underlying broad, electron-scattering broadened emission. Similarly, inadequate setting of the ``continuum'' emission level may also lead to an inadequate interpretation of the broadening of the narrow emission line component. This issue may impact data obtained with echelle spectra.

A second property is the blueshift of narrow emission features. This blueshift may be small early on in lines forming over a relatively extended volume of the CSM like H$\alpha$. Lines of higher ionization like C\four\,5801.31\,\AA\ or N\four\,7122.98\,\AA\ show a strong blueshift, which results from the fact that these lines form close to the photosphere where the temperature and ionization are higher. Consequently, they are more strongly affected by optical-depth effects. As the unshocked CSM becomes ionized, optically-thick, and radiatively accelerated, the blueshift increases and the bulk of the ``narrow'' line emission occurs blueward of the rest wavelength.

A narrow-line emission may be present until late times if the distant CSM is dense enough. In our simulations, we adopt a low outer CSM mass loss rate of 10$^{-6}$\,\msunyr\ so that a narrow emission component is essentially lacking at the end of our simulations at 15\,d (except in our models with the highest mass loss rate of 0.1\,\msunyr\ or extent of 10$^{15}$\,cm). Such a detection of a narrow H$\alpha$ line up until a month after explosion has been made for SN\,1998S by \citet{fassia_98S_01} and revealed typical absorption at 50\,\kms\ from the rest wavelength not just of H$\alpha$ but of multiple H\one\ or He\one\ lines. It suggests a gradual decline of the CSM density, perhaps with mass loss rates of 10$^{-4}$\,\msunyr\ at large distances, as well as an original CSM velocity probably close to 50\,\kms\ since this distant material should experience very modest radiative acceleration. We leave to future work the modeling of this more distant material.

Our simulations indicate that radiative acceleration of the unshocked CSM may lead to velocities as high as several 1000\,\kms. Such values remain smaller than the post-shock velocities, which in our standard explosion model are systematically greater than 5000\,\kms, whereas in the absence of CSM these velocities reach about 10000\,\kms. This indicates that the so-called ``intermediate-width'' emission component at around 1000\,\kms\ cannot be associated with post-shock gas -- it is emission from unshocked CSM that has either been radiatively accelerated or electron-scattering broadened. 

There is at present limited high-resolution data that our simulations can be compared to. High-resolution spectra of SN\,1998S at early times \citep{leonard_98S_00,fassia_98S_01} are comparable with our models with extended, dense CSM like models mdot0p1 or rcsm1e15. The high resolution observations of SN\,2024ggi at about 1--1.5\,d after explosion reported by \citet{pessi_24ggi_24} and \citet{shrestha_24ggi_24} are in good agreement with our model rcsm4e14 and analogs in terms of the narrow line evolution and blueshift. Unfortunately, these observations do not cover until later times. The adopted SN redshift by \citet{pessi_24ggi_24} suggests a complete offset of all lines relative to their rest wavelength by about 70\,\kms, which is in conflict with our model predictions -- models predict emission line blueshifts but with emission present at the rest wavelength and little emission redward of the rest wavelength. The observations of SN\,2023ixf reported by \citet{smith_23ixf_23} do show similar patterns with a progressive broadening of the H$\alpha$ emission over the first two weeks in a similar fashion to what is obtained here. There is also a clear narrow emission that persists at later times, as reported for SN\,2023ixf (see, e.g., Fig.~4 in \citealt{jacobson_galan_23ixf_23}) or for SN\,2024ggi \citep{wynn_24ggi_24,zhang_24ggi_24}. A detailed confrontation of the simulations presented in this work to observed high-resolution spectra of SNe II with interaction is deferred to a forthcoming study.

Overall, SN\,2020pni probably offers the most complete spectral evolution of narrow emission line components \citep{terreran_20pni_22}, reflecting that obtained in the present simulations, namely with a narrow component broadening with time until a few days and then becoming narrower at later times as the narrow-line formation region shifts to more distant locations where radiative acceleration was much weaker. This bell-shape morphology of the width of narrow-line emission (with possible absorption) is a fundamental signature of radiative acceleration in extended dense CSM surrounding exploding RSG stars.

\begin{acknowledgements}

LD thanks Wynn Jacobson-Gal\'an for fruitful discussions. This work was supported by the ``Programme National Hautes Energies'' of CNRS/INSU co-funded by CEA and CNES. This work was granted access to the HPC resources of TGCC under the allocation 2023 -- A0150410554 on Irene-Rome made by GENCI, France. This research was supported by the Munich Institute for Astro-, Particle and BioPhysics (MIAPbP) which is funded by the Deutsche Forschungsgemeinschaft (DFG, German Research Foundation) under Germany's Excellence Strategy  -- EXC-2094 -- 390783311. This research has made use of NASA's Astrophysics Data System Bibliographic Services.

\end{acknowledgements}

\appendix
\section{Additional illustrations}

\begin{figure}
\centering
\includegraphics[width=\hsize]{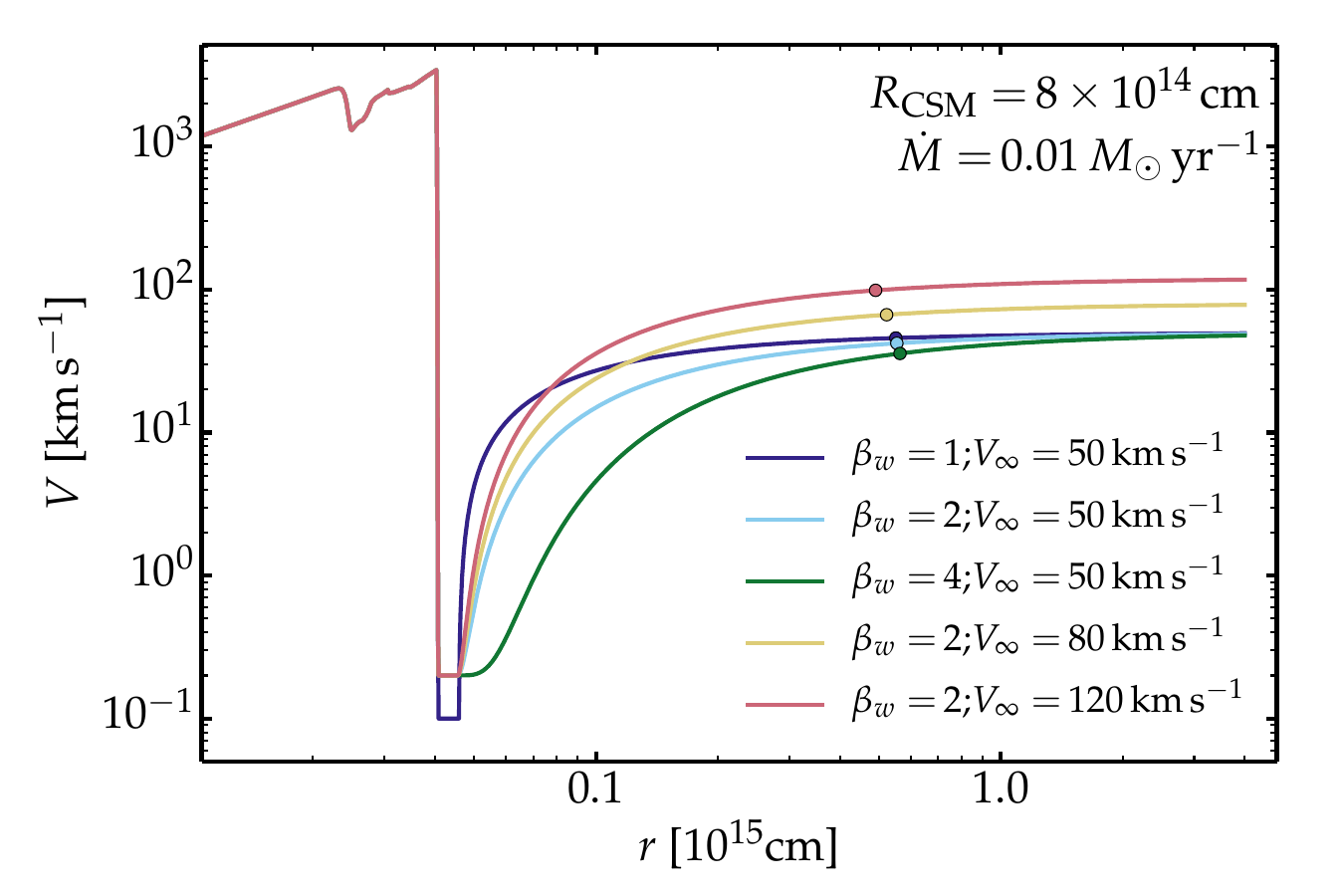}
\caption{Illustration of the velocity profile versus radius for the models discussed in Section~\ref{sect_dep_wind} and whose density profiles are shown in the bottom panel of Fig.~\ref{fig_init}.
\label{fig_init_wind_vel}
}
\end{figure}

We show additional illustrations. Fig~\ref{fig_init_wind_vel} shows the wind velocity profile adopted, varying in \bw\ and \vinf. The velocity at the hydrostatic base is adjusted in all cases so that the density gradient is constant at $R_\star$ (see discussion in Section~\ref{sect_setup}).

\begin{figure*}
   \centering
    \begin{subfigure}[b]{0.33\textwidth}
       \centering
       \includegraphics[width=\textwidth]{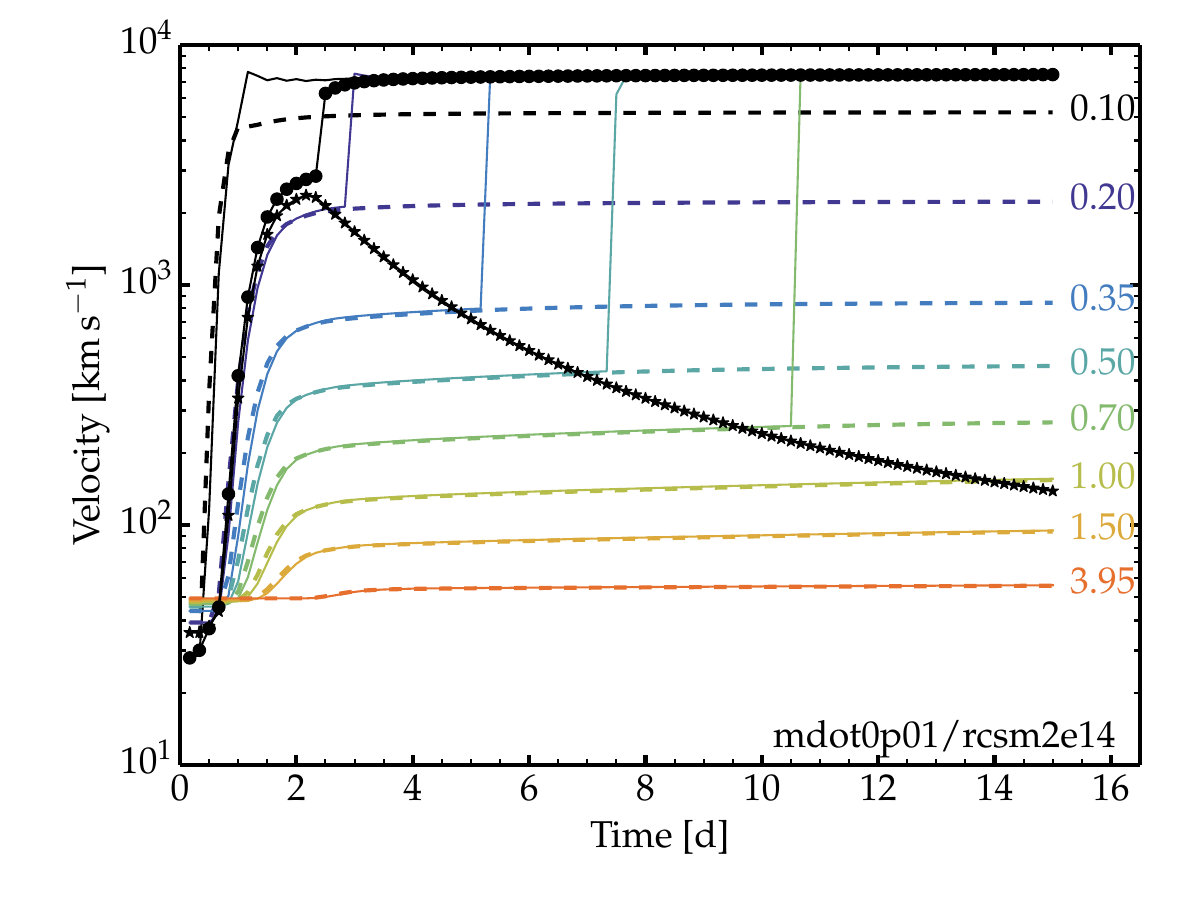}
    \end{subfigure}
    \hfill
    \begin{subfigure}[b]{0.33\textwidth}
       \centering
       \includegraphics[width=\textwidth]{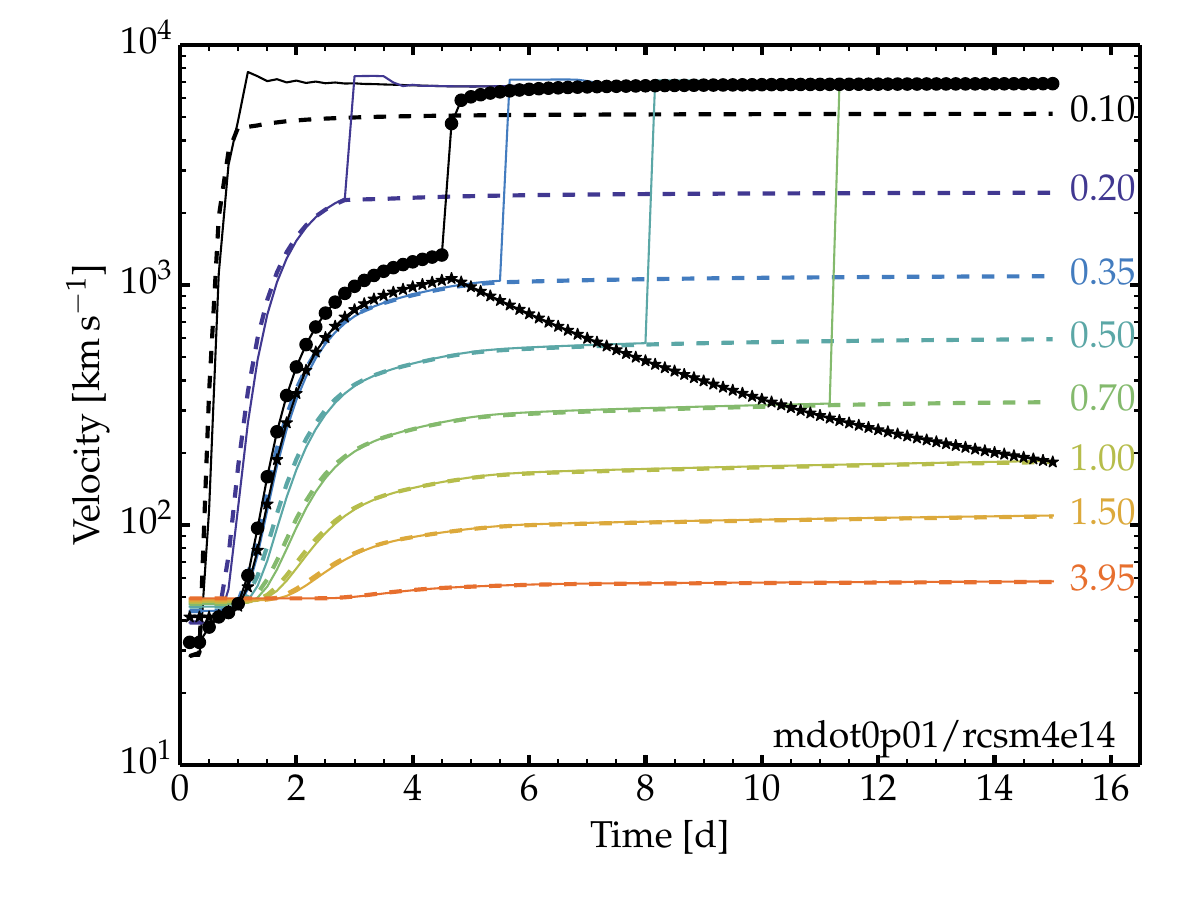}
    \end{subfigure}
    \hfill
    \begin{subfigure}[b]{0.33\textwidth}
       \centering
       \includegraphics[width=\textwidth]{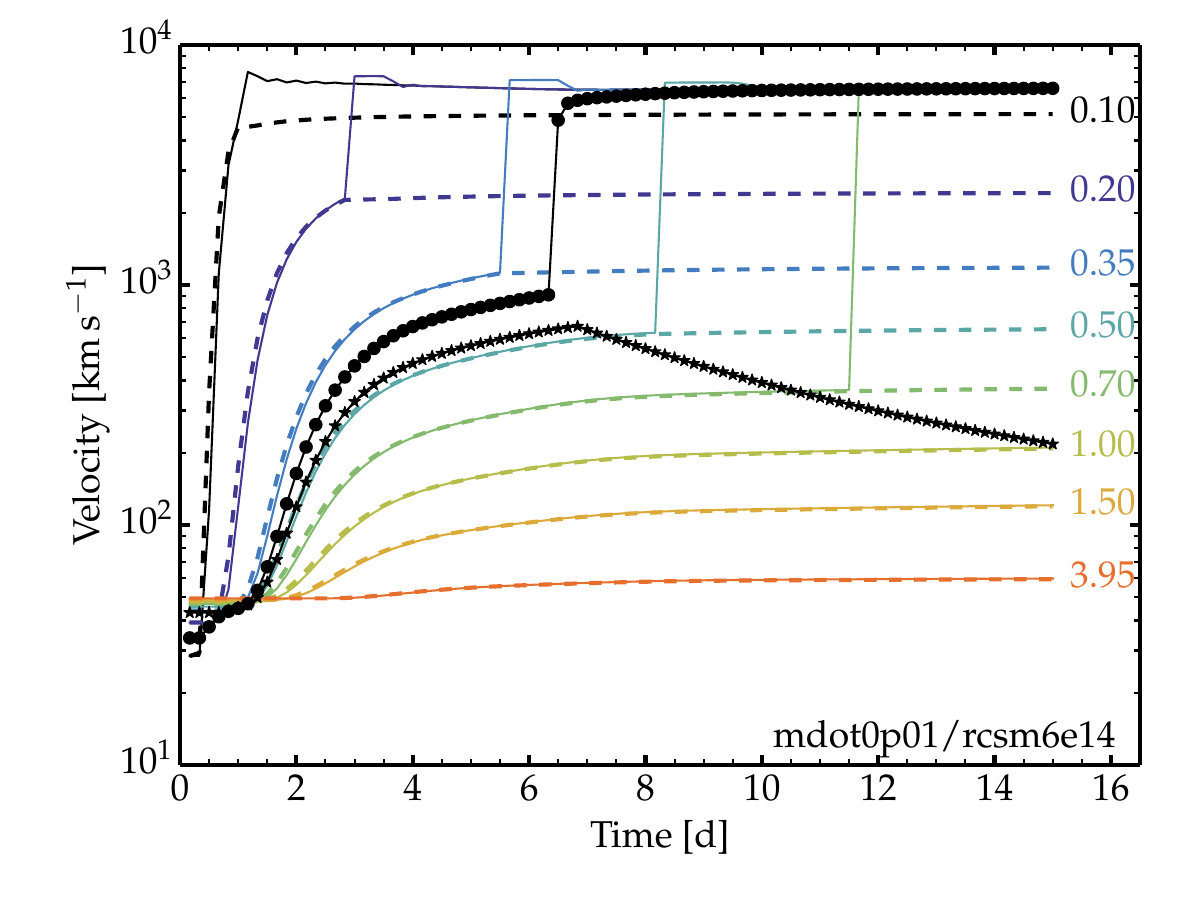}
    \end{subfigure}
    \hfill
    \begin{subfigure}[b]{0.33\textwidth}
       \centering
       \includegraphics[width=\textwidth]{evol_v_at_r_mdot0p01_rcsm8e14.pdf}
    \end{subfigure}
    \hfill
    \begin{subfigure}[b]{0.33\textwidth}
       \centering
       \includegraphics[width=\textwidth]{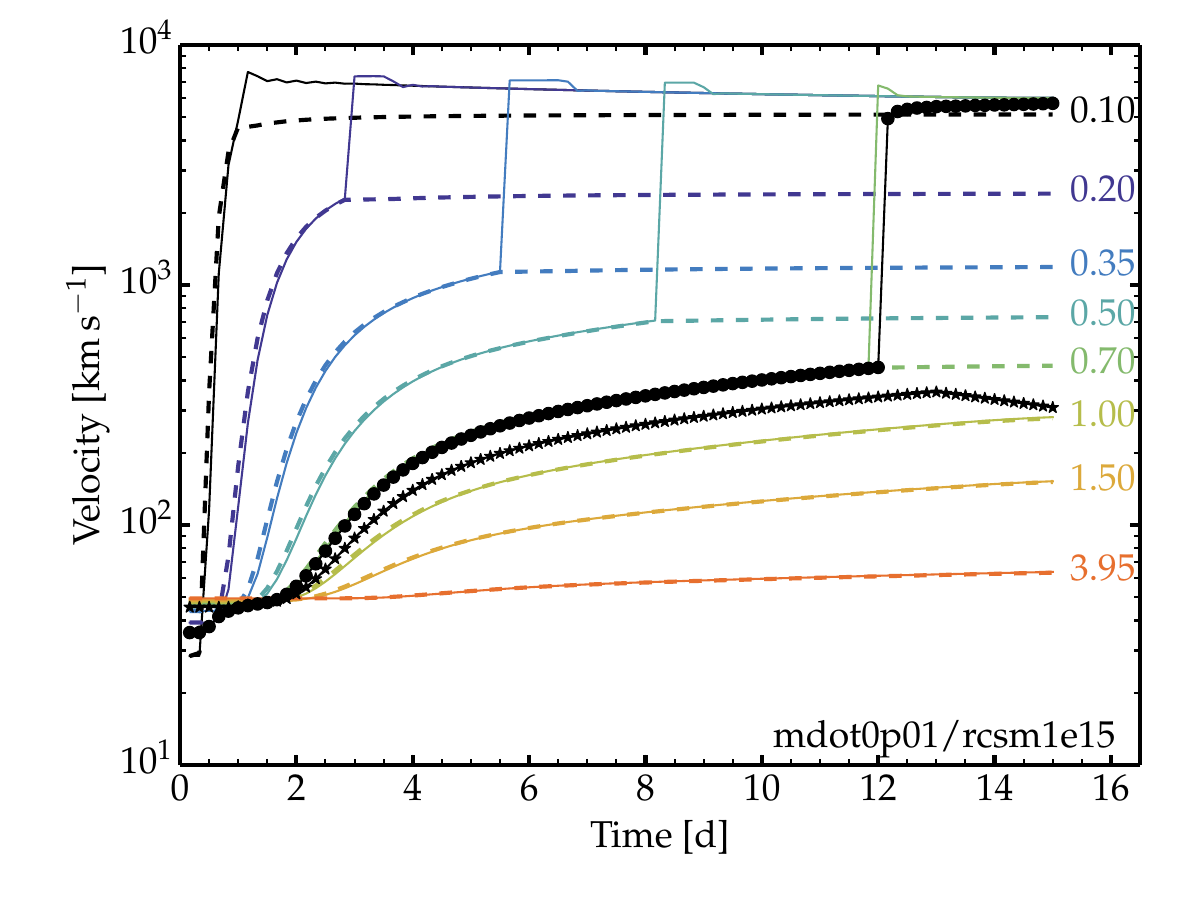}
    \end{subfigure}
    \hfill
    \begin{subfigure}[b]{0.33\textwidth}
       \centering
       \includegraphics[width=\textwidth]{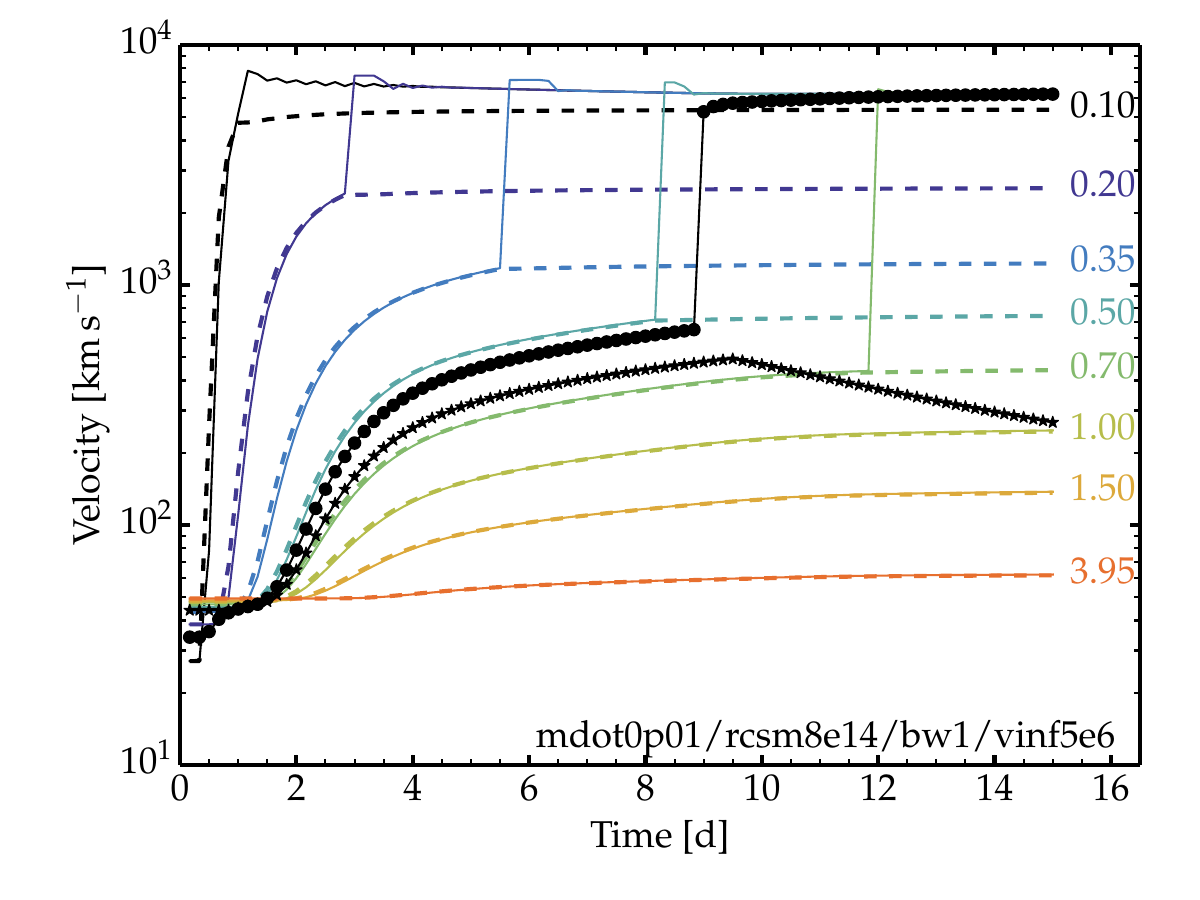}
    \end{subfigure}
    \hfill
    \begin{subfigure}[b]{0.33\textwidth}
       \centering
       \includegraphics[width=\textwidth]{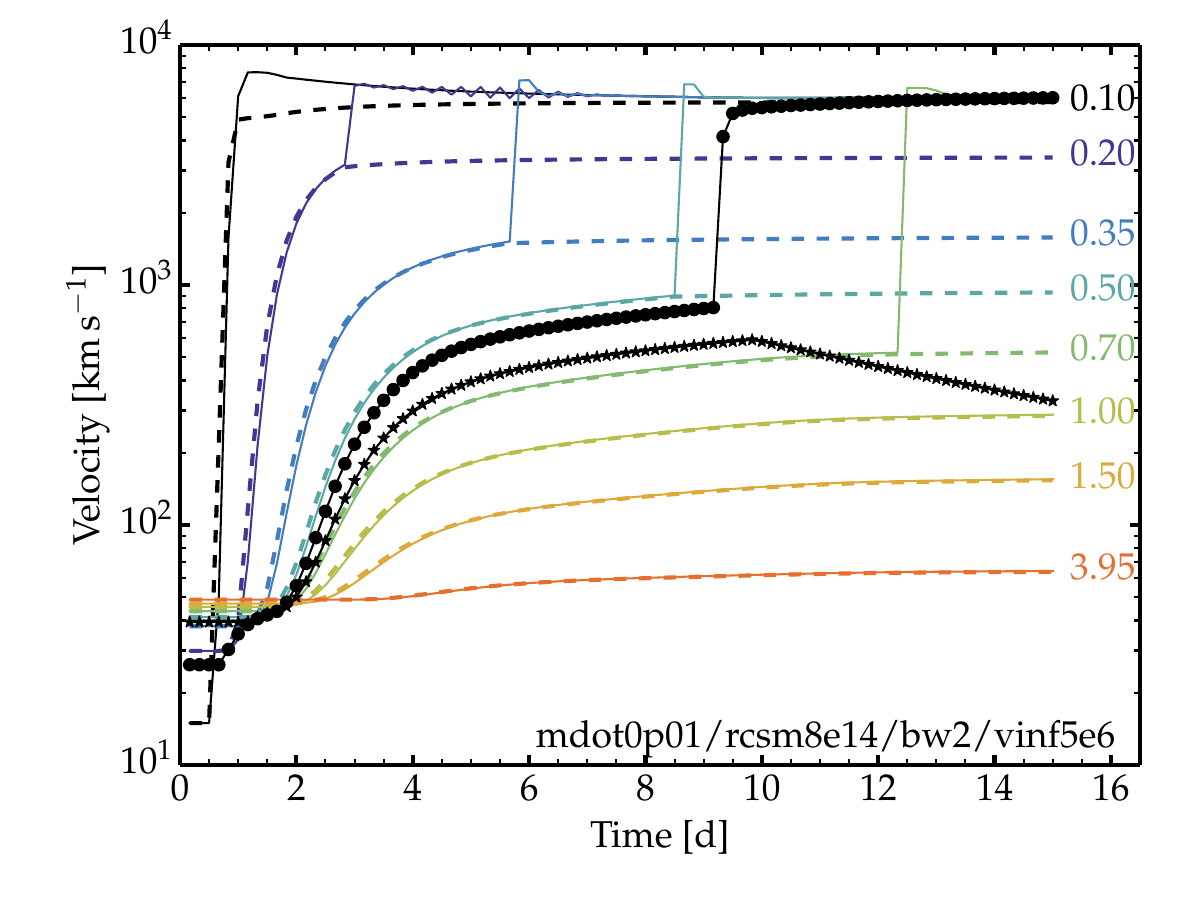}
    \end{subfigure}
    \hfill
    \begin{subfigure}[b]{0.33\textwidth}
       \centering
       \includegraphics[width=\textwidth]{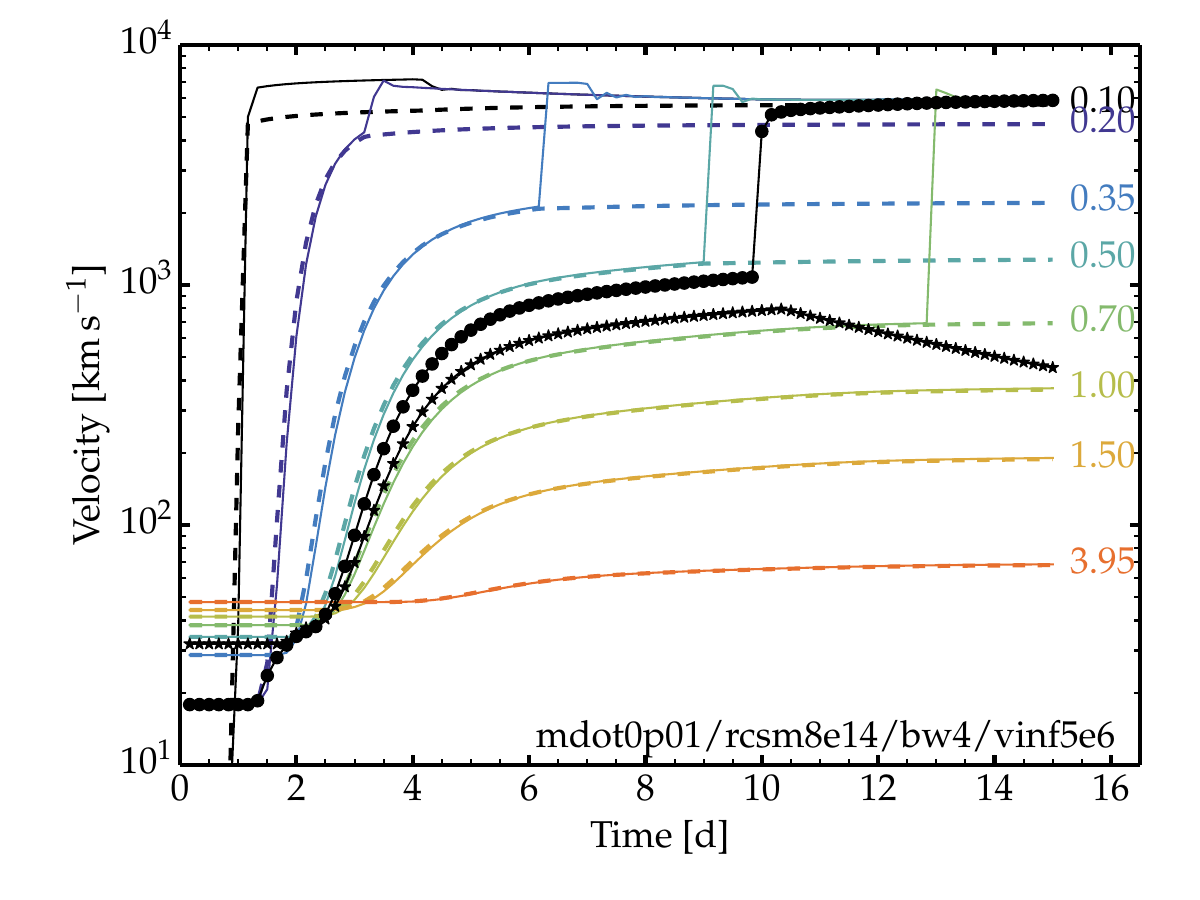}
    \end{subfigure}
    \hfill
    \begin{subfigure}[b]{0.33\textwidth}
       \centering
       \includegraphics[width=\textwidth]{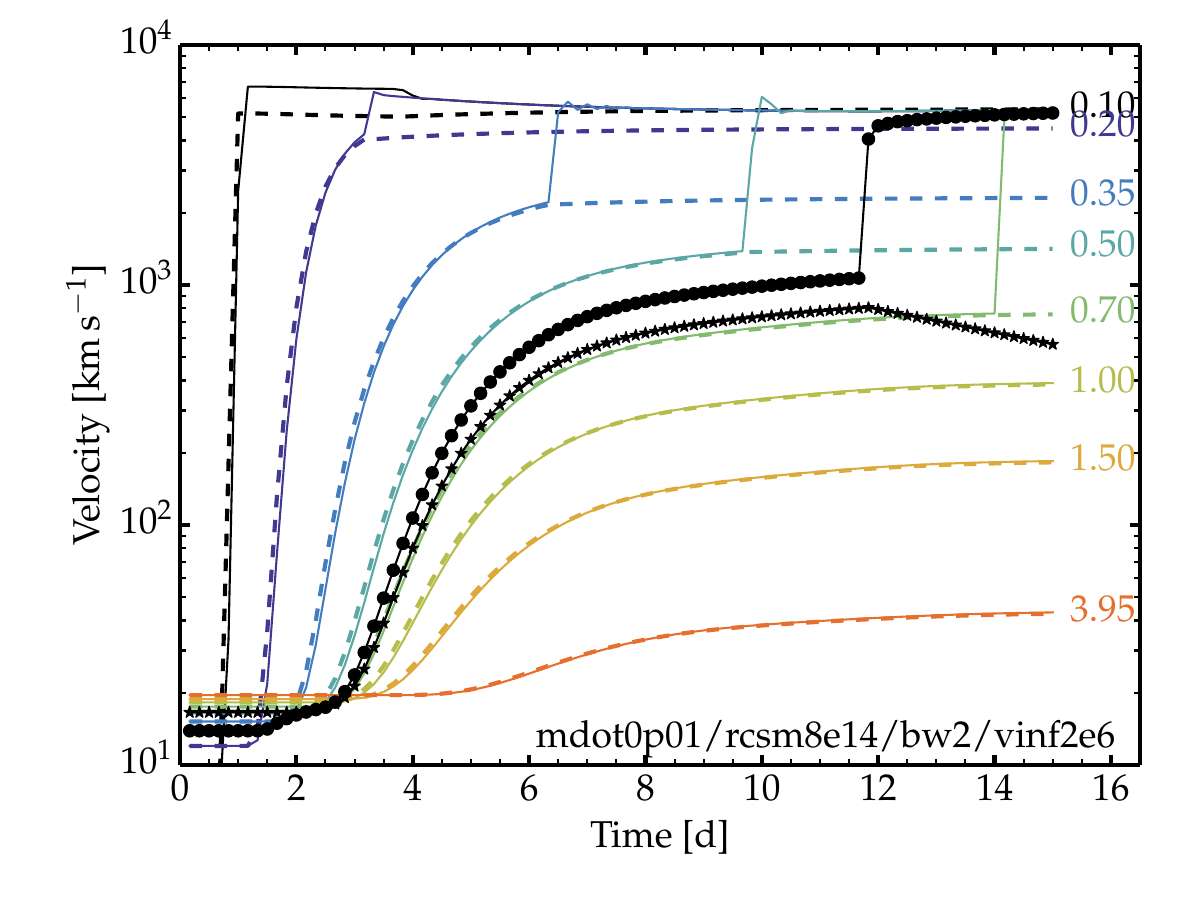}
    \end{subfigure}
    \hfill
    \begin{subfigure}[b]{0.33\textwidth}
       \centering
       \includegraphics[width=\textwidth]{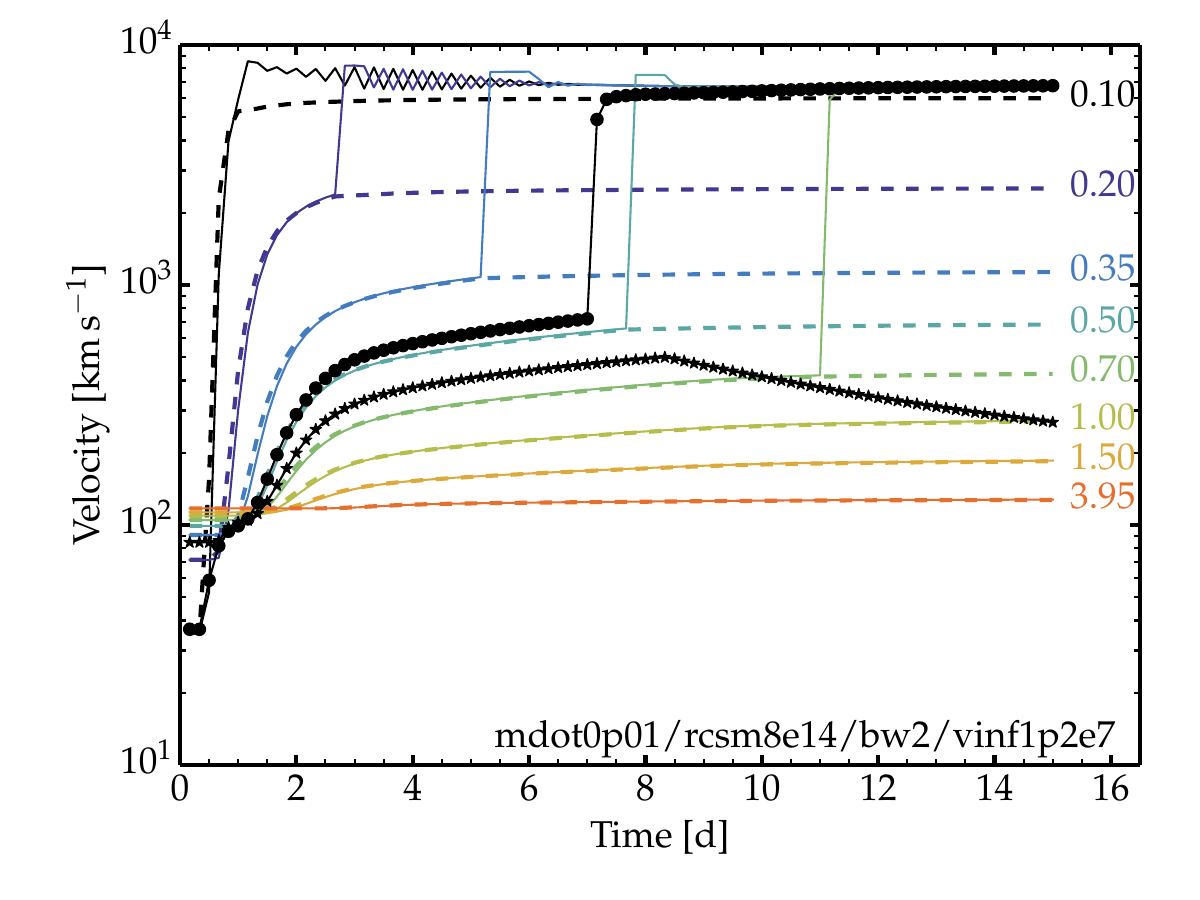}
    \end{subfigure}
    \hfill
    \begin{subfigure}[b]{0.33\textwidth}
       \centering
       \includegraphics[width=\textwidth]{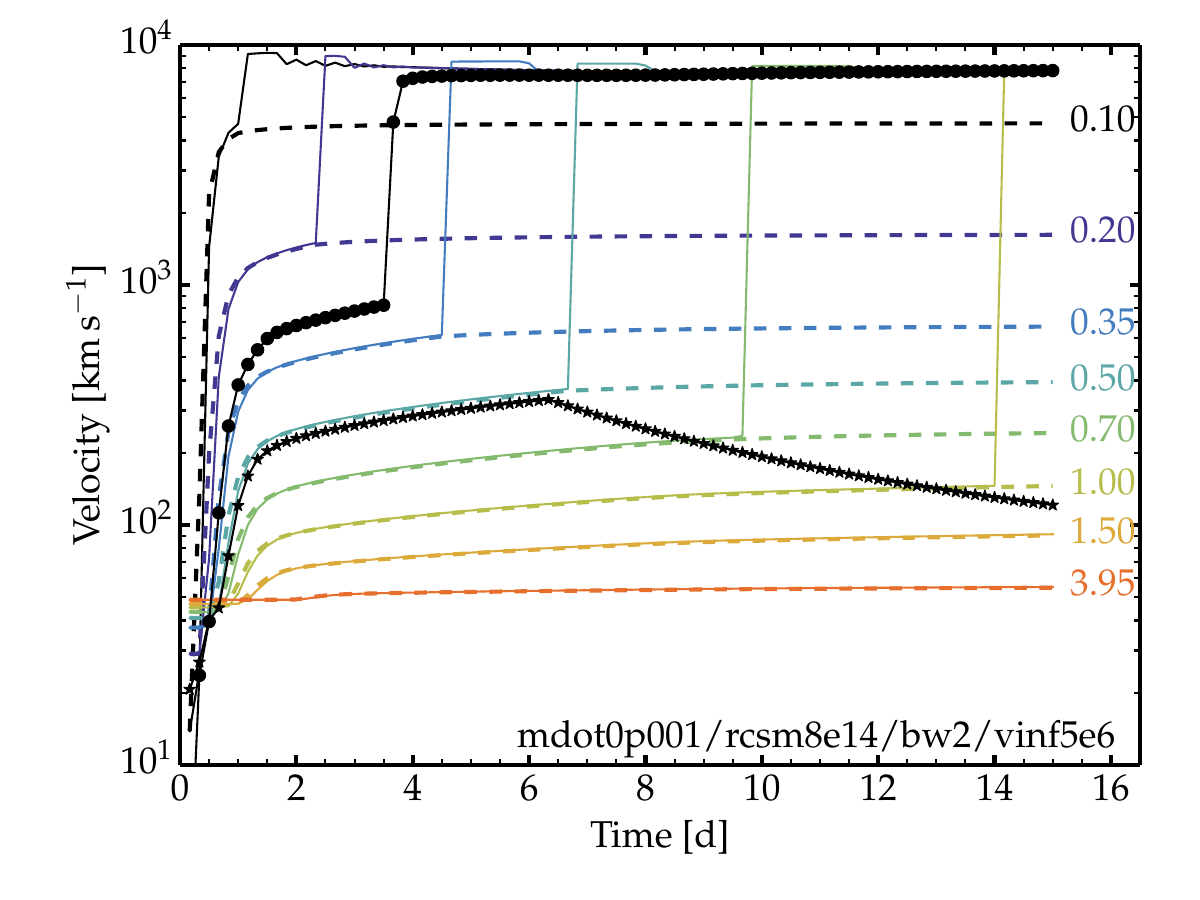}
    \end{subfigure}
    \hfill
    \begin{subfigure}[b]{0.33\textwidth}
       \centering
       \includegraphics[width=\textwidth]{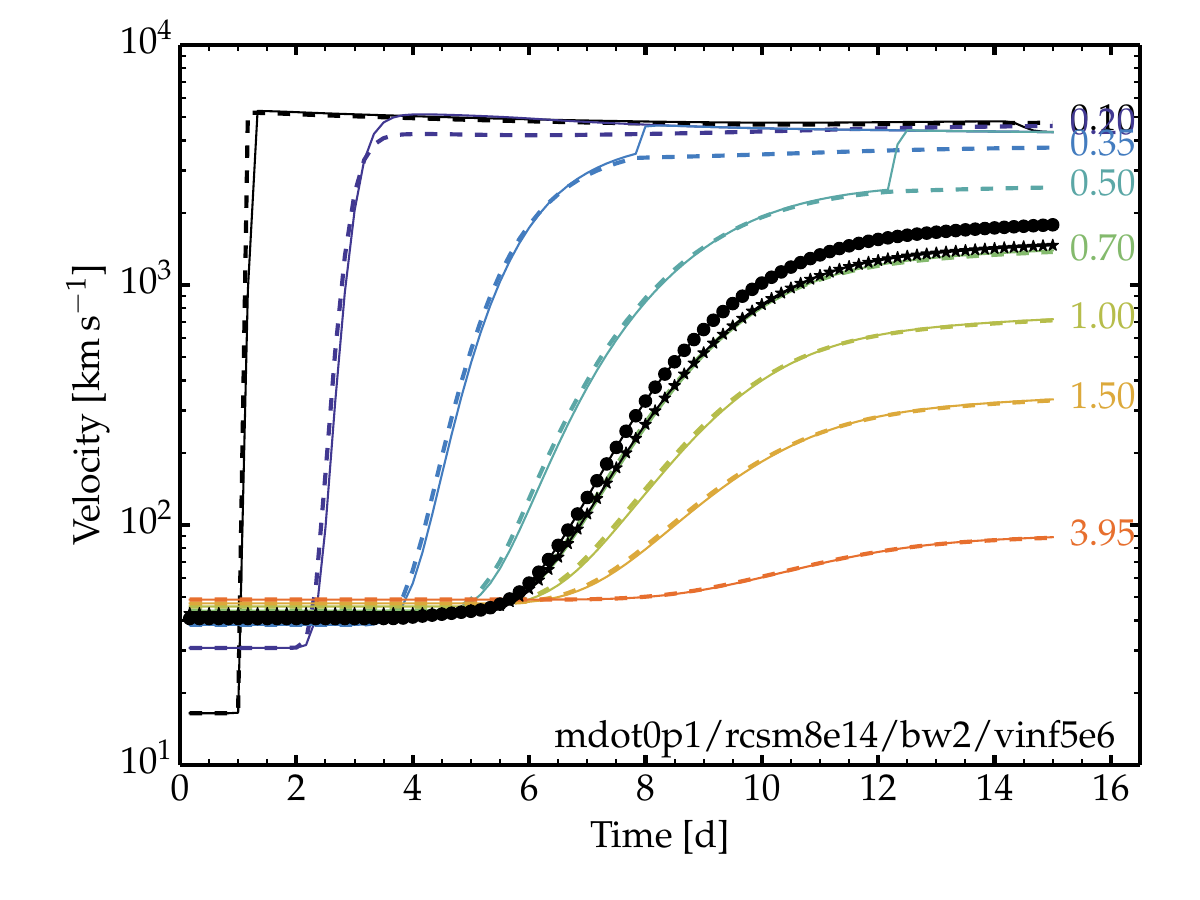}
    \end{subfigure}
\caption{Same as the bottom-left panel of Fig.~\ref{fig_evol_v_at_r} but for the entire model set computed with the radiation hydrodynamics code \heracles. The model paramters are discussed in Section~\ref{sect_init} and in Table~\ref{tab_init}.
\label{fig_evol_v_at_r_all}
}
\end{figure*}

Figure~\ref{fig_evol_v_at_r_all} is a duplicate of the bottom panel of Fig.~\ref{fig_evol_v_at_r} but now showing results for all \heracles\ simulations performed in this study.
We also show additional spectral evolution in the H$\beta$ and He\two\,5411\,\AA\ region for model mdot0p01/rcsm8e14, to complement the information shown in Section~\ref{sect_ref_lines}.

\begin{figure}
\centering
\includegraphics[width=\hsize]{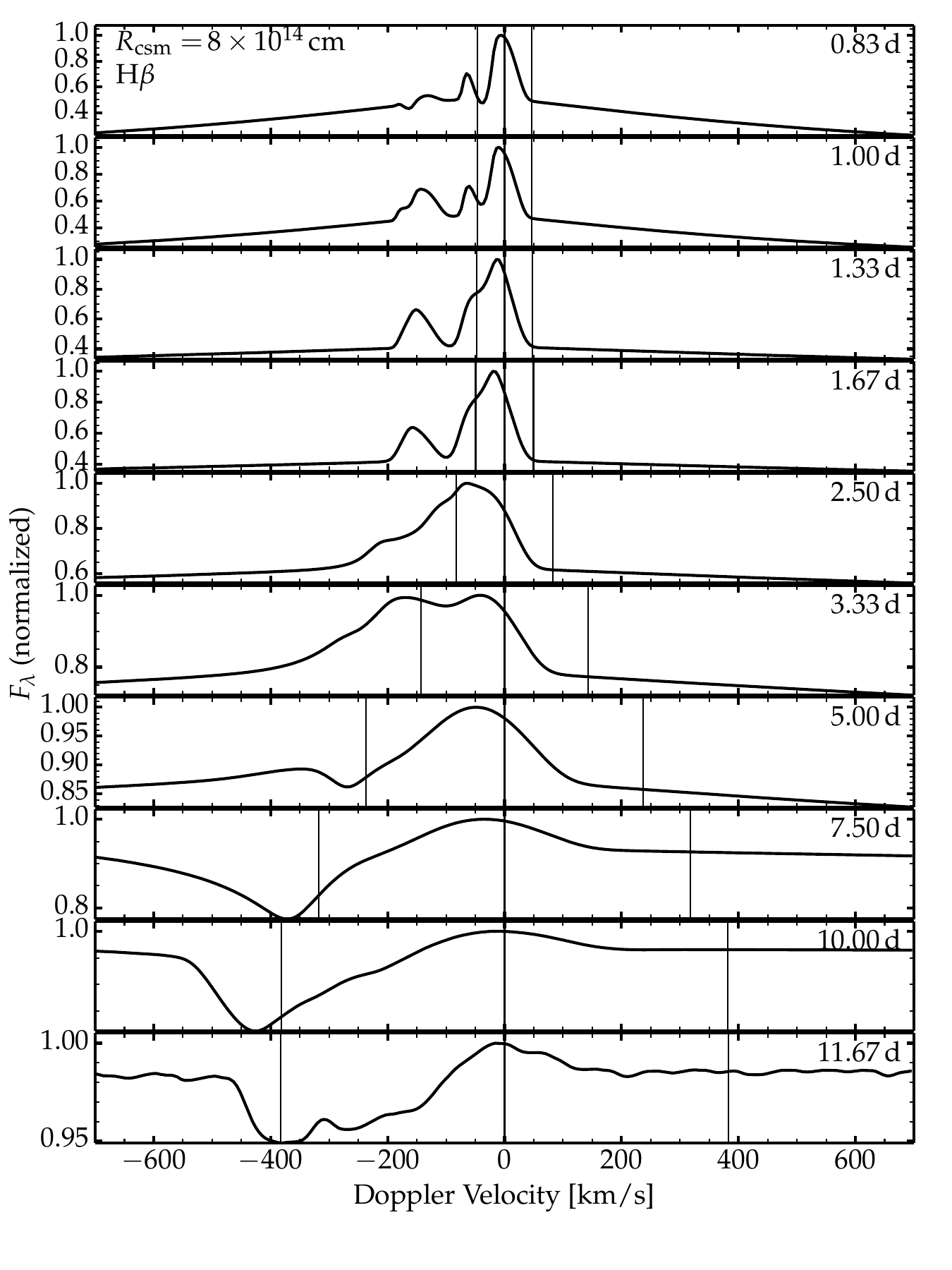}
\caption{Evolution of the spectral region centered on the rest wavelength of H$\beta$ for model mdot0p01/rcsm8e14 (see Fig.~\ref{fig_6562} for the results obtained for H$\alpha$).
\label{fig_4861}
}
\end{figure}

\begin{figure}
\centering
\includegraphics[width=\hsize]{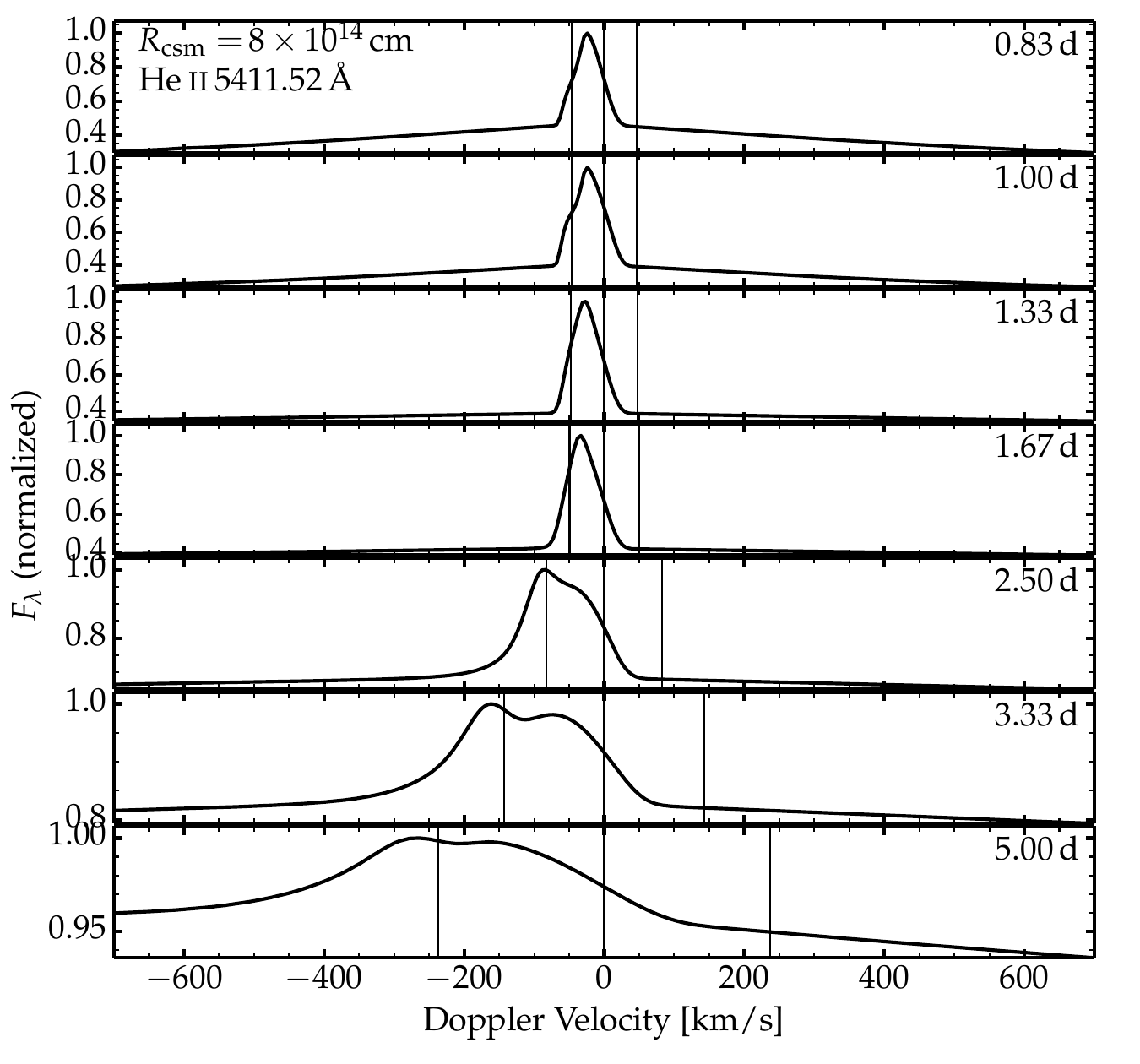}
\caption{Evolution of the spectral region centered on the rest wavelength of He\two\,5411.52\,\AA\ for model mdot0p01/rcsm8e14. This line disappears quickly after 5\,d, and thus earlier than the stronger He\two\,4685.70\,\AA\ shown in Fig.~\ref{fig_4685}. 
\label{fig_5411}
}
\end{figure}

\begin{figure*}
\centering
\includegraphics[width=\hsize]{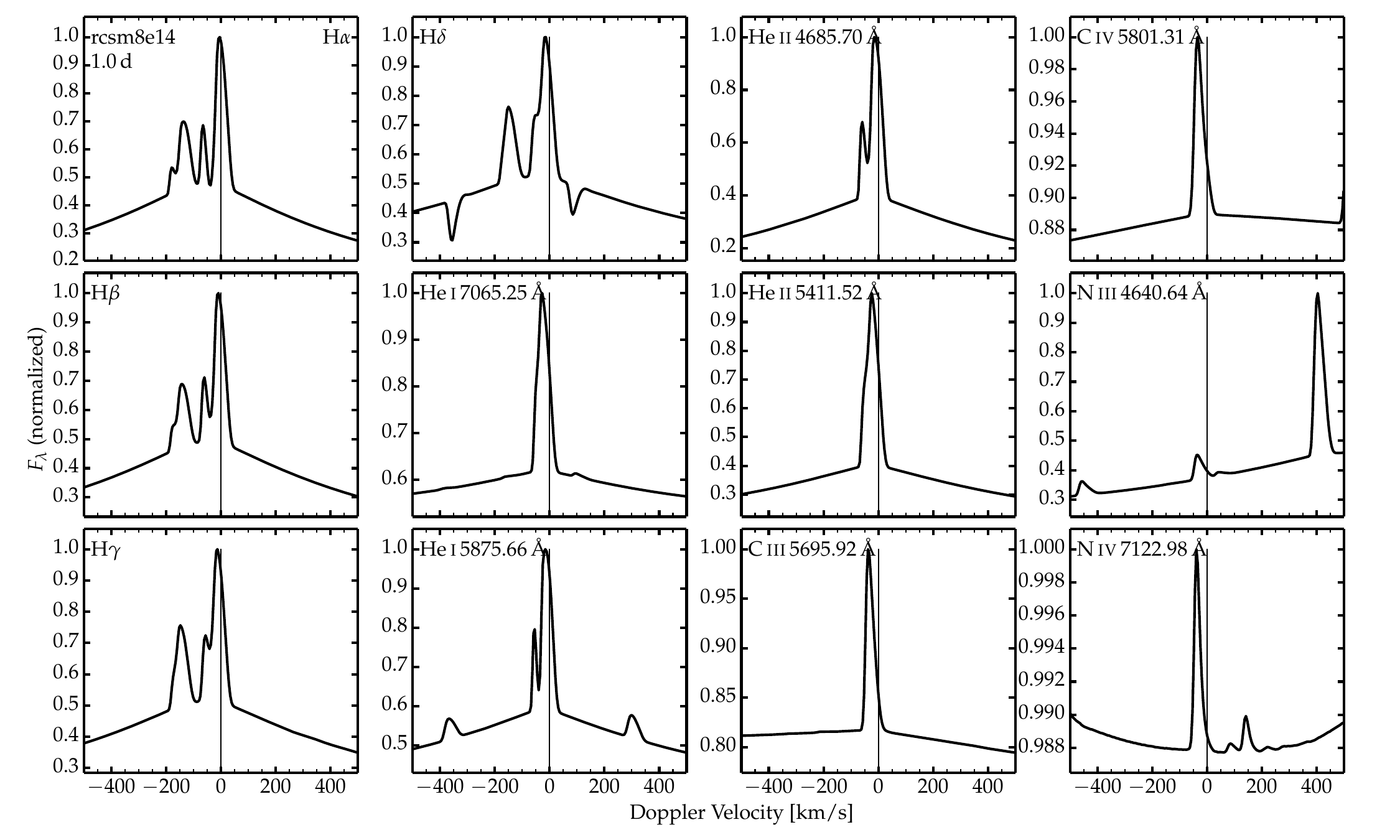}
\caption{Mosaic showing various lines of H\one, He\one, He\two, C\three, C\four, and N\four\ for the reference model mdot0p01/rcsm8e14 at 1\,d. In all cases, the xaxis covers from $\pm$\,500\,\kms\ but the yaxis covers a wide range of normalized fluxes, with very strong narrow emission lines like H$\alpha$ and much weaker ones like N\four\,7122.98\,\AA.
\label{fig_r1w6b_1d}
}
\end{figure*}

\begin{figure}
\centering
\includegraphics[width=\hsize]{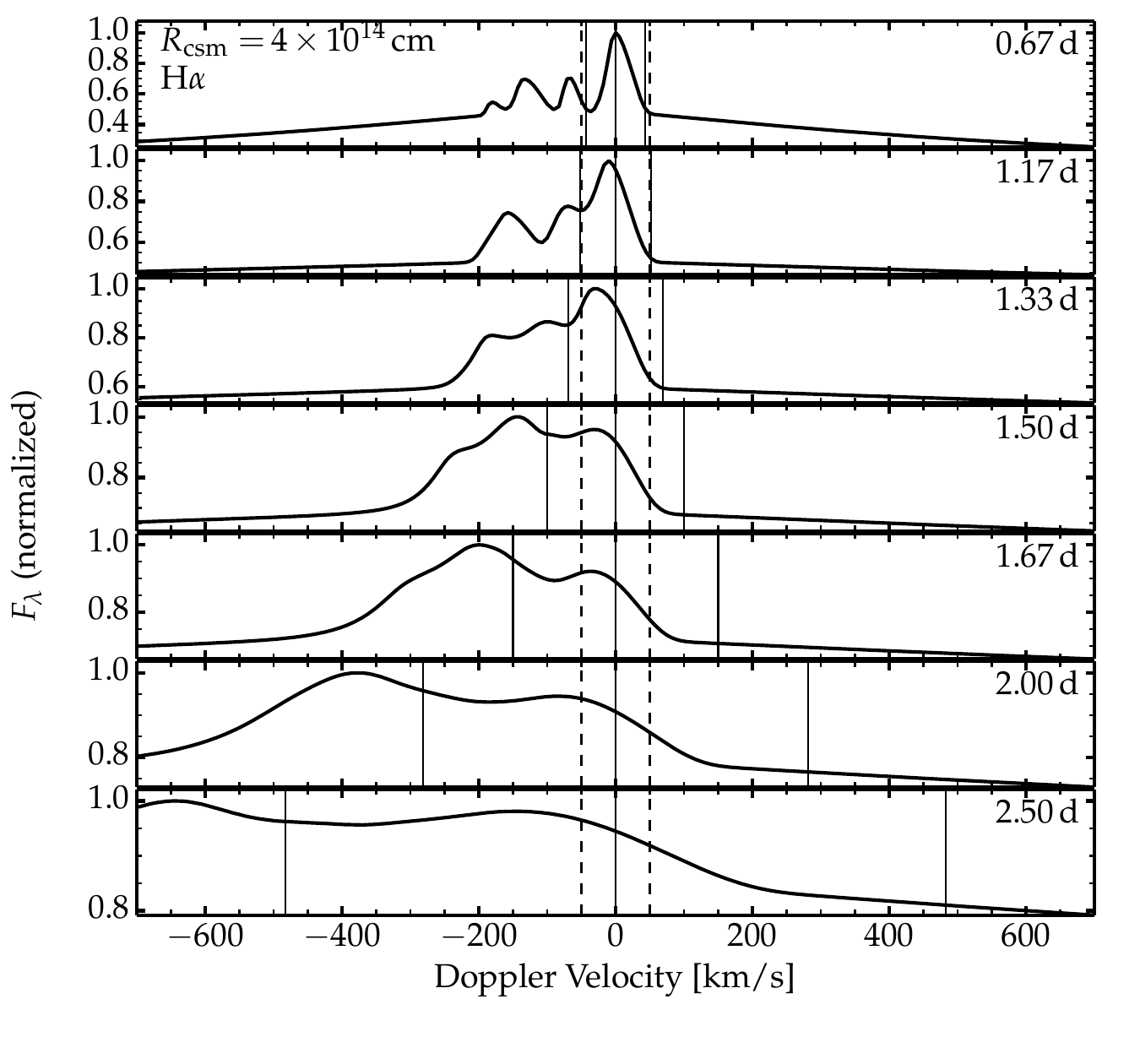}
\caption{Same as Fig.~\ref{fig_rcsm2e14_halpha}, but now for the model counterpart with a greater \rcsm\ value of $4 \times 10^{14}$\,cm.
\label{fig_rcsm4e14_halpha}
}
\end{figure}

\end{document}